\documentclass[12pt]{article}
\usepackage{graphicx}
\usepackage{verbatim}
\usepackage{amsmath,amsthm,amsfonts}
\usepackage{amssymb}

\usepackage{color}
\usepackage{hyperref}

\theoremstyle{definition}

\newtheorem{def-theorem}[theorem]{Definition-Theorem}

\newcommand{\D}{\mathrm{d}}

\newcommand{\AdS}{\mathrm{AdS}}

\DeclareMathOperator{\sech}{sech}

\DeclareMathOperator{\arccot}{arccot}

\DeclareMathOperator{\arctanh}{arctanh}

\DeclareMathOperator{\sign}{sign}

\newcommand{\be}{\begin{equation}}
\newcommand{\ee}{\end{equation}}
\newcommand{\bea}{\begin{eqnarray}}
\newcommand{\eea}{\end{eqnarray}}
\newcommand{\beas}{\begin{eqnarray*}}
\newcommand{\eeas}{\end{eqnarray*}}
\newcommand{\ba}{\begin{array}}
\newcommand{\ea}{\end{array}}

\DeclareMathOperator{\sgn}{sgn}

\begin{document}

\begin{titlepage}
\hfill
\vbox{
    \halign{#\hfil         \cr
           } % end of \halign
      }  % end of \vbox
\vspace*{15mm}
\begin{center}
{\Large \bf Black Hole Microstate Cosmology}

\vspace*{8mm}
\vspace*{1mm}
Sean Cooper${}^a$, Moshe Rozali${}^a$, Brian Swingle${}^b$ \\ Mark Van Raamsdonk${}^a$, Christopher Waddell${}^a$, David Wakeham${}^a$
\vspace*{0.5cm}
\let\thefootnote\relax\footnote{seancooper@phas.ubc.ca, rozali@phas.ubc.ca, bswingle@umd.edu, mav@phas.ubc.ca, \\ cwaddell@phas.ubc.ca, daw@phas.ubc.ca}

{\it ${}^a$ Department of Physics and Astronomy,
University of British Columbia\\
6224 Agricultural Road,
Vancouver, B.C., V6T 1Z2, Canada \\
${}$ \\
${}^b$Condensed Matter Theory Center, Maryland Center for Fundamental Physics,\\
Joint Center for Quantum Information and Computer Science,\\
and Department of Physics, University of Maryland, College Park, MD 20742, USA
}

\vspace*{0.5cm}
%%\maketitle
\end{center}
\begin{abstract}
In this note, we explore the possibility that certain high-energy holographic CFT states correspond to black hole microstates with a geometrical behind-the-horizon region, modelled by a portion of a second asymptotic region terminating at an end-of-the-world (ETW) brane. We study the time-dependent physics of this behind-the-horizon region, whose ETW boundary geometry takes the form of a closed FRW spacetime. We show that in many cases, this behind-the-horizon physics can be probed directly by looking at the time dependence of entanglement entropy for sufficiently large spatial CFT subsystems. We study in particular states defined via Euclidean evolution from conformal boundary states and give specific predictions for the behavior of the entanglement entropy in this case. We perform analogous calculations for the SYK model and find qualitative agreement with our expectations.

A fascinating possibility is that for certain states, we might have gravity localized to the ETW brane as in the Randall-Sundrum II scenario for cosmology. In this case, the effective description of physics beyond the horizon could be a big bang/big crunch cosmology of the same dimensionality as the CFT. In this case, the $d$-dimensional CFT describing the black hole microstate would give a precise, microscopic description of the $d$-dimensional cosmological physics.

\end{abstract}

\end{titlepage}

\tableofcontents

\section{Introduction}

The AdS/CFT correspondence is believed to provide a non-perturbative description of quantum gravity for spacetimes which are asymptotic to Anti-de Sitter space. For a holographic CFT defined on a spatial sphere, typical pure states with large energy expectation value correspond to microstates of a large black hole in AdS. Simple observables in the CFT can be used to probe the exterior geometry of this black hole, revealing the usual AdS Schwarzschild metric with a horizon. However, what lies beyond the horizon for such states and how this is encoded in the CFT is still a significant open question.

Classically, a static (eternal) black hole solution can be extended to include a second full asymptotically AdS region. In this classical picture, the horizon is not distinguished by any local physics, so a conventional expectation is that black hole microstate geometries should include at least some of the behind-the-horizon region from the maximally extended geometry.\footnote{ Some authors have argued that quantum effects should modify these expectations: the ``fuzzball'' proposal \cite{Lunin:2001jy, Lunin:2002qf, Mathur:2005zp, Skenderis:2008qn} suggests that microstate geometries are actually horizonless, while proponents of the ``firewall'' scenario \cite{Almheiri:2013hfa, Almheiri:2012rt} argued that consistency with unitarity and the equivalence principle imply that the geometry must end in some type of singularity at or just beyond the horizon. But many authors have given counter-arguments suggesting a more conventional picture.} On the other hand, including the full second asymptotic region is tantamount to introducing the degrees of freedom of a second CFT, so it is very plausible that single-CFT microstate geometries have at most a part of the second asymptotic region in common with the maximally extended spacetime.

\begin{figure}
\centering
\includegraphics[width=50mm]{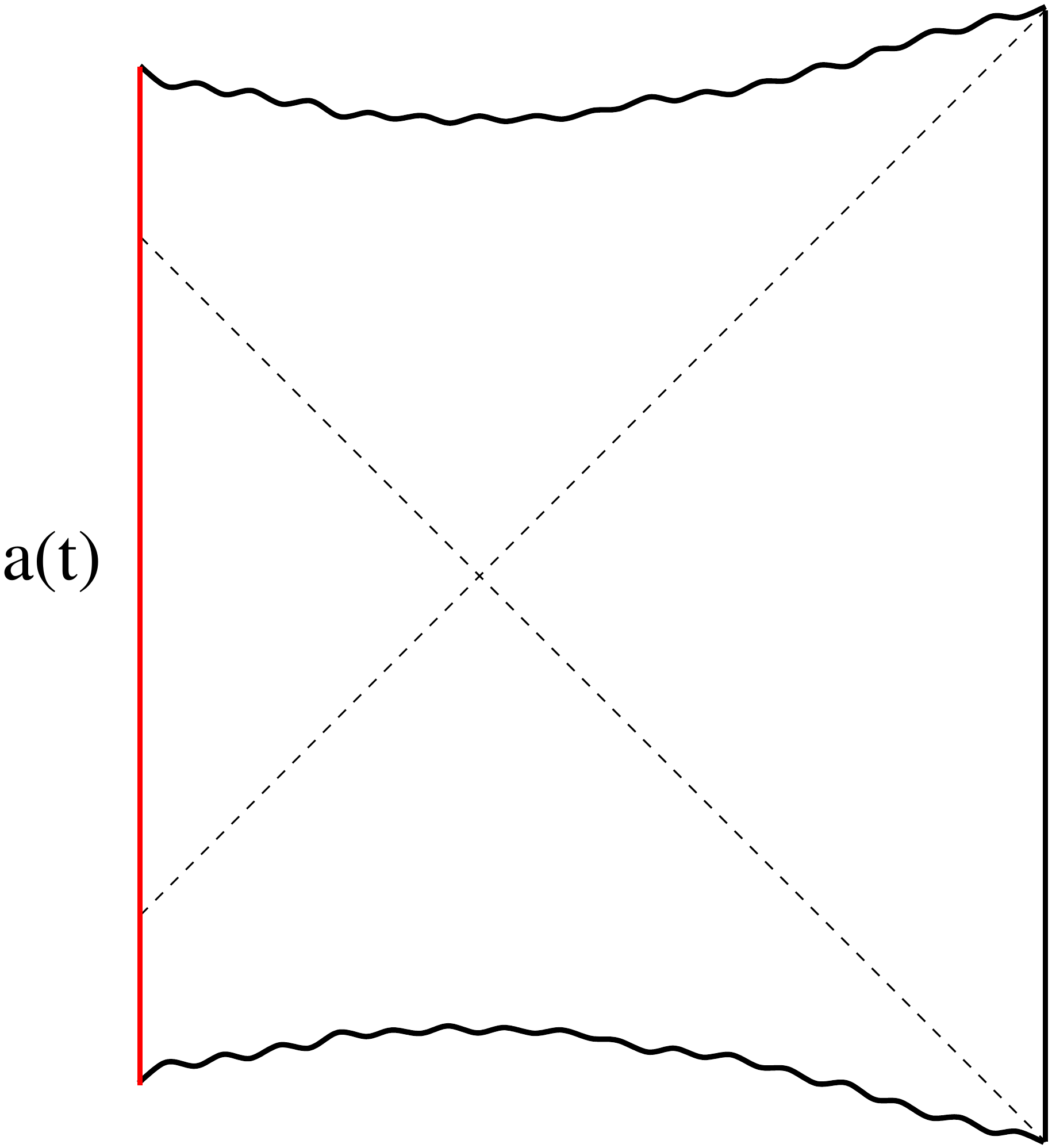}
\caption{Penrose diagram for spacetimes associated with certain black hole microstates. The spacetime terminates on the left with an effective end-of-the-world brane (shown in red on the left) whose worldvolume geometry is a four-dimensional FRW big bang/big crunch cosmology. For certain brane trajectories, the physics of the left region would correspond to a Randall-Sundrum II cosmology, with gravity localized on the brane. If there are CFT states that realize this scenario, the CFT would provide a complete microscopic description of this cosmology.}
\label{fig:ETW}
\end{figure}

In this paper, following \cite{Kourkoulou:2017zaj} and \cite{Almheiri:2018ijj}, we will explore the possibility that for certain CFT states, the corresponding black hole geometry is captured by the Penrose diagram in Figure \ref{fig:ETW}.\footnote{The recent paper \cite{deBoer:2018ibj} that appeared during the course of our work also considered black hole microstate geometries, describing a picture somewhat different from the one in Figure \ref{fig:ETW}. However, \cite{deBoer:2018ibj} were discussing typical black hole microstates, while we are focusing on more specific states, so there is no conflict.} Here, the geometry on the right side is the AdS-Schwarzschild black hole exterior. On the left, instead of the full second asymptotic region that would be present in the maximally extended black hole geometry, we have a finite region terminating on an end-of-the-world (ETW) brane (shown in red in Figure \ref{fig:ETW}). In the microscopic description, this brane could involve some branes from string/M theory theory or could correspond to a place where the spacetime effectively ends due to a degeneration of the internal space (as in a bubble of nothing geometry \cite{Witten:1981gj}). In this note we mainly make use of a simple effective description of the ETW brane, which we describe in detail below.

In order to decode the physics of these microstate spacetimes from the microscopic CFT state, we need to understand the CFT description of physics behind the black hole horizon. This is a notoriously difficult problem; the present understanding is that decoding local physics behind the horizon requires looking at extremely complicated operators in the CFT and furthermore that the operators needed depend on the particular CFT state being considered \cite{Papadodimas:2012aq, Papadodimas:2013jku, Harlow:2014yoa, Papadodimas:2015jra}.\footnote{For recent discussions of state dependence and bulk reconstruction of black hole interiors from the quantum error correction perspective, see \cite{Hayden:2018khn, Almheiri:2018xdw}.}

\begin{figure}
\centering
\includegraphics[width=80mm]{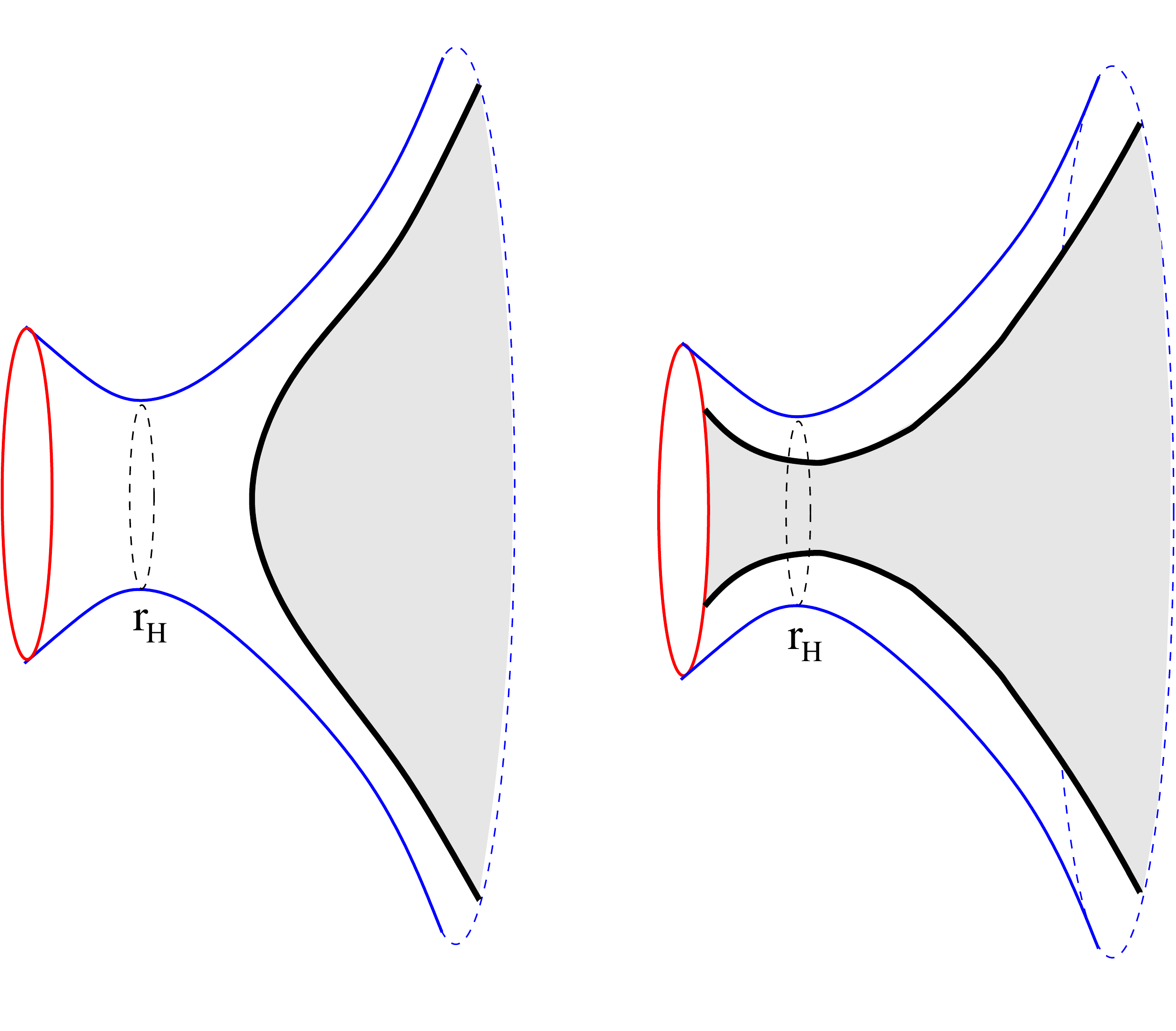}
\caption{Two possibilities for extremal surfaces and associated entanglement wedges (shaded) for ball-shaped boundary regions. The extremal surface on the right has the topology of $S^{d-2}$ times an interval, so is connected for $d > 2$.}
\label{fig:EEchoices}
\end{figure}

Fortunately, we will see that in many cases, entanglement entropy in the CFT can probe the geometry behind the horizon, and in particular can be used to inform us about the effective geometry of the ETW brane. To understand this, recall that for holographic theories, the entanglement entropy for a spatial region in the CFT corresponds to the area in the corresponding geometry of the minimal area extremal surface homologous to the region \cite{Ryu:2006bv, Rangamani:2016dms}. In the geometry of Figure \ref{fig:ETW}, we have extremal surfaces that remain outside the black hole horizon and extremal surfaces that penetrate the horizon and end on the ETW brane, as shown in Figure \ref{fig:EEchoices}. We find that if the black hole is sufficiently large, the behind-the-horizon region is not too large, and the CFT region is large enough, the extremal surfaces penetrating the horizon can have the minimal area for some window of boundary time $[-t_E,t_E]$, where $t_E$ depends on the size of the region being  considered. During this time, the entanglement entropy is time-dependent and directly probes the geometry of the ETW brane. This was observed for a simple case in \cite{Hartman2013}.\footnote{Various other works have considered the entanglement entropy in black hole geometries with a time-dependent exterior, such as the Vaidya geometry (see, for example, \cite{AbajoArrastia:2010yt}). In these cases, the entanglement entropy can also probe behind the horizon.}

Our investigations were motivated by the work of \cite{Kourkoulou:2017zaj} in the context of the SYK model, a simple toy model for AdS/CFT. Here, Kourkoulou and Maldacena argued that for states $e^{-\beta H}|B \rangle$ arising via Euclidean evolution of states $|B \rangle$ with limited entanglement, the corresponding $AdS_2$ black hole microstate take a form similar to that shown in Figure \ref{fig:ETW}. This work was generalized to CFTs in \cite{Almheiri:2018ijj}, where the states $|B \rangle$ were taken to be conformally invariant boundary states of the CFT.\footnote{The states $e^{-\beta H}|B \rangle$ in this case have been considered in the past by Cardy and collaborators \cite{Cardy:2015xaa}, \cite{Cardy:2017ufe} as time-dependent states used to model quantum quenches.} In that case, the corresponding geometries were deduced by making use of a simple ansatz discussed by Karch and Randall \cite{Karch:2000gx}, and by Takayanagi \cite{Takayanagi:2011zk} for how to holographically model conformally invariant boundary conditions in CFTs. The resulting geometries again take the form shown in Figure \ref{fig:ETW}, with the trajectory of the ETW brane depending on properties of the CFT boundary state. We review the construction of these states and their corresponding geometries in section 2, generalizing the calculations to higher dimensions. We make use of this particular set of geometries for our detailed calculations since they are simple to interpret holographically, but we expect that the qualitative picture of Figure \ref{fig:ETW} should hold in a more complete holographic treatment of Euclidean-time-evolved CFT boundary states, and perhaps for a more general class of states.

\begin{figure}
\centering
\includegraphics[width=100mm]{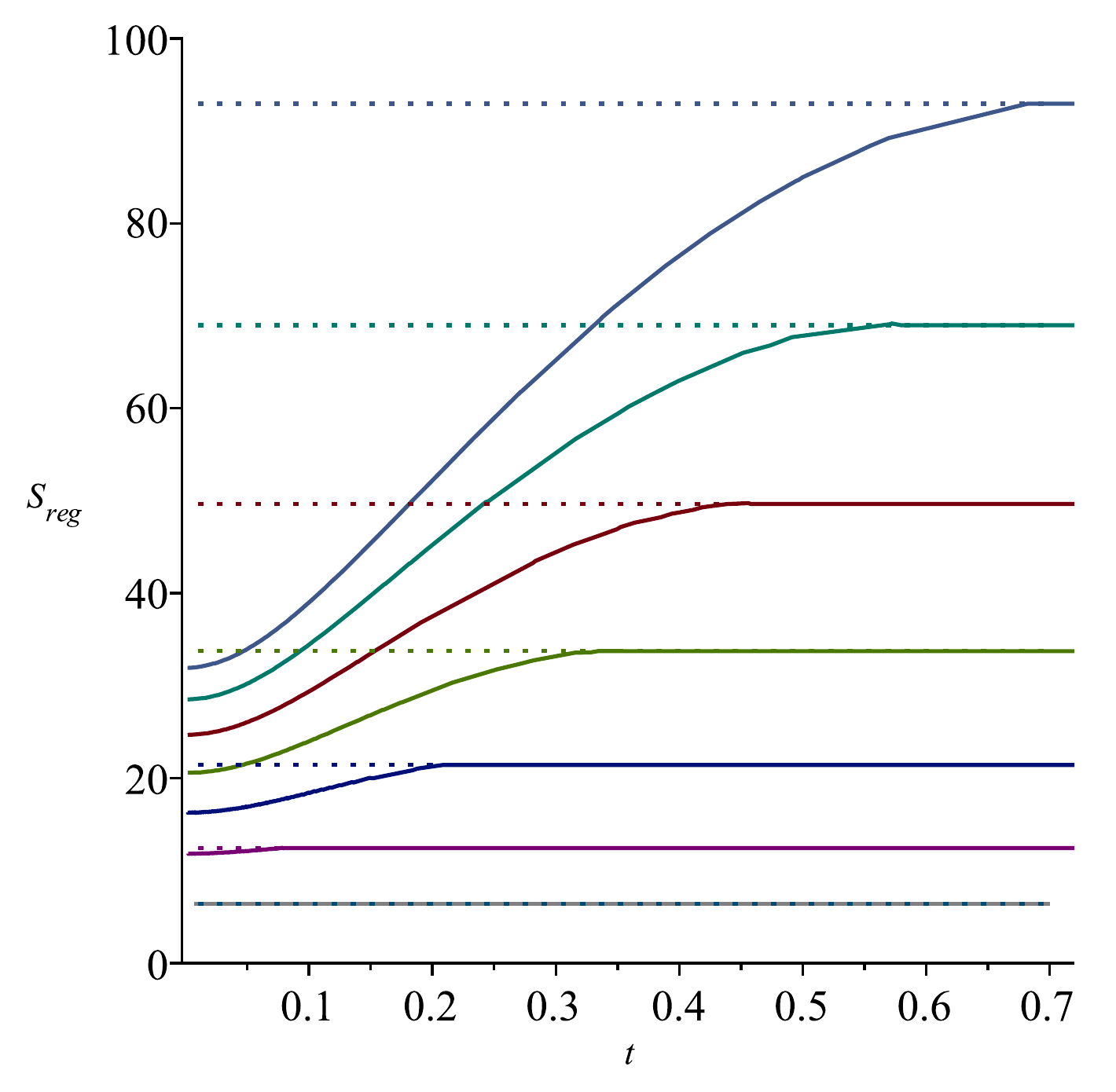}
\caption{Time-dependence of subsystem entanglement entropy for a five-dimensional black hole microstate modeled by a constant tension ETW brane behind the horizon. Curves from bottom to top correspond to successively larger ball-shaped subsystems on the sphere. For large enough subsystems, the minimal area extremal surfaces probe behind the horizon for an interval of time.}
\label{fig:EE5}
\end{figure}

Our calculations of entanglement entropy for these states are described in detail in section 3. As an example of the results, figure \ref{fig:EE5} shows the entanglement entropy for ball-shaped regions in a particular five-dimensional black hole geometry with constant-tension ETW brane behind the horizon. For small subsystems or late times, the RT surfaces stay outside the horizon and the entanglement entropy is time-independent. However, for large enough subsystems, there is an interval of time where the minimal-area extremal surfaces probe behind the horizon and end on the ETW brane. Thus, the entanglement entropy gives a direct probe of behind-the-horizon physics.

The ansatz of Karch/Randall/Takayanagi, in which boundaries in the asymptotic region are extended into the bulk along a dynamical ETW brane of a fixed tension, is the simplest proposal that reproduces expected properties of boundary CFT entanglement entropy via a holographic calculation. For specific microstates of specific CFTs, the detailed microstate geometry is more complicated and the ETW brane will have a more specific microscopic description, but it is plausible that the qualitative picture is similar. Thus, our results for the behavior of entanglement entropy using the simple ansatz can be viewed as a prediction for the qualitative behaviour of entanglement entropy in actual Euclidean-time-evolved boundary states of holographic CFTs. This can be tested by direct calculation for specific states; obtaining results similar to the ones we find based on the above described simple ansatz would provide a check that our general picture is viable.

As a warm-up for such a direct test, we perform an analogous calculation in a generalization of the SYK model, a coupled-cluster model which includes both all-to-all within-cluster interactions and spatially local between-cluster interactions. Here, the states we consider are analogs of those of \cite{Kourkoulou:2017zaj} extended to include the physics of spatial locality, where in place of the boundary state $|B \rangle$, we have states which are eigenstates of a collection of spin operators formed from pairs of fermions. We numerically calculate the entanglement entropy as a function of time for subsets of various numbers of fermions (as a model for CFT spatial regions on varying size) for a single SYK cluster and for two coupled SYK clusters. We find that the dependence of entanglement entropy on time and on the fraction of the system being considered is qualitatively similar to our predictions for holographic CFT states (compare figure \ref{fig:syk_S_time} with figure \ref{fig:EE5}), but (as expected) without the sharp features observed in the holographic case. We also give analytical large-$N$ arguments that apply to many clusters, where direct numerical calculation is not possible. These calculations are described in detail in section 4.

It is noteworthy that imaginary time-evolved product states have also been considered in the condensed matter literature. For example, they were proposed as tools to efficiently sample from thermal distributions of spin chains. In that context, they were named minimally entangled typical thermal states (METTS), with the expectation that they would be only lightly entangled~\cite{White:2009metts,Stoudenmire:2010metts}. Interestingly, we find that such states are generically highly entangled, unlike what was seen for simple gapped spin chains~\cite{White:2009metts,Stoudenmire:2010metts}. One can argue that the low entanglement observed in the finite-size gapped spin chain occurs because of the strong microscopic-scale energy gap. To better understand the holographic and SYK results in some simple models, and with this quantum matter background in mind, we also give some additional results for spin/qubit models in Appendices~\ref{app:entgrowth} and \ref{app:sqrt}.

We also consider in section 5 the calculation of holographic complexity \cite{Susskind:2014rva, Stanford:2014jda, Brown:2015bva} (both the action and volume versions). These provide additional probes of the behind-the-horizon physics, though their CFT interpretation is less clear. We find interesting differences in behavior between the action and volume versions. While both show the expected linear growth at late times, the volume-complexity increases smoothly from the time-symmetric point $t=0$, while the action-complexity has a phase transition that separates the late-time growth from an earlier period where the action-complexity is constant.

In section 6, we point out a Rindler analog of our construction in 2+1 dimensions, where the maximally extended black hole geometry is replaced with empty AdS space divided into complementary Rindler wedges and the microstates are particular states of a CFT on a half-sphere with BCFT boundary conditions.
Since the BTZ geometry is obtained as a quotient of pure $\text{AdS}_3$, we can unwind the compact direction and reuse the results of section 3 to determine when knowledge of a boundary subsystem grants access to the region behind the Rindler horizon.

\subsection*{Black hole microstate cosmology}

An interesting feature of the geometries we consider is that the geometry on the left side can be thought of as an asymptotically AdS spacetime (the second asymptotic region of the maximally extended geometry) cut off by a UV brane. This is reminiscent of the Randall-Sundrum II scenario for brane-world cosmology. In that case, we have gravity localized on the brane; that is, the physics on the brane can be described (in the case where the full spacetime is $d+1$-dimensional) over a large range of scales by $d$-dimensional gravity coupled to matter.\footnote{Via another application of the AdS/CFT correspondence, some of the matter, dual to the gravitational physics in the partial second asymptotic region, should be described by a cutoff $d$-dimensional conformal field theory.}

Whether or not we have an effective four-dimensional description for physics in the second asymptotic region will depend on the details of the microstate geometry, in particular on the size of the black hole relative to the AdS scale and to the ETW brane trajectory. These in turn depend on the details of the state we are considering. If there exist states for which the conditions for localized gravity are realized, the effective description of the physics beyond the black hole horizon would correspond to $d$-dimensional FRW cosmology, where the evolution of the scale factor corresponds to the evolution of the proper size of the ETW brane in the full geometry. This evolution corresponds to an expanding and contracting FRW spacetime which classically starts with a big bang and ends with a big crunch, though we expect that the early and late time physics does not have a good $d$-dimensional description.

Since the states we are describing are simply specific high-energy states in our original CFT, the original CFT should provide a complete microscopic description of this cosmological physics. A very optimistic scenario is that for the right choice of four-dimensional CFT (or other non-conformal holographic theory) and black hole microstate, the effective four-dimensional description of the dynamics of the ETW brane could match with the cosmology in our universe. In this case, the CFT itself could be supersymmetric\footnote{Perhaps it could even be ${\cal N} = 4$ supersymmetric Yang-Mills Theory.}; the effective theory on the ETW brane will be related to the choice of state in the CFT and need not have unbroken supersymmetry. The small cosmological constant would be explained by having a large central charge in the CFT together with some properties of the CFT state we are considering.

Even if the relevant cosmologies turn out not to be realistic, it is intriguing that CFTs could provide a microscopic description of interesting cosmological spacetimes, since the usual applications of AdS/CFT describe spacetimes whose asymptotics are static.\footnote{There have been many other approaches to describing cosmological physics using holography. For examples, see \cite{Banks:2001px,  Banks:2018ypk, Strominger:2001pn, Alishahiha:2004md, Freivogel:2005qh, McFadden:2009fg}.} Understanding how to generalize AdS/CFT to provide a non-perturbative formulation of quantum gravity in cosmological situations is among the most important open questions in the field, so it is very interesting to explore whether the scenario we describe can be realized in microscopic examples.

In section 7, we give a more detailed review of Randall-Sundrum II cosmology and the conditions for localizing gravity. We then explore whether these conditions can be met in the simple class of geometries with a constant tension ETW brane. Our analysis suggests that realizing the localized cosmology requires considering a black hole which is much larger than the AdS scale, and an ETW brane tension that is sufficiently large. Unfortunately, while the Lorentzian geometries corresponding to these parameters are sensible, our analysis in section 2 suggests that for CFT states corresponding to these parameter values, a different branch of solutions for the dual gravity solution may be preferred. However, a more complete holographic treatment for the BCFT physics will be required in order to reach a more decisive conclusion.

Finally, in section 8, we comment on various possible generalizations and future directions.

\section{Microstates with behind-the-horizon geometry}

In this section, we describe a specific class of CFT excited states which describe certain black hole microstates when the CFT is holographic. For these states, it is possible to plausibly describe the full black hole geometry, at least approximately. These states were suggested and studied  in the context of the SYK model by \cite{Kourkoulou:2017zaj}, and later studied directly in the context of holographic CFTs in \cite{Almheiri:2018ijj}. Simple specific examples of these states and the corresponding geometries have been discussed earlier, for example in \cite{Louko1999b, Maldacena2001, Hartman2013}. The microstate geometries will be time-dependent and hence ``non-equilibrium''; for a different construction of non-equilibrium microstates with geometry behind the horizon, see \cite{Papadodimas:2017qit}.
 In this section, we will review and generalize those discussions, starting with the definition of the CFT states and then moving to the geometrical interpretation. We will make use of this specific construction in the remainder of the paper in order to have an example where we can do explicit calculations.

\subsection{CFT states}

The states we consider, suggested in \cite{Kourkoulou:2017zaj}, have two equivalent descriptions. First, consider the thermofield double state of two CFTs (on $S^d$) which we will call the left and right CFTs,
\be
|\Psi^\beta_{TFD} \rangle = {1 \over Z_\beta} \sum e^{-\beta E_i \over 2} |E_i \rangle_L \otimes |E_i \rangle_R \; .
\ee
For high enough temperatures, this corresponds to the maximally extended AdS-Schwarzschild black hole geometry. Now consider projecting this state onto some particular pure state $|B \rangle$ of the left CFT.  This could be the result of measuring the state on the left. We will be more specific about the pure state $|B \rangle$ later on. The result is a pure state of the right CFT given by
\be
|\hat{\Psi}^\beta_{B} \rangle = {1 \over Z_\beta} \sum e^{-\beta E_i \over 2} \langle B |E_i \rangle  |E_i \rangle \; .
\ee
We can think of this state as the result of measuring the state of the left CFT. If this measurement corresponds to looking at the state of local (UV) degrees of freedom, we might expect that the effects on the corresponding geometry propagate inwards causally (forward and backward, since we will be considering time-symmetric states) from near the left boundary, so that the geometry retains a significant portion of the second asymptotic region. This motivates considering states $|B \rangle$ with no long-range entanglement.

We can also consider a closely related state $|\Psi_{B}^\beta \rangle$ obtained by complex conjugation of the coefficients in the superposition,
\bea
|\Psi^\beta_{B} \rangle &=& {1 \over Z_\beta} \sum e^{-\beta E_i \over 2} \langle E_i |B \rangle  |E_i \rangle \cr
&=& {1 \over Z_\beta} \sum e^{-\beta E_i \over 2}   |E_i \rangle \langle E_i |B \rangle \cr
&=& {1 \over Z_\beta} e^{- \beta H / 2} |B \rangle
\; .
\label{EucEv}
\eea
We recall that the operation $|\hat{\Psi}^\beta_{B}\rangle  \to |\Psi^\beta_{B} \rangle $ is anti-linear and anti-unitary and corresponds to the operation of time-reversal. For example, given any Hermitian ${\cal O}$ we have that
\be
\langle \Psi^\beta_{B} (t) | {\cal O} |\Psi^\beta_{B} (t) \rangle = \langle \hat{\Psi}^\beta_{B} (-t) | {\cal O} |\hat{\Psi}^\beta_{B} (-t) \rangle \; .
\ee
In our case, we will consider states which are time-reversal symmetric, so the two definitions are equivalent.

\begin{figure}
\centering
\includegraphics[width=40mm]{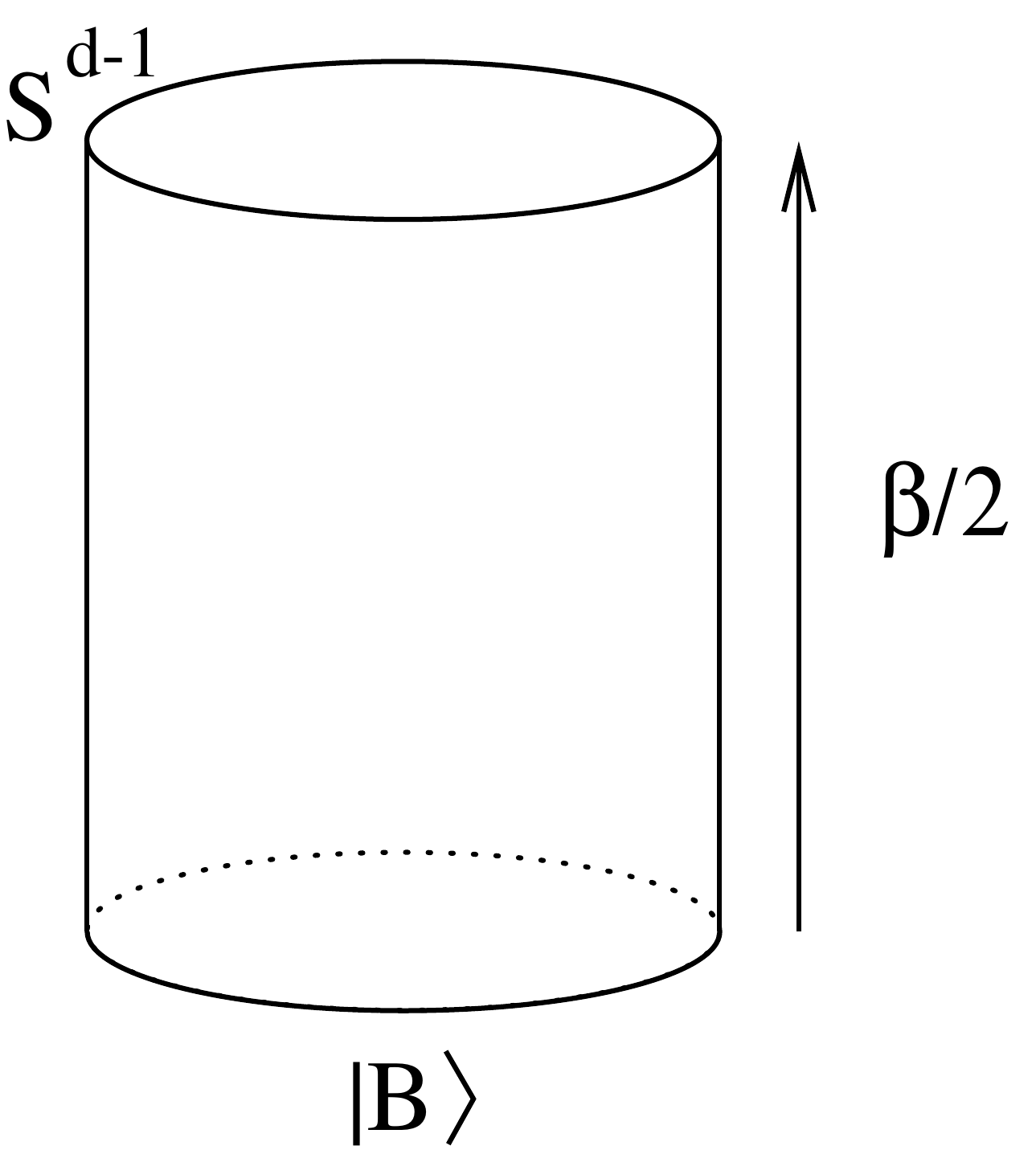}
\caption{Path integral description of black hole microstates $|\Psi^\beta_{B} \rangle$.}
\label{fig:PathInt}
\end{figure}

We see from (\ref{EucEv}) that the states $|\Psi^\beta_{B} \rangle$ correspond to starting from a state $|B \rangle$ and having a finite amount of Euclidean evolution. These states are naturally defined by a Euclidean path integral as shown in Figure \ref{fig:PathInt}. Since the CFT path integral for holographic theories maps onto the gravity path integral, we will be able to make use of the AdS/CFT correspondence to deduce the corresponding geometries if we can choose states $|B \rangle$ for which we can understand a gravity prescription for dealing with the boundary condition at the initial Euclidean time.

\vspace{5pt}
\subsubsection*{Euclidean evolution of CFT boundary states}

In the CFT context, a nice class of states to consider for the states $|B \rangle$ are certain {\it boundary states} of the CFT, as suggested in \cite{Kourkoulou:2017zaj} and explored in detail in \cite{Almheiri:2018ijj}. For any CFT, we can ask whether it is possible to define the theory on a manifold with boundary. In general, there will be a family of distinct theories corresponding to different allowed boundary conditions. Some of these boundary conditions are special in the sense that they preserve some of the conformal symmetry of the theory; specifically, the vacuum state of the CFT on a half space with such a boundary condition would preserve $\mathrm{SO}(d-1,2)$ of the $\mathrm{SO}(d,2)$ conformal symmetry.

For each of these allowed boundary conditions, we can associate a boundary state $|B \rangle$ for the CFT on $S^{d-1}$ by saying that choosing this state in (\ref{EucEv}) is equivalent to the state obtained from the Euclidean path integral with our chosen boundary condition at $\tau = - \beta/2$. The boundary state itself (equal to $|\Psi^\beta_{B} \rangle$ in the limit $\beta \to 0$) is singular and has infinite energy. It also can be understood to have no long range entanglement, as we motivated above \cite{Miyaji:2014mca}. However, the Euclidean evolution suppresses the high-energy contributions to give a state with finite energy. The states $|\Psi^\beta_{B} \rangle$ are generally time-dependent and were considered by Cardy and collaborators in studying quantum quenches \cite{Cardy:2015xaa, Calabrese:2016xau, Cardy:2017ufe}.

For our purposes, the boundary states are interesting since now the description of our states is completely in terms of a Euclidean path integral with a specific boundary condition for the CFT at $\tau = -\beta/2$.

\subsection{Holographic model}

\begin{figure}
  \centering
  \vspace{10pt}
  \includegraphics[scale=0.55]{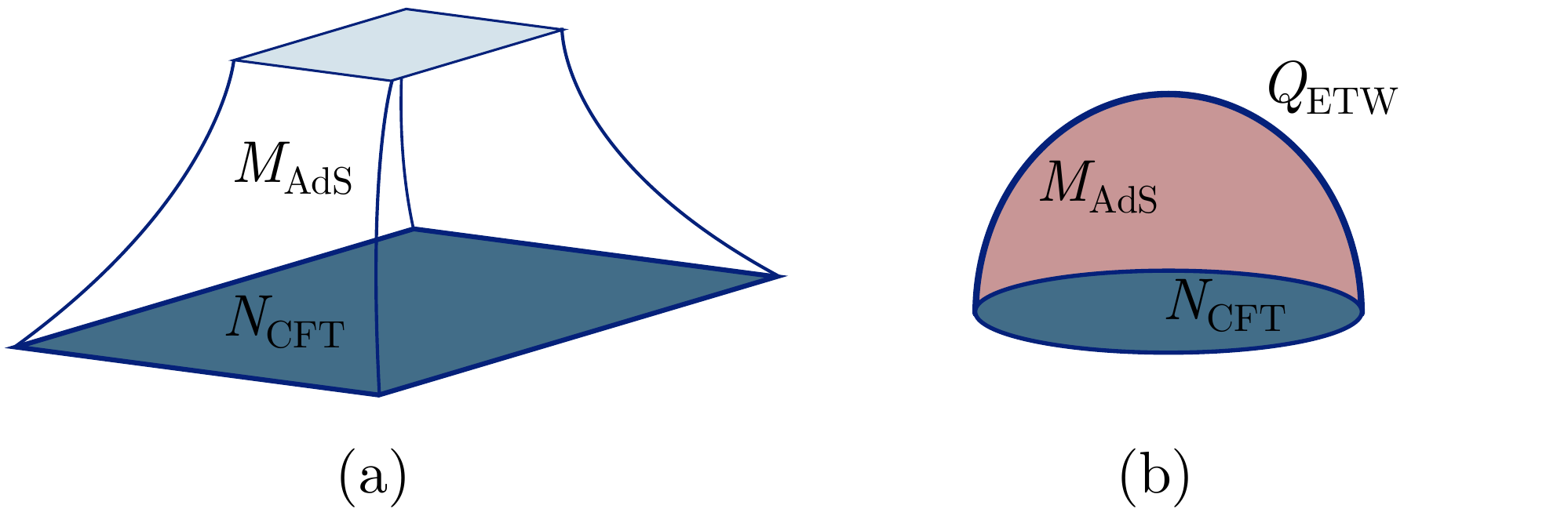}
  \caption{(a) The AdS/CFT correspondence, with an asymptotically AdS bulk $M_\text{AdS}$
    and an asymptotic boundary $N_\text{CFT} = \partial
    M_\text{AdS}$. (b) The AdS/BCFT correspondence. We add a boundary
    to the CFT, whose holographic ``image'' is the ETW brane $Q$.}
  \label{ads-bcft}
\end{figure}

In \cite{Karch:2000gx} and \cite{Takayanagi:2011zk, Fujita:2011fp}, these boundary conditions were discussed in the context of AdS/CFT. These references proposed that the gravitational dual for a CFT with boundary should be some asymptotically AdS spacetime with a dynamical IR boundary that forms an extension of the CFT boundary into the bulk, as depicted in Figure \ref{ads-bcft}. For simplicity, the physics of this boundary was modelled by an end-of-the-world brane with constant tension, and a Neumann boundary condition ensuring that no energy/momentum flows through the brane. A refined proposal for how to treat the boundary conditions was presented recently in \cite{Miao2017}, but for the cases we consider, the proposals are equivalent.

It is convenient to introduce a dimensionless tension parameter $T$ defined so that the stress-energy tensor on the ETW brane is
\be
\label{defT}
8 \pi G T_{ab} = (1-d) T g_{ab}/L_\mathrm{AdS} \; ,
\ee
where $T$ can be positive or negative. The parameter $T$ is related to properties of the boundary state; we will review the physical significance of this parameter in the CFT below. The gravitational action including bulk and boundary terms is then given as
\be
 I_\text{bulk} + I_\text{ETW} = \frac{1}{16\pi G}\int_{N_{AdS}} \D^{d+1}x\, \sqrt{-g}(R-2\Lambda) + \frac{1}{8\pi G}\int_{Q_\text{ETW}} \D^{d-1}y\, \sqrt{-h} (K - (d-1)T)\,.
\ee
With this simple model, various expected properties of boundary CFT were shown to be reproduced via gravity calculations. In \cite{Takayanagi:2011zk} and \cite{Fujita:2011fp}, the boundary conditions were taken as spatial boundary conditions for a CFT on an interval or strip, but we can apply the same model in our case with a past boundary in Euclidean time.

For general holographic BCFTs, we expect that the boundary action would be more complicated; it could include general terms involving intrinsic and extrinsic curvatures, sources for various bulk fields, and additional fields localized to the boundary. However, for this this paper, we will focus on studying the simple one-parameter family of models as proposed in \cite{Karch:2000gx, Takayanagi:2011zk}.

\subsubsection*{Relation between tension and boundary entropy in 1+1 dimensions}

The significance of the tension parameter $T$ may be understood most simply for the case of 1+1 dimensional conformal field theories. In that case, each conformally invariant boundary condition may be characterized by a parameter $g$ that can be understood as a boundary analogue of the central charge \cite{Cardy:1989ir,Affleck:1991tk}. We can define $g$ by
\be
g = \langle 0 | B \rangle
\ee
which has the interpretation of the disk partition function, computed with the boundary conditions associated with $|B \rangle$. Along boundary RG flows (defined by deforming a BCFT by some boundary operator), the parameter $g$ always decreases \cite{Friedan:2003yc}. This parameter $g$ also appears in the expression for the vacuum entanglement entropy for the CFT on a half line \cite{Calabrese:2004eu}. The entanglement entropy for an interval of length $L$ including the boundary is given in general by
\be
S(L) = {c \over 6} \log\left({L \over \epsilon} \right) + \log(g) \; .
\ee
Here, the second term is known as the {\it boundary entropy} and in general can have either sign.

Using the holographic prescription, Takayanagi computed both the disk partition function and the entanglement entropy for intervals on a half line, showing that in both cases, the holographic calculation matches with the CFT result if the tension parameter is related to the boundary entropy by
\be
\log g = {L_\mathrm{AdS} \over 4 G} {\rm arctanh}(T) \; .
\ee
Thus, larger values of the tension correspond to larger boundary entropy, or more degrees of freedom associated with the boundary. We expect that this qualitative relationship also holds in higher dimensions.

\begin{figure}
\centering
\includegraphics[width=90mm]{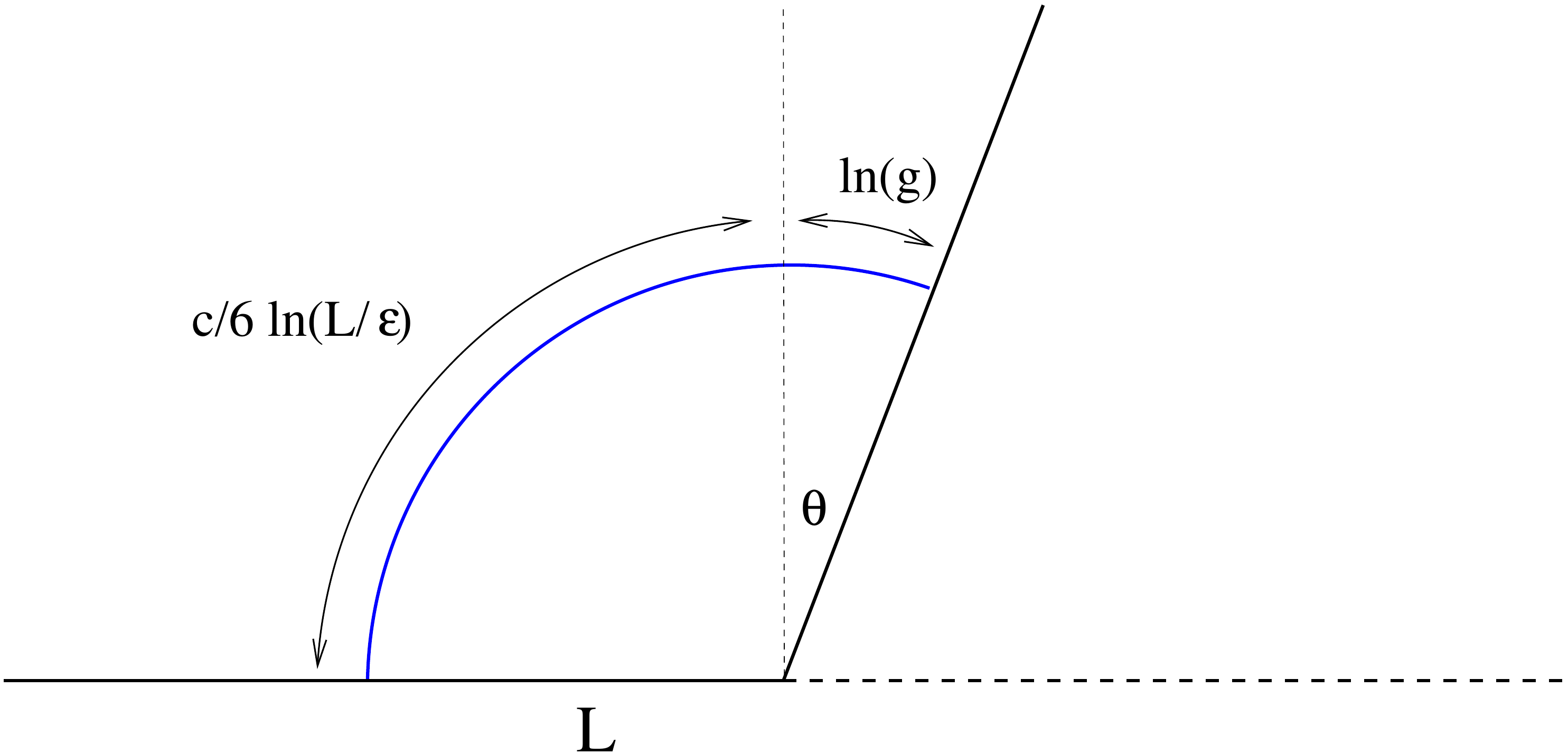}
\caption{Calculation of entanglement entropy for an interval of length $L$ including the boundary in the vacuum state of a holographic BCFT. The geometry is locally Poincar\'e-AdS, with the ETW brane at a constant angle $\theta = \arcsin(T)$. The boundary entropy is the $x > 0$ portion of the RT area.}
\label{fig:BoundEnt}
\end{figure}

Geometrically, the tension parameter $T$ determines the angle at which the ETW brane intersects the boundary, via $T = \sin(\theta)$; this also holds in higher dimensions \cite{Fujita:2011fp}. As an example, Figure \ref{fig:BoundEnt} depicts the calculation of entanglement entropy for an interval of including the boundary in the vacuum state of a holographic BCFT.

\vspace{5pt}
\subsection{Microstate geometries from Euclidean-time-evolved boundary states}

We now make use of the simple holographic BCFT recipe to deduce the microstate geometries associated with Euclidean-time-evolved boundary states
\be
|\Psi \rangle = e^{- \tau_0 H} | B \rangle.
\ee
This was already carried out for 1+1 dimensional CFT states in \cite{Almheiri:2018ijj}. We review their calculations and generalize to higher dimensions.

We are considering a CFT on a spatial $S^{d-1}$ with the state prepared by a Euclidean path integral with boundary conditions in the Euclidean past at $\tau = -\tau_0$. We would like to work out a Lorentzian geometry dual to our state. We start by noting that $t=0$ correlators in our state $|\Psi^{\tau_0}_B \rangle$ may be computed via the Euclidean path integral on $S^{d-1}$ times an interval of Euclidean time $\tau \in [-\tau_0, \tau_0]$, with operators inserted at $\tau = 0$. Holographically, this can be computed using the extrapolate dictionary as a limit of bulk correlators in a Euclidean geometry with boundary $S^{d-1} \times [-\tau_0, \tau_0]$ that is determined by extremizing the gravitational action with appropriate boundary terms for the ETW brane. This geometry is time-reversal symmetric. To find the Lorentzian geometry associated with our state, we take the $\tau = 0$ bulk slice as the initial data for our Lorentzian solution (which will also be time-reversal symmetric).

\begin{figure}
\centering
\includegraphics[width=50mm]{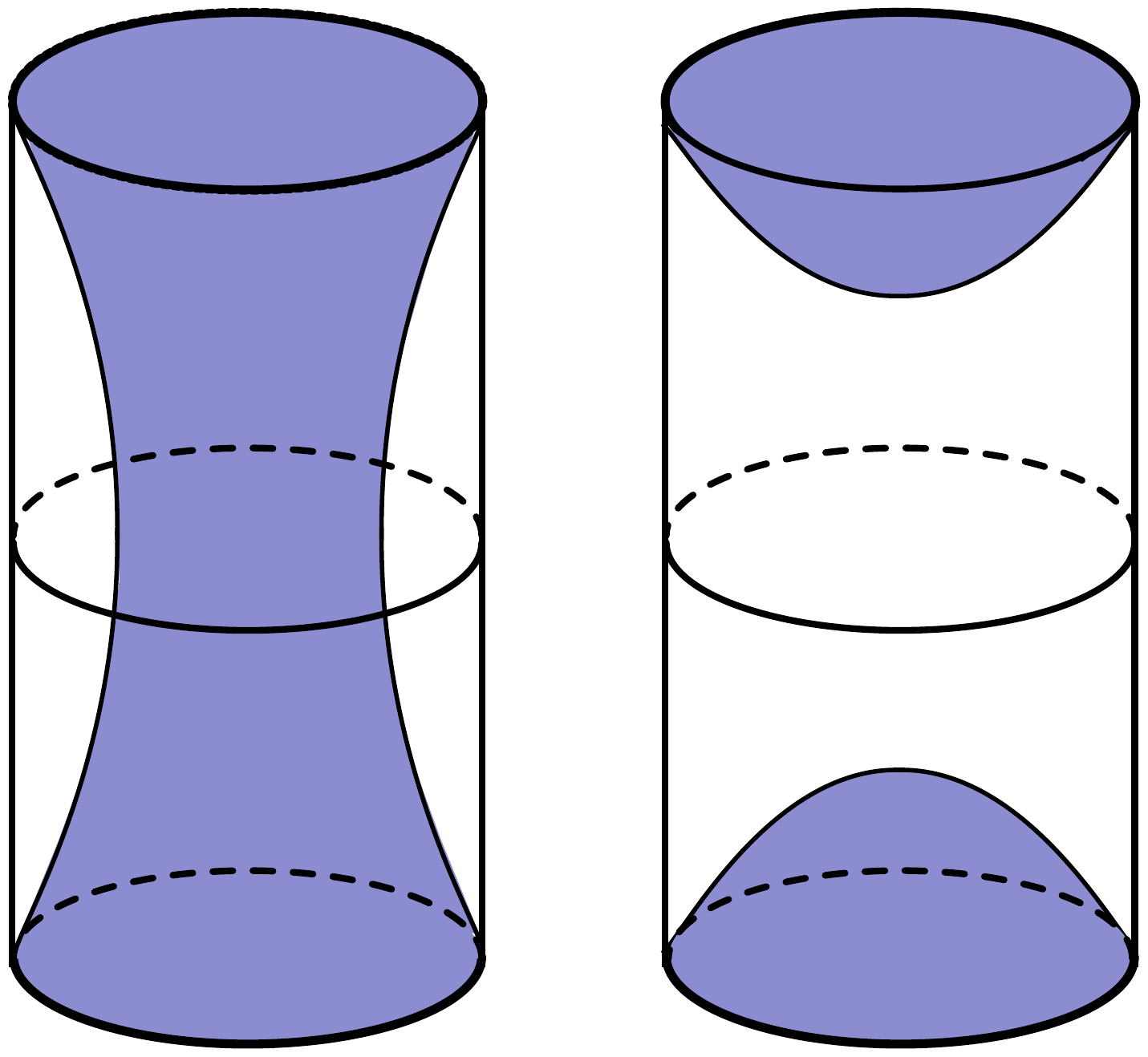}
\caption{Euclidean gravity solutions corresponding to the CFT path integral for $\langle B | e^{- \beta H} | B \rangle$. The boundary geometry is a cylinder $S^d \times [-\tau_0,\tau_0]$. The phase with a connected ETW brane configuration (left), dominant for small $\tau_0$, gives rise to a Lorentzian black hole geometry.}
\label{fig:HighLow}
\end{figure}

There are two possible configurations of the ETW brane in the Euclidean solution, depending on the values of $T$ and $\tau_0$, as shown in Figure \ref{fig:HighLow}. The configuration which dominates the gravitational path integral is the one with lower action. For some values of $T$ we can have a transition between these solutions analogous to the Hawking-Page transition. Above a critical value $\tau_*(T)$, the lower action configuration is a portion of Euclidean AdS, and the Lorentzian solution will be pure AdS with a small amount of quantum matter (as we have for the dual of a finite temperature CFT below the Hawking-Page transition). For $\tau_0 < \tau_*(T)$, the Lorentzian solution corresponds to a part of the AdS-Schwarzschild geometry. For $T > 0$, this includes the full exterior solution plus spacetime behind the horizon terminating with the ETW brane.

In appendix A, we present a detailed derivation of the Euclidean and Lorentzian solutions corresponding to the Euclidean-time-evolved boundary states; here, we summarize the basic results.

\subsubsection{Euclidean solutions}

We begin by describing the Euclidean solutions for each of the phases. In each case, the boundary geometry is taken to be a sphere $S^{d-1}$ with unit radius times an interval $[-\tau_0, \tau_0]$. For the case $d=2$, our calculation is actually equivalent to a calculation in \cite{Fujita:2011fp}, who considered the Euclidean solutions associated with the path integral for a BCFT defined on an interval (i.e. with two boundaries) at finite temperature. In that case, the interval $[-\tau_0, \tau_0]$ represented the spatial direction, while the $S^1$ was the thermal circle.

Since the states we consider preserve spherical symmetry, the relevant geometries will also be spherically symmetric, and must therefore locally be described by the Euclidean AdS-Schwarzschild geometry,
\be
ds^2 = f(r) d\tau^2 + {dr^2 \over f(r)} + r^2 d \Omega_{d-1}^2
\ee
with
\be
f(r) = {r^2 \over  L_{AdS}^2} + 1 - {r_H^{d-2} \over r^{d-2}}\left({r_H^2 \over L_{AdS}^2} + 1 \right) \; .
\ee
Here, the value of $r_H$ will depend on which phase we are in and on the values of $\tau_0$ and $T$. The periodicity of $\tau$ (for $r_H > 0$) is determined by smoothness at $r=r_H$ to be
\be
\label{beta}
\beta = {4 \pi r_H L_{AdS}^2 \over (d-2) L_{AdS}^2 + d r_H^2} \; .
\ee
This relates the inverse black hole temperature to $r_H$.

\subsubsection*{Black hole phase}

We will mainly be interested in the ``black hole'' phase in which ETW brane is connected and takes the form shown on the left in Figure \ref{fig:HighLow}. Describing the spherically symmetric brane embedding by $r(\tau)$ we find that the equations of motions for the brane imply that the trajectory obeys
\be
\label{Eeom}
{dr \over d \tau} = {f(r) \over T r} \sqrt{f(r) - T^2 r^2} \; .
\ee
Solutions that are symmetric about $\tau = 0$ will have ${dr \over d \tau} = 0$ for $\tau = 0$, with $r$ equal to some minimum value $r_0$ determined in terms of $T$ and $r_H$ by
\be
f(r_0) = T^2 r_0^2 \; .
\ee
This gives the maximum ETW brane radius in the Lorentzian solution. As we increase $T$, the ratio $r_0/r_H$ increases monotonically from 1 at $T=0$. In $d=2$, we have simply
\be
{r_0 \over r_H} = {1 \over \sqrt{1-T^2}} \; ,
\ee
while in higher dimensions, we will see below that this ratio reaches a finite maximum value.

The brane locus is then given by
\be
\label{taur2}
\tau(r) = \int_{r_0}^r d\hat{r} { T \hat{r} \over f(\hat{r}) \sqrt{f(\hat{r}) - T^2 \hat{r}^2}} \; .
\ee

\begin{figure}
\centering
\includegraphics[width=120mm]{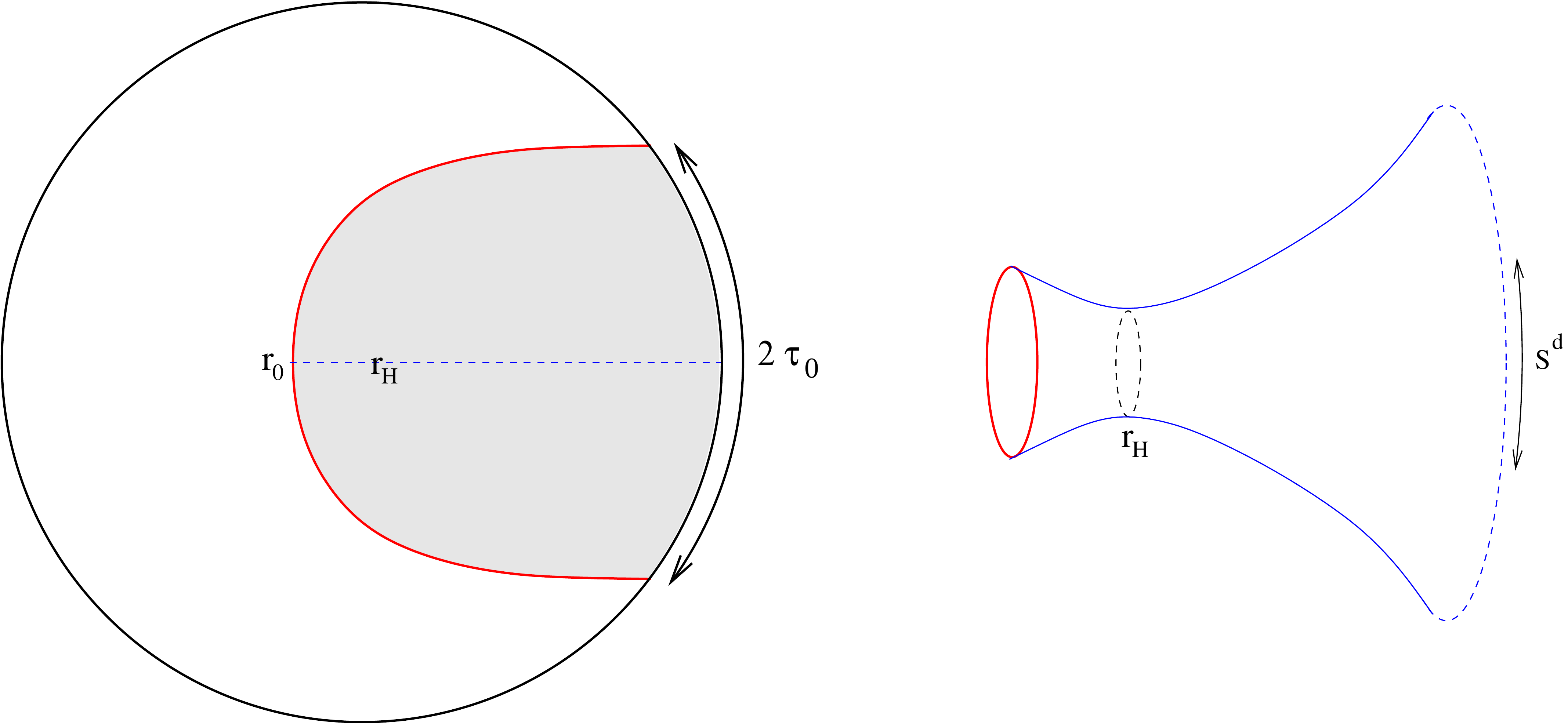}
\caption{Euclidean geometry associated with a $T>0$ state. Left: ETW brane trajectory on $r,\tau$ plane, with $r = r_H$ at the center and $r = \infty$ represented as the boundary of the disk. We have a $S^d$ of radius $r$ associated with each point. Right: spatial geometry fixed by time-reflection symmetry (blue dashed line on the left). This provides the initial data for the Lorentzian solution.}
\label{fig:EuclideanTrajectory2}
\end{figure}

A typical solution for $T>0$ is depicted in Figure \ref{fig:EuclideanTrajectory2}. On the left, the full disk represents the $r, \tau$ coordinates of the Euclidean Schwarzschild geometry, with $r$ ranging from $r_H$ at the center to infinity at the boundary. We have an $S^d$ of radius $r$ associated with each point. The ETW brane bounds a portion of the spacetime (shaded) that gives the Euclidean geometry associated with our state. This has a time-reflection symmetry about the horizontal axis. The invariant co-dimension one surface (blue dashed line) gives the $t=0$ geometry (depicted on the right) for the associated Lorentzian solution. In this picture, the minimum radius sphere corresponds to the black hole horizon, so we see that the ETW brane is behind the horizon.

For $T<0$, we obtain the same trajectories, but the geometry corresponds to the unshaded part, and the ETW brane from the initial data slice is outside the horizon.

For a given $r_H$ and $T$, the Euclidean preparation time $\tau_0$ associated with the solution corresponds to half the range of $\tau$ bounded by the ETW brane at the AdS boundary. This is given explicitly by
\be
\tau_0 = {2 \pi r_H \over d r_H^2 + (d-2)} - \int_{r_0}^\infty dr { T r \over f(r) \sqrt{f(r) - T^2 r^2}}\,.
\ee
For a specified tension $T$ and preparation time $\tau_0$, the temperature of the corresponding black hole is determined implicitly by this equation. There can be more than one pair $r_H$ that gives the same $\tau_0$ for fixed $T$, but in this case, the solution with smaller $r_H$ is never the minimum action solution.

For $d = 2$, we find that for every value of $T$ and $r_H$, the ETW brane trajectory meets the boundary of the $(r,\tau)$ disc at antipodal points, so the black hole temperature is very simply related to the Euclidean preparation time,
\be
\tau_0 = {\beta \over 4} = {\pi \over 2 r_H} \; .
\ee
In this case, the ETW brane radius on the initial data slice is
\be
r_0 = {r_H \over \sqrt{1-T^2}} \; ,
\ee
so the region behind the horizon can become arbitrarily large as we take $T \to 1$.

For $d > 2$ we find that Euclidean solutions in this phase exist only for a portion of the $\tau_0-T$ plane, shown for $d=4$ in figure \ref{fig:Ttau}. In particular, we have some maximum value $T_{max}$ above which there are no Euclidean solutions with a connected ETW brane (corresponding a Lorentzian black hole geometry).

For $d=3$, we find $T_{\rm max} \approx .95635$. This leads to a maximum value of $(r_0/r_H)_{\text{max}} \approx 2.2708$ for the ratio of the ETW brane radius to the horizon radius.

For $d=4$, we find that the large $r_H$ limit of $T_*$ is $T_{\rm max} \approx 0.79765$. This leads to a maximum value of $(r_0/r_H)_{\text{max}} \approx 1.2876$ for the ratio of the ETW brane radius to the horizon radius.

For $T > T_*(r_H)$, the corresponding Euclidean solutions are not sensible since the ETW brane overlaps itself, as shown on the left in Figure \ref{fig:EuclideanOverlaps}. In this case, the thermal AdS geometry (with disconnected ETW branes bounding the Euclidean past and future in the Euclidean solution) is apparently the only possibility.

\subsubsection*{Pure AdS phase}

For any value of $\tau_0$ and $T>0$, we can also have a Euclidean solution where the ETW brane has two disconnected components as shown on the right in figure \ref{fig:HighLow}. The Euclidean geometry is a portion of pure Euclidean $AdS$ (described by the metric above $f(r) = r^2 + L_{AdS}^2$) bounded by the two branes. We can parameterize the brane embedding by $\tau(r)$ with $\tau(\infty) = \pm \tau_0$ for the upper and lower brane respectively. The equations determining the brane location are the same as in the previous case since the geometry takes the same form, so we find that the brane embedding is given by
\be
\label{taur3}
\tau(r) - \tau_0 = \int_{r}^\infty d\hat{r} { T \hat{r} \over f(\hat{r}) \sqrt{f(\hat{r}) - T^2 \hat{r}^2}} \; ,
\ee
with $f(r) = r^2 + 1$. Integrating, we find (in any dimension)
\be
\tau(r) - \tau_0 = {\rm arcsinh}\left({T \over \sqrt{r^2 +1} \sqrt{1 - T^2}}\right)
\ee
The negative $\tau$ component of the ETW brane is obtained via $\tau \to - \tau$.

\subsubsection*{Comparison of the gravitational actions}

In order to determine which type of solution leads to the classical geometry associated with our state for given $(\tau_0, T)$, we need to compare the gravitational action for solutions from the two phases. For $d=2$, this calculation was carried out in \cite{Fujita:2011fp} (section 4) while studying the Hawking-Page type transition for BCFT on an interval. Our calculations in Appendix A generalize this to arbitrary dimensions. In order to compare the actions, we need to regularize; in each case, we can integrate up to the $r$ corresponding to $z=\epsilon$ in Fefferman-Graham coordinates and then take the limit $\epsilon \to 0$ after subtracting the actions for the two phases.

\begin{figure}
  \vspace{10pt}
  \centering
  \includegraphics[scale=0.30]{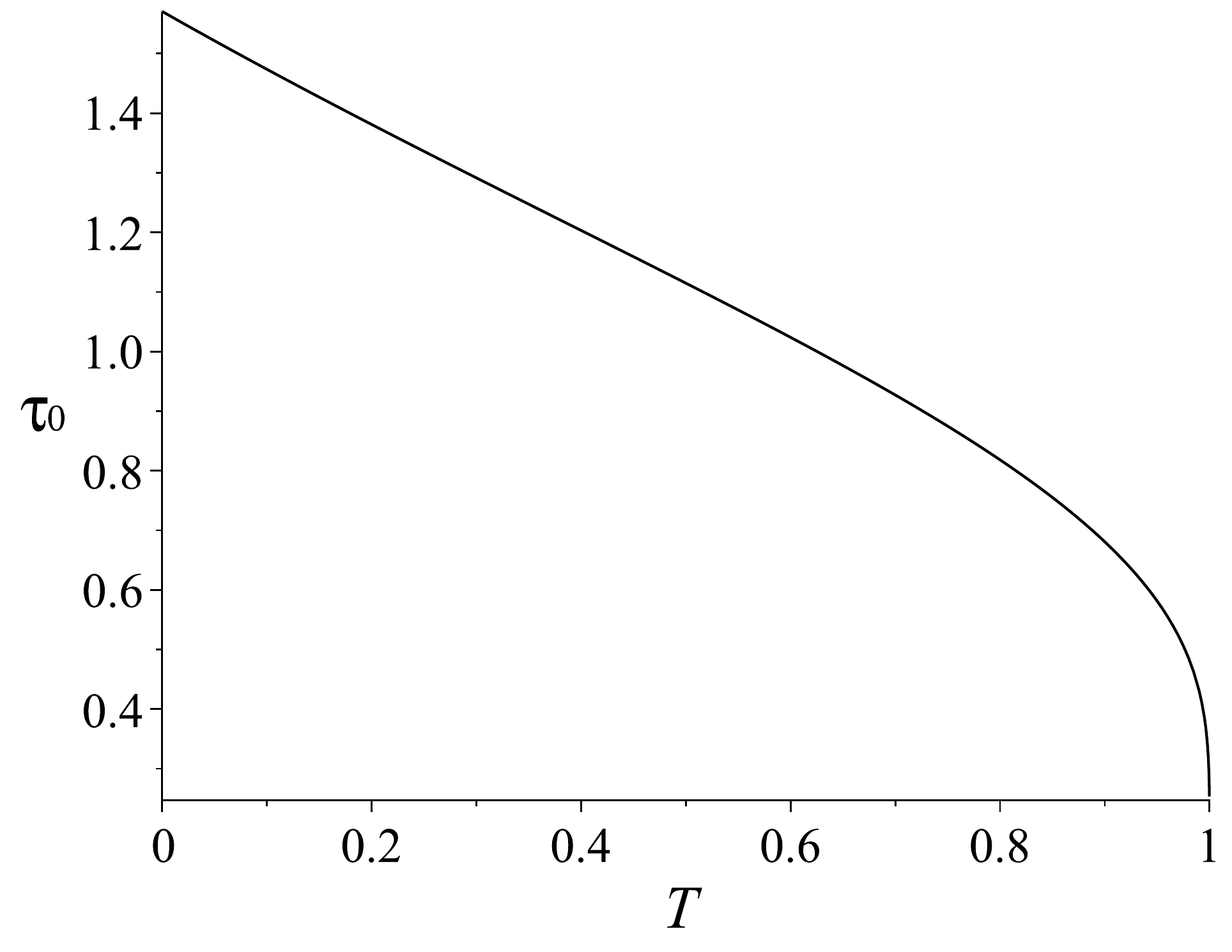} \hspace{10pt}
  \includegraphics[scale=0.50]{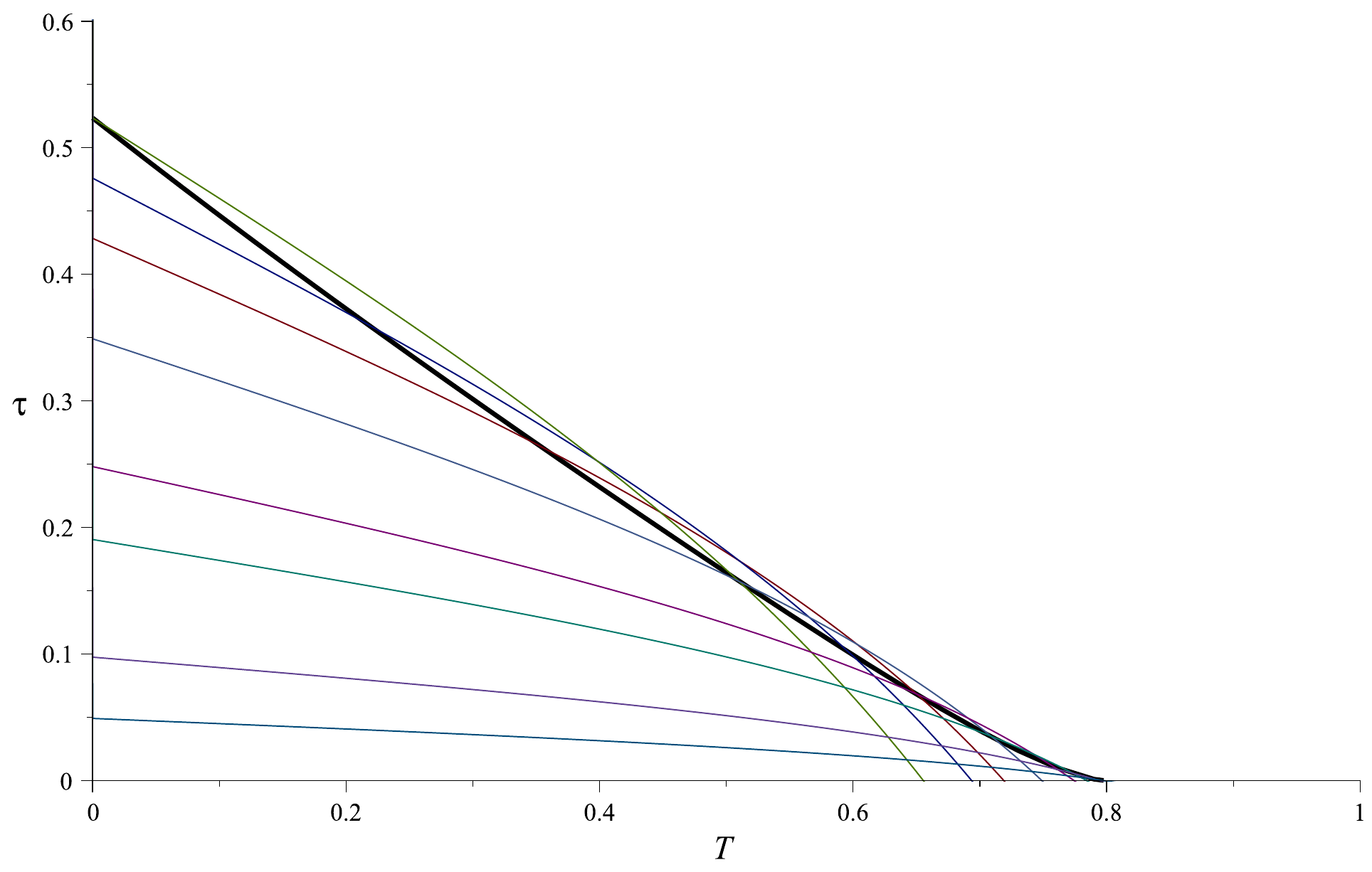}
  \caption{Critical value of $\tau_0$ vs $T$ for $d=2$ (left) and $d=4$ (right). The thick curve on the right shows the phase boundary below which the black hole phase dominates. The other curves on the right show $\tau_0(T)$ for fixed values of $r_H$, equal to 1,1.25,1.5,2,3,4,8, and 16 from top to bottom on the left. Where the curves overlap in the black hole phase region, the value of $r_H$ for the physical solution is always the larger one.}
  \label{fig:Ttau}
\end{figure}

As examples, we find that for $d=2$, we have
\be
\lim_{\epsilon \to 0} (I_E^{AdS}(T,\tau_0,\epsilon) - I_E^{BH}(T,\tau_0,\epsilon))  = {1 \over 2G} \left[ - {\rm arctanh}(T) - {\tau_0 \over 2} + {\pi^2 \over 8 \tau_0} \right] \; .
\ee
Thus, our states (for a CFT on a unit circle) correspond to bulk black holes when
\be
\tau_0 < - {\rm arctanh}(T) + \sqrt{{\pi^2 \over 4} + {\rm arctanh}^2(T)} \; .
\ee
This phase boundary is shown in Figure \ref{fig:Ttau}. Our result agrees with the calculation of \cite{Fujita:2011fp} (reinterpreted for our context).

For $d=4$, the action difference is given in equation (\ref{DeltaI4}) in the appendix. The resulting phase boundary is shown in figure \ref{fig:Ttau}; the critical $\tau_0$ decreases from $\pi/6$ at $T=0$ to 0 at $T=T_{max}$. We see that for $T>0$, the black hole solutions typically have lower action when they exist.

It is somewhat surprising that the black hole phase never dominates (and doesn't even exist) for any value of $\tau_0$ above $T=T_{max}$, since taking $\tau_0$ sufficiently small would be expected to lead to a state of arbitrarily large energy, which should correspond to a black hole in the Lorentzian picture. One possible resolution to this puzzle is that among the possible conformally invariant boundary conditions for holographic CFTs, there may not exist examples that correspond to $T>T_c$ in our models. Our Euclidean gravity results could be seen as a prediction of some constraints on the possible boundary conditions for holographic CFTs (and specifically on a higher-dimensional analogue of boundary entropy).

\begin{figure}
\centering
\includegraphics[width=100mm]{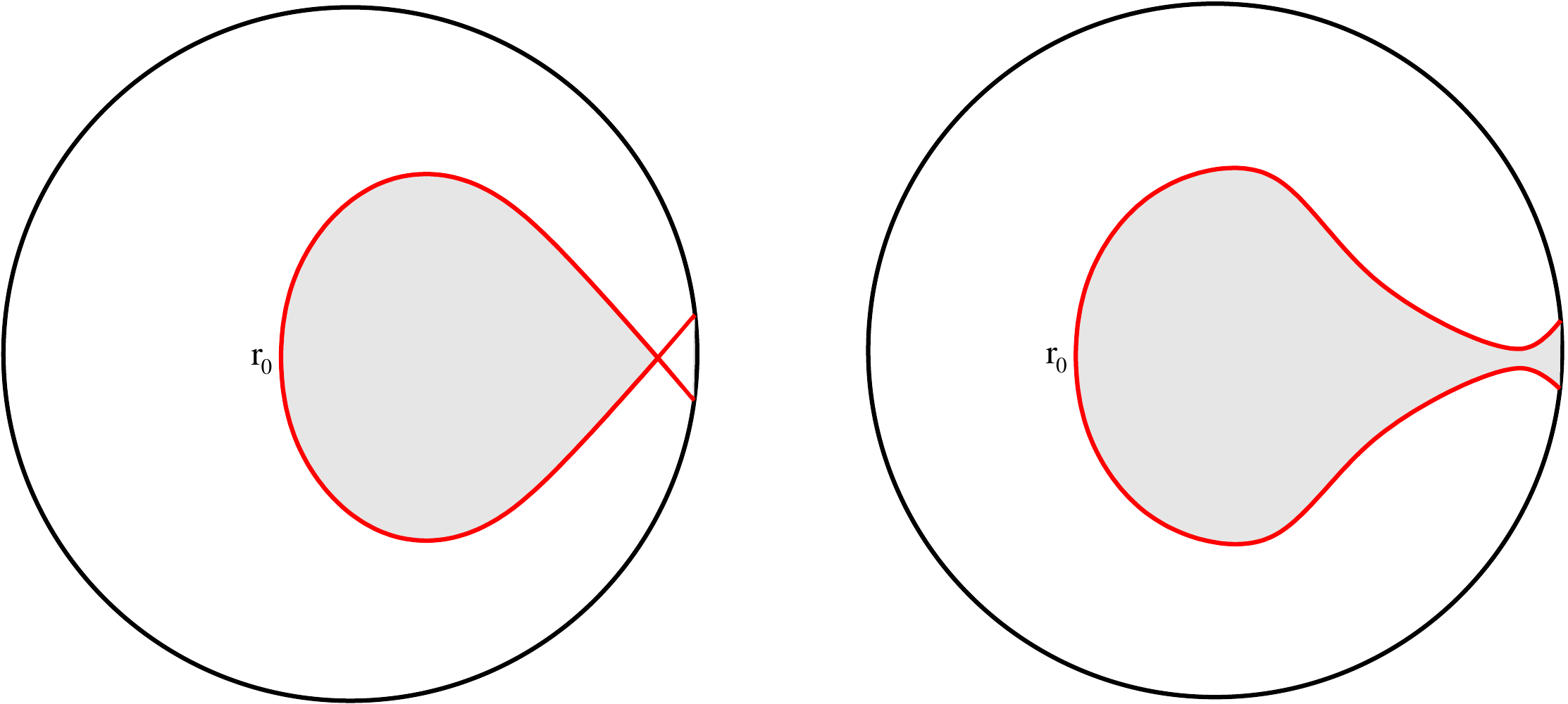}
\caption{Left: Euclidean ETW brane trajectories for $d > 2$ and $T_*(r_H) < T < T_{crit}$. The naive ETW brane trajectory overlaps itself. Right: a possible alternative picture in a more complete holographic model with self-interactions of the ETW brane. }
\label{fig:EuclideanOverlaps}
\end{figure}

Alternatively, the simple prescription of holographically modelling the CFT boundary by introducing a bulk ETW brane with some constant tension may not be adequate to model boundary conditions which naively correspond to larger values of $T$. For example, about $T_*$, solving the equations to determine the Euclidean trajectory naively gives a result that folds back on itself. But a more complete model of the ETW brane physics would presumably include interactions of the brane with itself that invalidate our naive analysis. For example, an effective repulsion could turn a naively unphysical solution into a physical one, as shown in Figure \ref{fig:EuclideanOverlaps}.

\subsubsection{Lorentzian geometries}

To find the Lorentzian geometries associated with our states, we use the $\tau = 0,\pi$ slice of the Euclidean geometry as initial data for Lorentzian evolution. The resulting geometry is a portion of the maximally extended black hole geometry, with one side truncated by a dynamical ETW brane. These Lorentzian geometries parallel earlier results on domain walls and thin shells in AdS \cite{Kraus:1999it, Fidkowski2003, Freivogel2005}.\footnote{Indeed, the Neumann condition reduces to the thin shell junction condition where the extrinsic curvature on the ``excised'' side of the brane vanishes.}

For $T > 0$, we will see that the brane emerges from the past singularity, expands into the second asymptotic region and collapses again into the future singularity. For $T<0$ we have an equivalent ETW brane trajectory but on the other side of the black hole, so that the brane emerges from the horizon, enters the right asymptotic region, and falls back into the horizon.

Using Schwarzschild coordinates to describe the portion of the ETW brane trajectory in one of the black hole exterior regions, the brane locus is given by the analytic continuation of the Euclidean trajectory,
\be
\label{LorTraj}
t(r) = \int_{r_0}^r d\hat{r} { T \hat{r} \over f(\hat{r}) \sqrt{T^2 \hat{r}^2 - f(\hat{r})}} \; .
\ee
For example, in $d=2$, we obtain
\be
\label{brane2D}
\cosh(t r_H) \sqrt{{r^2 \over r_H^2} - 1} = {T \over \sqrt{1-T^2}} \; .
\ee

\begin{figure}
  \vspace{10pt}
  \centering
  \includegraphics[scale=0.20]{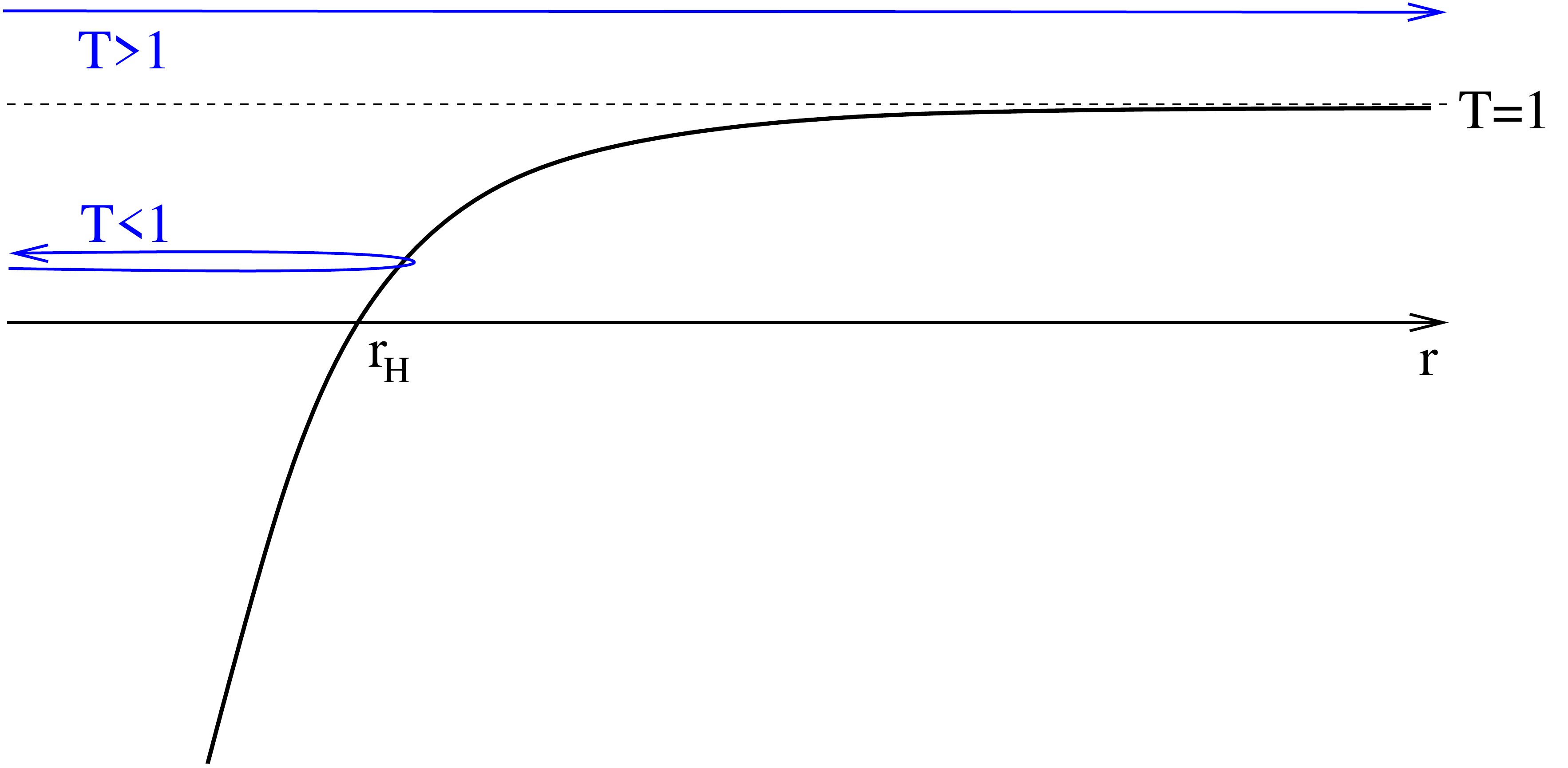} \hspace{10pt}
  \includegraphics[scale=0.20]{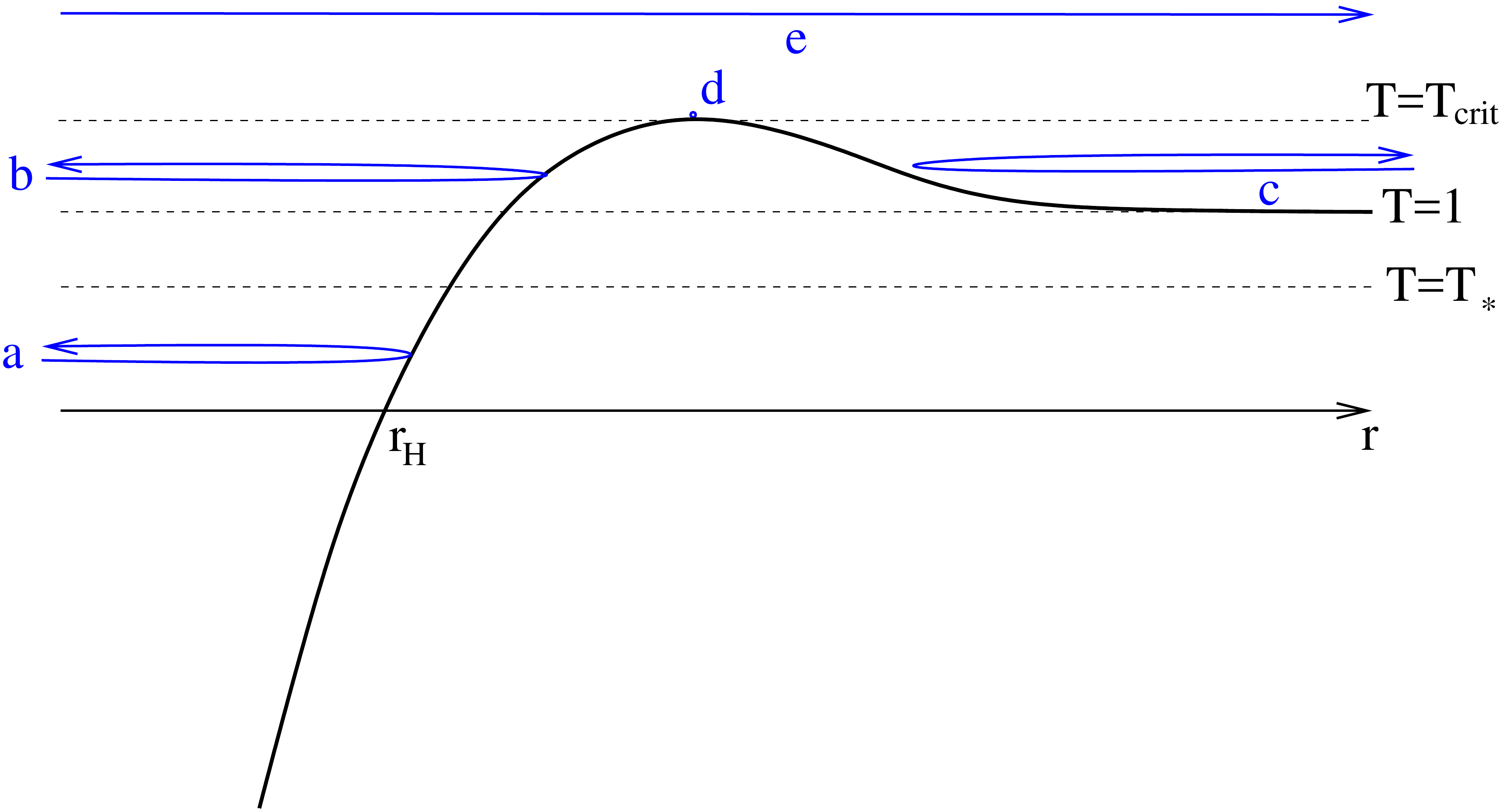}
  \caption{Effective potential $V(L)$ and types of Lorenzian ETW brane trajectories for $d=2$ (left) and $d > 2$ (right).}
  \label{fig:Veff}
\end{figure}

To understand the behaviour of the brane in the full spacetime, it is convenient to rewrite the equation in terms of the proper time $\lambda$ on the brane, related to Schwarzschild time by
\be
{\D t \over  \D \lambda} = \gamma = \sqrt{\frac{f(r)}{f(r)^2 - {\dot{r}}^2}} \;.
\ee
We then find that the coordinate-independent equation of motion for the brane relating the proper radius $r$ to the proper time $\lambda$ is simply
\begin{equation}
  \label{EOM-ETW}
\dot{r}^2 + [f(r) - T^2 r^2] \; ,
\end{equation}
where now the dot indicates a derivative with respect to proper time. In terms of $L = \log(r)$, this becomes simply
\be
\dot{L}^2 + V(L) = T^2
\ee
where
\be
V(L) = {f(r) \over r^2} = 1 + e^{- 2 L} - e^{-d (L - L_H)}(1 + e^{- 2 L_H}) \; .
\ee
So the trajectory $L(\lambda)$ is that of a particle in a one-dimensional potential $V(L)$ with energy $T^2$. These potentials take the form shown in Figure \ref{fig:Veff}.

Considering general values of $T$, we can have five classes of trajectories (two for $d=2$), as shown on the right in Figure \ref{fig:Veff}. However, all of our time-symmetric Euclidean solutions in the black hole phase correspond to values $T<1$ (corresponding to case a) in figure \ref{fig:Veff}) for which the Lorentzian trajectory starts at $r=0$, increases to $r=r_0$ and decreases back to $r=0$. Thus, the brane emerges from the past singularity, reaches a maximum size $r_0$, and shrinks again to $r=0$ at the future singularity.

Using the proper time parametrization, the world-volume metric for the brane takes the close FRW form
\be
ds^2 = - d \lambda^2 + r^2(\lambda) d \Omega^2 \; ,
\ee
where the scale factor $r(t)$ is determined from equation (\ref{EOM-ETW}). The entire trajectory covers some finite amount of proper time given by
\be
\lambda_{\mathrm{tot}} = 2 \int_0^{r_0} {dr \over \sqrt{T^2 r^2 - f(r)}} \; .
\ee
For $d=2$, the explicit scale factor in the world-volume metric is
\be
r(\lambda) = {r_H \over 1-T^2}\cos(\lambda \sqrt{1-T^2})
\ee
and the total proper time for the evolution (in units with $L_{AdS} = 1$) is
\be
\lambda_{\mathrm{tot}}^{d=2} = {\pi L_{AdS} \over \sqrt{1-T^2}}\,.
\ee
For $d=4$, the scale factor is
\be
\label{FRW4}
r(\lambda) = {1 \over \sqrt{2(1-T^2)}} \left[\cos(2 \sqrt{1-T^2} \lambda ) \sqrt{1 + 4(1-T^2)r_H^2(1 + r_H^2)} - 1\right]^{1 \over 2}
\ee
and the total proper time for the evolution is
\be
\label{Life4}
\lambda_{\mathrm{tot}}^{d=4}  = { 1 \over \sqrt{1-T^2}} \arccos \left({1 \over \sqrt{1 + 4(1-T^2)r_H^2(1 + r_H^2)}}\right) \; .
\ee
The $d=3$ results are given in terms of elliptic integrals.

We briefly discuss the remaining trajectories in appendix A, in case they may be relevant to some other class of CFT states. In section \ref{sec:cosmo} we discuss the possibility that for certain parameter ranges, we can have gravity localized to the ETW brane, so that the FRW metrics here would represent cosmological solutions of an effective $d$-dimensional theory of gravity.

\section{Probing behind the horizon with entanglement}

In this section, we consider the holographic calculation of entanglement entropy for CFT states whose dual geometries are captured by Figure \ref{fig:ETW}. We will continue to use the simple model of a spacetime terminating with an ETW brane, but we expect the same qualitative conclusions when the ETW brane is replaced by a more complete microscopic description.
We begin by considering a general behind-the-horizon ETW brane trajectory $r(t)$ symmetric about $t=0$ with maximum radius $r(0) = r_0$.

We will consider the entanglement entropy for ball-shaped regions on the sphere as a function of size and of CFT time. As depicted in Figure \ref{fig:EEchoices}, we have extremal surfaces that stay outside the horizon, but we can also have extremal surfaces that enter the horizon and end on the ETW brane.\footnote{We recall that the topological constraint on the extremal surfaces is that they are homologous to the boundary region under consideration. This means that the surface together with the boundary region form the boundary of some portion of a spatial slice of the bulk spacetime. The relevant regions in the two cases are shown as the shaded regions in Figure \ref{fig:EEchoices}. In the case where the extremal surfaces go behind the horizon and terminate on the ETW brane, this region includes part of the ETW brane. We emphasize that this is not part of the extremal surface and its area should not be included in the holographic calculation of entanglement entropy.} Depending on the value of time and the ball size, we can have transitions between which type of surface has least area. In the phase where the exterior surface has less area, the CFT entanglement entropy will be time-independent (at leading order in large $N$), while in the other phase, we will have time dependence inherited from the time-dependent ETW brane trajectory. In our examples below, we will find that in favorable cases, the minimal area surface for sufficiently large balls goes behind the horizon during some time interval $[-t_0,t_0]$ which increases with the size of the ball.

We now turn to the details of the holographic calculation of entanglement entropy given some ETW brane trajectory $r(t)$.
This was calculated for the $T = 0$ case in \cite{Hartman2013}. Similar methods were used in slightly more exotic geometries, and reaching different conclusions, in \cite{Faraggi2007}.

\subsubsection*{Exterior extremal surfaces}

First, consider the exterior extremal surfaces, working in Schwarzschild coordinates. Let $\theta_0$ be the angular size of the ball, such that $\theta_0 = \pi/2$ corresponds to a hemisphere.

Since the exterior geometry is static, the extremal surface lives in a constant $t$ slice, and we can parameterize it by $r(\theta)$. In terms of this, the area is calculated as
\be
\label{exterior}
{\rm Area} = \omega_{d-2} \int d \theta r^{d-2} \sin^{d-2} \theta \sqrt{r^2 + {1 \over f(r)} (r')^2} \; .
\ee
where $\omega_{d-2}$ is the volume of a $d-2$-dimensional sphere.

Extremizing this action, we obtain equations of motion that can be solved numerically (or analytically in the $d=2$ case --- see below).

To obtain a finite result for entanglement entropy, we can regulate by integrating up to some fixed $r_{\text{max}}$ corresponding to $z = \epsilon$ in Fefferman-Graham coordinates, subtracting off the vacuum entanglement entropy (calculated in the same way but with $f(r) = r^2 + 1$), and then taking $\epsilon \to 0$.

\subsubsection*{Interior extremal surfaces}

To study extremal surfaces that pass through the horizon, it is convenient to work in a set of coordinates that cover the entire spacetime. In this case, we parameterize the surfaces by a time coordinate and a radial coordinate, which are both taken to be functions of an angle $\theta$ on the sphere.

The only new element here is that the extremal surfaces intersect the ETW brane, and we need to understand the appropriate boundary conditions here. Since we are extremizing area, our extremal surfaces must intersect the ETW brane normally, so that a variation if the intersection locus does not change the surface area to first order.

\subsubsection*{Criterion for seeing behind the horizon with entanglement}

When the behind-the-horizon extremal surfaces have less area, the CFT entanglement is detecting a difference between our state and the thermal state. We expect that this is most likely to happen for $\theta = \pi/2$, where we are looking at the largest possible subsystem, and for $t=0$, since at other times the state will become more thermalized.

For this case $\theta_0 = \pi/2$, $t=0$, the behind-the-horizon extremal surface remains at $\theta = \pi/2$ and $t=0$, extending all the way to the ETW brane on the far side of the horizon. This intersects the ETW brane normally by the time-reflection symmetry. In this case, we can calculate the regulared areas explicitly as
\be
{\rm Area}_\text{int}(\theta=\pi/2,t=0,r_0) = \omega_{d-2} \int_{r_H}^{r_{\text{max}}} dr {r^{d-2} \over \sqrt{f(r)}} + \omega_{d-2} \int_{r_H}^{r_0} dr {r^{d-2} \over \sqrt{f(r)}} \; .
\ee
When this area is greater than the area of the exterior extremal surface corresponding to $\theta = \pi/2$, we expect that the entanglement entropy will always be calculated in terms of the exterior surfaces. Thus, we have a basic condition
\be
\label{Econstraint}
{\rm Area}_\text{ext}(\pi/2) > {\rm Area}_\text{int}(\theta=\pi/2,t=0,r_0)
\ee
for when entanglement will tell us something about the geometry behind the horizon. This is more likely to be satisfied for smaller values of $r_0$ (ETW brane not too far past the horizon). It can fail to be satisfied even for $r_0 = r_H$ if the black hole is too small, so below some minimum value $r_H^{\text{min}}$, all minimal area extremal surfaces probe outside the horizon.

For $d = 2$, we will see below that the constraint (\ref{Econstraint}) gives explicitly that
\be
(r_H^{\text{min}})^{d=2} = {2 L_{AdS} \over \pi} {\rm arcsinh}(1)
\ee
and that for larger $r_H$, the maximum brane radius must satisfy
\be
{r_0 \over r_H} \le {1 \over 2} \left( \sinh \left( {r_H \pi \over 2 L_{AdS}} \right) + \sinh^{-1} \left( {r_H \pi \over 2 L_{AdS}} \right) \right)\,.
\ee
in order that we can see behind the horizon with entanglement.

\subsection{Example: BCFT states for $d=2$}

In this section, we work out the explicit results for $d=2$ where the CFT lives on a circle. We calculate the entanglement entropy $S(\Delta \theta, t)$ for an interval of angular size $\Delta \theta$ on the circle, as a function of CFT time $t$. We find that having access to large enough subsystem of the CFT allows us to probe behind the horizon, and thus renders the microstates distinguishable, in broad qualitative agreement with \cite{Bao:2017guc}.

\subsubsection*{Exterior extremal surfaces}

First consider the exterior surfaces, which we parameterize by $r(\theta)$. Since the integrand in (\ref{exterior}) does not depend explicitly on $\theta$, the extremizing surfaces must satisfy
\be
r' {\delta {\cal L} \over \delta r'} - {\cal L} = {\rm constant}
\ee
Calling this constant $r_0$ (this represents the minimum value of $r$ on the trajectory, where $r' = 0$), we get
\be
r' = \pm {r \over r_0 L} \sqrt{(r^2 - r_H^2)(r^2 - r_0^2)} \; .
\ee
The solution, taking $\theta = 0$ to be the point where $r = r_0$, is given implicitly by
\be
\theta = - {L \over 2 r_H} \ln \left[ {-2 r_H^2 r_0^2 + r_H^2 r^2 + r^2 r_0^2 - 2 r_0 r_H \sqrt{(r^2 - r_0^2)(r^2 - r_H^2)} \over r^2 (r_0^2 - r_H^2) }\right] \; .
\ee
We will only need that
\be
\theta(r = \infty) = {L \over 2 r_H} \ln \left( {r_0 + r_H \over r_0 - r_H} \right) \; ,
\ee
so that
\be
{r_0 \over r_H} = \coth \left( {r_H \Delta \theta  \over 2 L} \right) \; .
\ee
The area of such a surface, regulating by integrating only up to $r_{\text{max}} = L/\epsilon$ is
\be
{\rm Area} (\Delta \theta) = 2 L \ln \left({2 L \over \epsilon r_H} \sinh(r_H \Delta \theta /2 L)\right)
\ee
where we have dropped terms of order $\epsilon$. Using $c = 3L/2G$, this gives entropy $S = \text{Area}/(4G)$ of
\be
S = {c \over 3} \ln \left({2 L \over \epsilon r_H} \sinh(r_H \Delta \theta / 2 L)\right) \; .
\label{EE1}
\ee
In terms of the CFT effective temperature $\beta$, we have $r_H/L = 2 \pi L_{CFT}/\beta$, so the result in terms of CFT parameters is
\be
S = {c \over 3} \ln \left({\beta \over \pi \epsilon L_{CFT} } \sinh( \pi L_{CFT} \Delta \theta / \beta)\right) \; .
\ee
where $L_{CFT}$ is the size of the circle on which the CFT lives.

For comparison, the area of a disconnected surface with two parts extending from the interval boundaries to the horizon via the geodesic path at constant $\theta$ and $t$ gives
\be
{\rm Area}_0 = 2 \int_{r_H}^{r_{\text{max}}} {dr \over \sqrt{f(r)}} = 2L\ln(2L/(\epsilon r_H)) \; .
\ee
This shows that regardless of what happens behind the horizon, the entanglement entropy of an interval with size $\Delta \theta$ will be calculated by an extremal surface outside the horizon if
\be
\sinh(r_H \Delta \theta /(2L)) \le 1 \; .
\ee
This will hold even for the largest interval $\Delta \theta = \pi$ if
\be
r_H/L \le {2 \over \pi} {\rm arcsinh}(1) \; .
\ee
Thus, we must have a sufficiently large black hole if the CFT entanglement entropy is going to have any chance of seeing behind the horizon.

\subsubsection*{Interior extremal surfaces}

Now we consider the extremal surfaces that enter the horizon and end on the ETW brane. Here, it is most convenient to use coordinates for which the maximally extended black hole spacetime takes the form
\be
\label{SYmetric}
ds_\text{BTZ}^2 = {1 \over \cos^2(y)} \left(-ds^2 + dy^2 + r_H^2 \cos^2(s)\, d \phi^2 \right)
\ee
where the coordinate ranges are $-\pi/2 \le s,y \le \pi/2$, with the horizons at $y = \pm s$. The coordinate transformations relating this to Schwarzschild coordinates are given in appendix A. Using these, the ETW brane trajectory is found to be simply
\be
y = -\arcsin(T) \; .
\ee

\begin{figure}
\centering
\includegraphics[width=70mm]{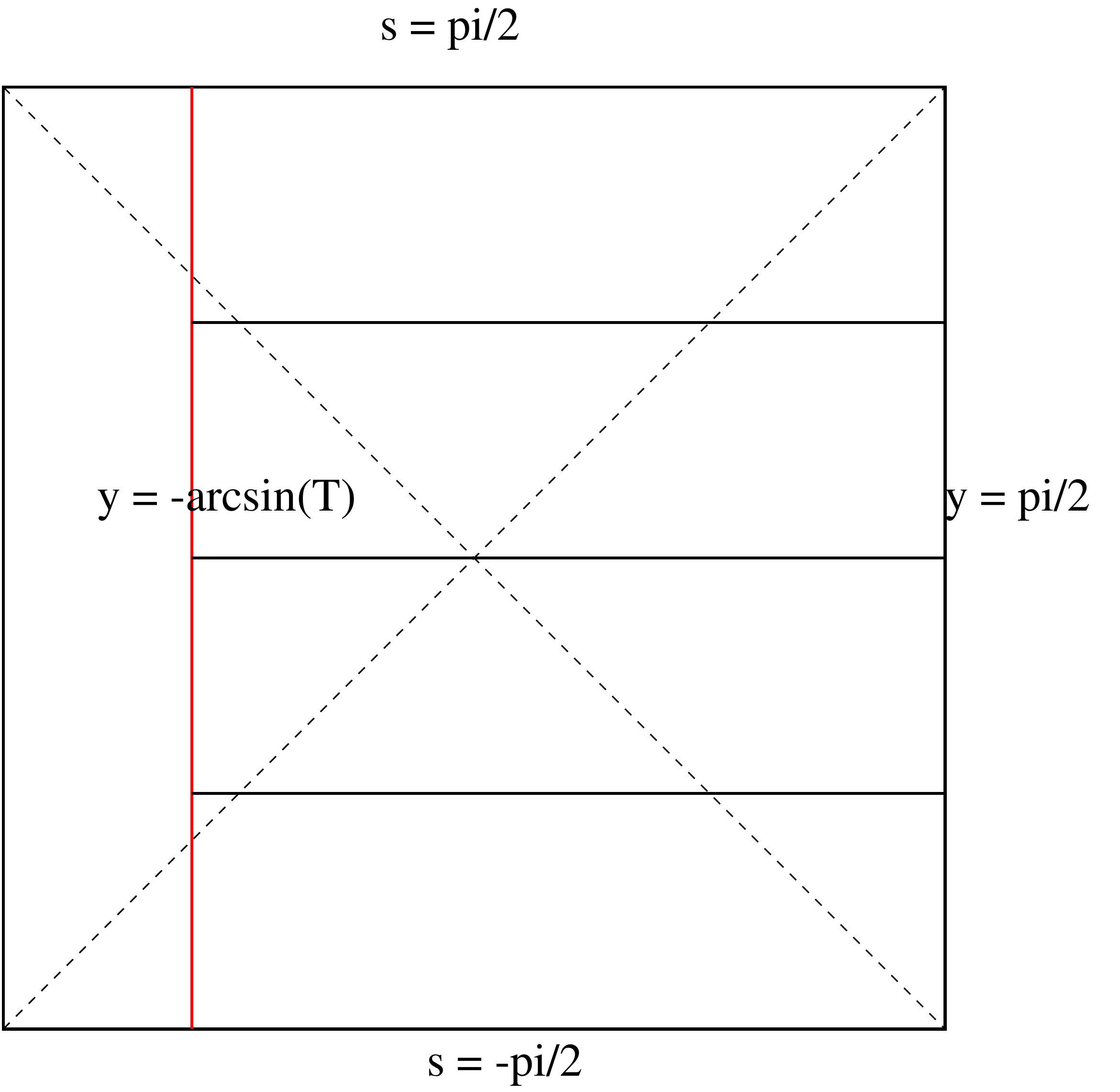}
\caption{BTZ black hole in $s,y$ coordinates, showing ETW brane (red) and various geodesics orthogonal to it. Geometry to the left of the ETW brane is excised.}
\label{fig:BHsy}
\end{figure}

We find that the general spacelike geodesics in this geometry take the form
\be
\sin(s_B - s_0) \sin(y) = \sin(s - s_0)
\ee
where the geodesic passes through $s_0$ at $y=0$ and ends on the AdS boundary ($y = \pi/2$) at $s_B$. The geodesics with fixed $s_B$ and different $s_0$ all end on the same point at the AdS boundary, but different points on the ETW brane. However, requiring that the surface extremize area also with respect to variations of this boundary point on the ETW brane implies that the geodesic should be normal to the ETW brane worldvolume. This gives the very simple class of geodesics
\be
s = s_0
\ee
which sit at fixed $\theta$ and $s$. The black hole geometry together with these geodesics is depicted in figure \ref{fig:BHsy}.

We can now evaluate the area of these extremal surfaces. We will evaluate the area up to the same regulator point $r_{\text{max}} = L/\epsilon$. This gives a maximum $y$ of
\be
y_{\text{max}} = \arctan \left( e^{-r_H t_0} \sqrt{r_{\text{max}}/r_H - 1 \over r_{\text{max}}/r_H + 1}\right) + \arctan \left( e^{r_H t_0} \sqrt{r_{\text{max}}/r_H - 1 \over r_{\text{max}}/r_H + 1}\right) \; ,
\ee
Note that this depends on the Schwarzschild time $t_0$. We have then
\bea
{\rm Area_{int}(\Delta \theta)} &=& 2 \int_{-\arcsin(T)}^{y_{\text{max}}} {dy \over \cos(y)} \cr
&=& 2 L \ln\left( {2L  \over \epsilon r_H} \right) + 2L \ln \left( \cosh \left({t_0 r_H \over L^2}\right) \sqrt{1 + T \over 1 - T} \right) \; .
\eea
where we have restored factors of $L$. The regulated entanglement entropy is then
\be
\Delta S = {c \over 3} \ln \left({2 L \over \epsilon r_H} \cosh (t_0 r_H/  L^2) \sqrt{1 + T \over 1 - T} \right) \; .
\label{EE2}
\ee
In terms of CFT parameters, this gives
\be
\Delta S = {c \over 3} \ln \left({\beta \over \epsilon \pi L_{CFT}} \cosh (2 \pi t_{CFT}/ \beta) \sqrt{1 + T \over 1 - T} \right) \; .
\ee

This gives less area than the exterior surface (so that entanglement entropy will probe the interior) when
\be
\sinh \left( {r_H \Delta \theta \over 2 L} \right) \ge \cosh \left({t_0 r_H \over L^2}\right) \sqrt{1 + T \over 1 - T} \; .
\label{EE3}
\ee
When this is satisfied, the entanglement entropy (times $4 G$) is given by the expression (\ref{EE2}) and is time-dependent but independent of the interval size.\footnote{If we express condition (\ref{EE3}) in terms of the radius $r$ of the ETW brane where we shoot out a normal geodesic, we obtain an even simpler condition $$\sinh\left(\frac{r_H\Delta\theta}{2L}\right) \geq \frac{r_H}{(1-LT)r}.$$} Otherwise, the entanglement entropy is time-independent but depends on the interval size and given by (\ref{EE1}).

\begin{figure}
\centering
\includegraphics[width=120mm]{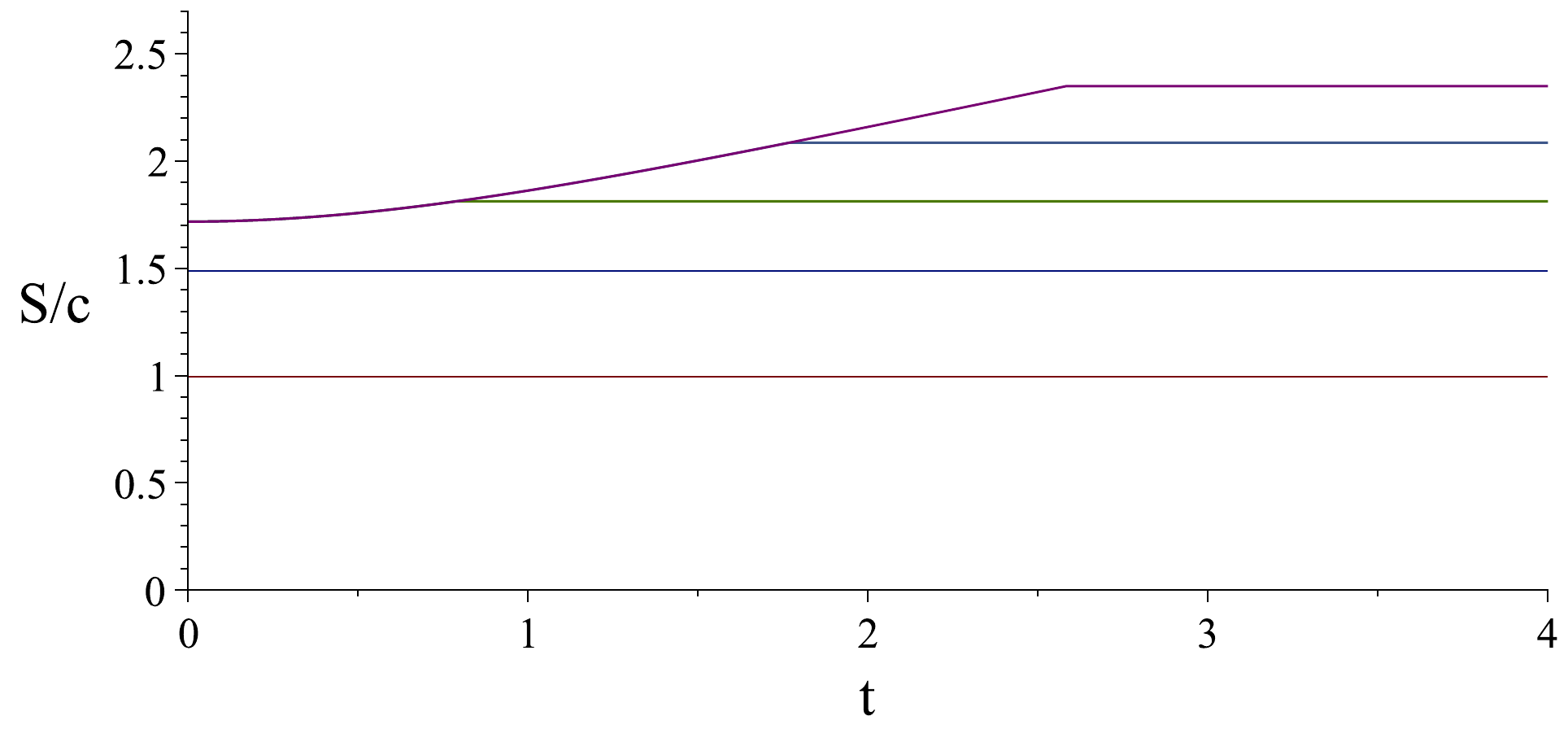}
\caption{Regulated entanglement entropy as a function of time for various interval sizes for $T=0.5$, $r_H = 2 L_{AdS}$, $\epsilon  = 0.01$. Plots from bottom to top show $\Delta \theta = \pi/16, \pi/4, \pi/2, 3 \pi/4, \pi$.}
\label{fig:EEsizes}
\end{figure}

\begin{figure}
\centering
\includegraphics[width=120mm]{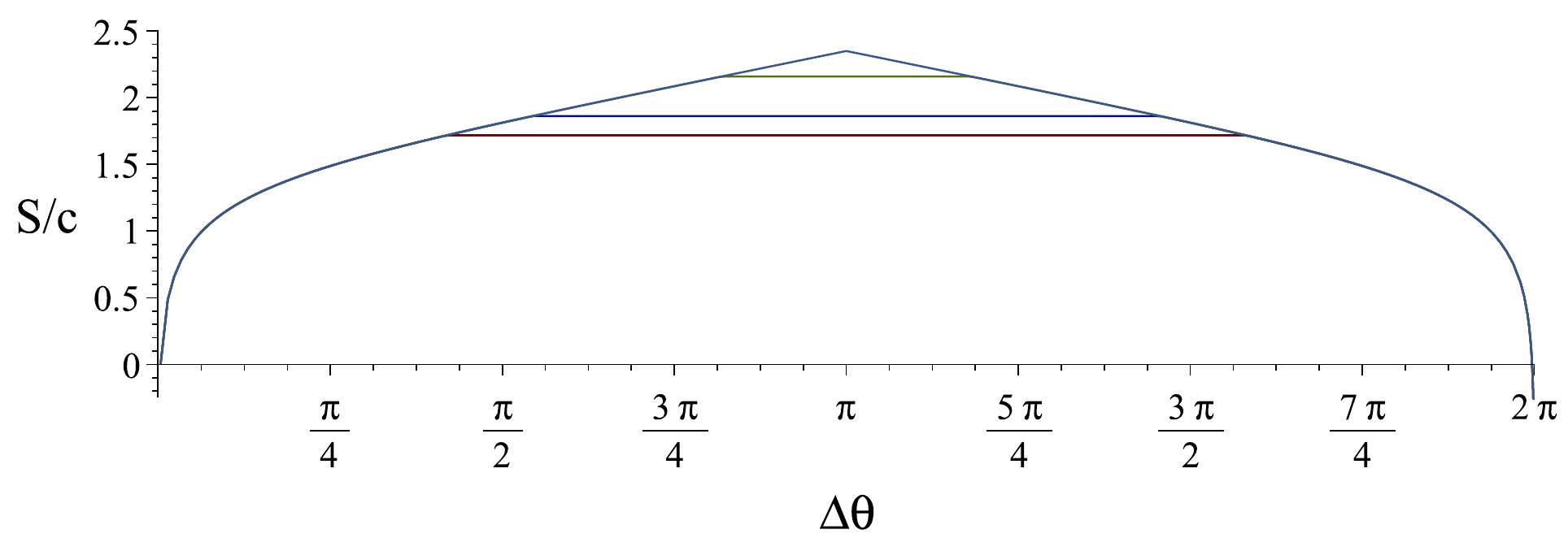}
\caption{Regulated entanglement entropy as a function of interval size for $T=0.5$, $r_H = 2 L_{AdS}$, $\epsilon  = 0.01$. Plots from bottom to top show successively later times starting at $t=0$.}
\label{fig:EEtimes}
\end{figure}

The entanglement entropy as a function of interval size for various times is shown in Figure \ref{fig:EEtimes}. The entanglement entropy as a function of time for various interval sizes is shown in Figure \ref{fig:EEsizes}. The fact that the entanglement entropies are independent of angle when the minimal-area extremal surfaces probe behind the horizon is a special feature of the $d=2$ case arising from the fact that these extremal surfaces have two disconnected parts, each at a constant angle. In higher dimensions, the corresponding surfaces are connected and we have non-trivial angular dependence for all angles.

\subsection{Results for $d=4$}

As another explicit example, we consider the case of a 4+1 dimensional black hole. In this case, the Lagrangian describing the extremal surfaces has an explicit angle dependence, and the surfaces must be found numerically.

\subsubsection*{Interior extremal surfaces}

The metric for the 4 + 1 dimensional Schwarzschild black hole in Schwarzschild coordinates is
\be
ds^2 = -f(r) dt^2 + {dr^2 \over f(r)} + r^2 d \Omega_3^2
\ee
where
\be
f(r) = {r^2 \over L^2} + 1 - {r_H^2 \over r^2}\left({r_H^2 \over L^2} + 1 \right) \;.
\ee
To switch to Kruskal type coordinates, we define
\be
\label{defXT}
t = {r_H \over 2 (2 r_H^2 + 1)} \ln \left({X+T \over X -T}\right) \qquad g(r) = {r_H \over 2 (2 r_H^2 + 1)} \ln(X^2 - T^2)
\ee
where
\be
g(r) = \int^r {dr \over f(r)} = {r_H \over 2 (2 r_H^2 + 1)}\log \left|{r_H - r \over r_H + r}\right| + {\sqrt{r_H^2 + 1} \over 2 r_H^2 + 1}\arctan{r \over \sqrt{r_H^2 + 1}} + C  \; .
\ee
Then the metric is
\bea
ds^2 &=& {r_H^2 \over (2 r_H^2 + 1)^2} e^{-{2 (2 r_H^2 + 1) \over r_H}g(r)}f(r) [-dT^2 + dX^2] + r^2 d \Omega^2 \cr
&=&  {r_H^2 \over (2 r_H^2 + 1)^2} {f(r) \over X^2 - T^2} [-dT^2 + dX^2] + r^2 d \Omega^2 \;
\eea
where $r$ is defined implicitly as a function of $X^2 - T^2$ by the second equation in (\ref{defXT}). Note that the zero at $r = r_H$ in $f(r)$ cancels the pole in the exponential factor, leaving a function that is regular at the horizon.

Changing the constant $C$ amounts to a rescaling of $X$ and $T$, so we can make a choice $C=0$. Then, the metric is
\be
ds^2 = B(r) (-dT^2 + dX^2) + r^2 d \Omega^2
\ee
with
\be
B(r) = {r_H^2 \over (2 r_H^2 + 1)^2} {(r + r_H)^2(r^2 + rH^2 + 1) \over r^2}e^{-{2 \sqrt{r_H^2 + 1} \over r_H} \arctan\left({r \over \sqrt{r_H^2 + 1}}\right)}
\ee
and $r$ defined in terms of $X^2 - T^2$ as
\be
X^2 - T^2 = {r - r_H \over r + r_H} e^{{2 \sqrt{r_H^2 + 1} \over r_H} \arctan\left({r \over \sqrt{r_H^2 + 1}}\right)} \equiv F(r)\;.
\ee
We would like to extremize the action
\be
S = 4 \pi \int dX r^2 \sin^2 \theta \sqrt{B(r) \left(1 - \left({dT \over dX} \right)^2 \right) +r^2 \left({d \theta \over dX} \right)^2} \equiv 4 \pi \int dX {\cal L}
\ee
for surfaces described by $T(X)$, $\theta(X)$, $r(X)$ with
\be
\label{constr}
X^2 - T^2 - F(r) =0 \; .
\ee
Introducing a Lagrange multiplier $4 \pi \Lambda$ for the constraint, this gives equations
\bea
{d \over d X} {\delta {\cal L} \over \delta T'} + 2 \Lambda T &=& 0 \cr
{d \over d X} {\delta {\cal L} \over \delta \theta'} - {\delta {\cal L} \over \delta \theta} &=& 0 \cr
{\delta {\cal L} \over \delta r} - \Lambda {d F \over dr} &=& 0 \; .
\eea
Eliminating $\Lambda$, and using (\ref{constr}) to get an equation for $r'$, we get
\bea
{d \over d X} {\delta {\cal L} \over \delta T'} + {2 T \over {d F \over dr}} {\delta {\cal L} \over \delta r}  &=& 0 \cr
{d \over d X} {\delta {\cal L} \over \delta \theta'} - {\delta {\cal L} \over \delta \theta} &=& 0 \cr
r' + {2 \over {dF \over dr}} ( T T' - X ) &=& 0 \; .
\eea
These differential equations can be solved numerically, along with the equation for the surface area
\be
A' = 4 \pi r^{2} \sin^{2} \theta \sqrt{B(r) \left(1 - \left({dT \over dX} \right)^2 \right) +r^2 \left({d \theta \over dX} \right)^2} \: ,
\ee
to determine the functions $(T(X), \theta(X), r(X), A(X))$. For initial conditions, we should again enforce normality of the extremal surface to the brane. One can use the brane equation of motion
\begin{equation}
    \dot{r}^{2} + [f(r) - r^{2} \tilde{T}^{2}] = 0
\end{equation}
to determine the brane trajectory, and select some initial coordinates $(t_{\textnormal{br}}, r_{\textnormal{br}}, \theta_{\textnormal{br}})$ on the brane. The Kruskal coordinate transformation in equation (\ref{defXT}) is then used to find the corresponding $T_{ \textnormal{br}} = T(t=t_{\textnormal{br}}, r=r_{\textnormal{br}}), X_{\textnormal{br}} = X(t=t_{\textnormal{br}}, r=r_{\textnormal{br}})$,
and we take initial conditions
\begin{gather}
    T(X_{\textnormal{br}}) = T_{\textnormal{br}} \: , \quad \theta(X_{\textnormal{br}}) = \theta_{\textnormal{br}} \: , \quad r(X_{\textnormal{br}}) = r_{\textnormal{br}} \: , \\ A(X_{\textnormal{br}}) = 0 \: , \quad
    T'(X_{\textnormal{br}}) = \frac{\sqrt{1 - \frac{f(r_{\textnormal{br}})}{r_{\textnormal{br}}^{2} \tilde{T}^{2}}}X_{\textnormal{br}} - T_{\textnormal{br}}}{\sqrt{1 - \frac{f(r_{\textnormal{br}})}{r_{\textnormal{br}}^{2} \tilde{T}^{2}}}T_{\textnormal{br}} - X_{\textnormal{br}}} \: .
\end{gather}
Provided that this extremal surface does not fall into the singularity, one can integrate up to some cutoff radius $r=r_{\textnormal{max}}$ near the SAdS boundary;
the result of this computation is a cutoff surface area $A^{\textnormal{I}}_{\textnormal{cutoff}}$, a boundary subregion size $\theta^{\textnormal{I}}_{\textnormal{b}}$, and a boundary Schwarzschild time $t^{\textnormal{I}}_{\textnormal{b}}$. (The superscript denotes that these quantities correspond to the interior surface.)

\begin{figure}
\centering
\includegraphics[width=60mm]{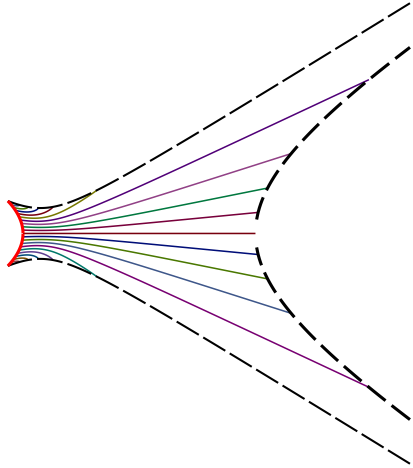}
\caption{Radial profiles of extremal surfaces in Kruskal coordinates $(T, X)$. Those surfaces emitted from the brane at sufficiently late or early times fall into the singularity. }
\label{fig:fallin}
\end{figure}

\subsubsection*{Exterior extremal surfaces}

The exterior extremal surface was computed in Schwarzschild coordinates; again, the geometry is static, so the surface lives in a constant $t$ slice, and one has action
\begin{equation}
    S = 4 \pi \int d \lambda \: r^{2} \sin^{2} \theta \sqrt{\frac{(r')^{2}}{f(r)} + r^{2} (\theta')^{2}} \equiv 4 \pi \int d \lambda \: \mathcal{L} \: .
\end{equation}
There is of course a reparametrization invariance; it is numerically desirable to consider the gauge
\begin{equation}
    M(\lambda) \equiv \frac{(r')^{2}}{f(r)} + r^{2} (\theta')^{2} = 1 \: .
\end{equation}
Substituting this constraint into the equations of motion, one arrives at
\begin{equation}
    2 r' f(r) \cos \theta \sqrt{1 - \frac{(r')^{2}}{f}} + r'' r f(r) \sin \theta + \Big( 3 f(r) - \frac{r}{2} \frac{df}{dr} \Big) (r')^{2} \sin \theta - 3 f(r)^{2} \sin \theta = 0 \: ,
\end{equation}
which can be integrated together with our constraint equation, and the equation for the surface area
\begin{equation}
    A' = 4 \pi r^{2} \sin^{2} \theta \: ,
\end{equation}
to determine the functions $(r(\lambda), \theta(\lambda), A(\lambda))$ given some initial conditions\footnote{The boundary angle $\theta^{\textnormal{E}}_{\textnormal{b}}$ turns out to be a smooth function of $r_{0}$; we can invert this function $\theta^{\textnormal{E}}_{\textnormal{b}}(r_{0})$ to find the appropriate initial condition $r_{0}$ such that $\theta^{\textnormal{E}}_{\textnormal{b}}(r_{0}) = \theta^{\textnormal{I}}_{\textnormal{b}}$. This is necessary in order to compare interior and exterior surfaces subtending the same boundary region.} $r(0) = r_{0}, \theta(0) = 0, A(0) = 0$. We can again integrate up to some radius $r_{\textnormal{max}}$ to find a cutoff area $A^{\textnormal{E}}_{\textnormal{cutoff}}$ and a boundary angle $\theta^{\textnormal{E}}_{\textnormal{b}}$.

\subsubsection*{Regularization of the surface area}

To understand the divergences appearing in the entanglement entropy, it is helpful to work out an explicit expression for the regularized entanglement entropy in the case of vacuum AdS. In this case, the area associated with extremal surfaces in the vacuum geometry may be calculated most easily by working in Poincare-coordinates where the extremal surfaces are hemispheres with some radius $R(\theta_0)$. Making the appropriate change of coordinates and integrating the area up to the value of $z$ that corresponds to $r = r_{\text{max}}$ gives
\be
A_\text{vac}(\theta^{\textnormal{E}}_{\textnormal{b}}) = 2 \pi [r_{\text{max}}^2 \sin^2 \theta^{\textnormal{E}}_{\textnormal{b}} - \ln(2 r_{\text{max}} \sin \theta^{\textnormal{E}}_{\textnormal{b}}) - {1 \over 2} \cos(2 \theta^{\textnormal{E}}_{\textnormal{b}})] + {\cal O}(r_{\text{max}}^{-2})
\ee
In performing numerical calculations, the divergent part of this can be subtracted from the cutoff areas of the extremal surfaces in the black hole geometry to give a finite result in the limit $r_{\textnormal{max}} \to \infty$.

\begin{figure}
\centering
\includegraphics[width=85mm]{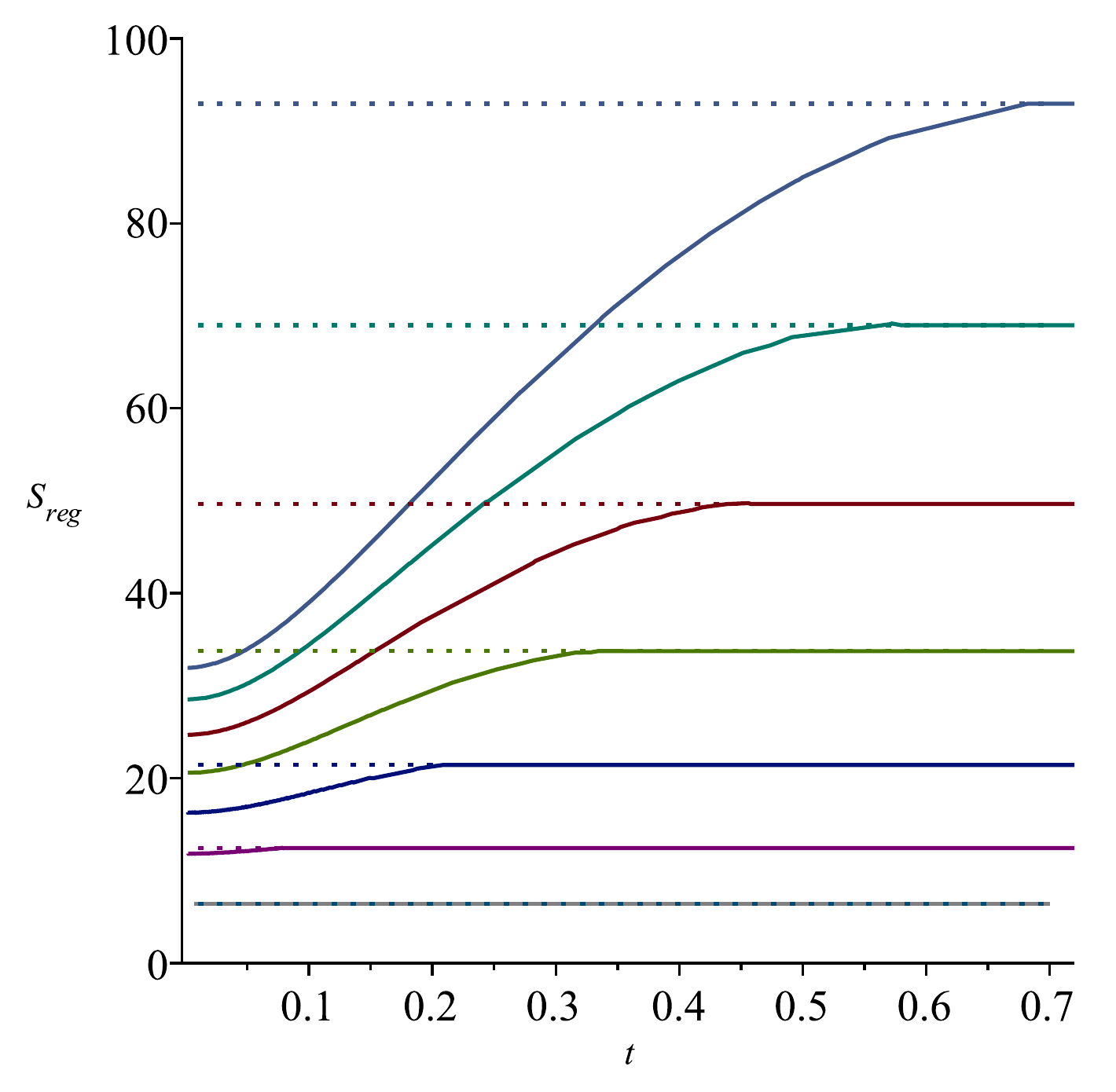}
\caption{Regulated entanglement entropy as a function of time for $\tilde{T}=0.5, r_{\textnormal{H}} = 3 L_{AdS}, r_{\textnormal{max}} = 100$. Plots from bottom to top show $\Delta \theta = 1.0, 1.2, 1.4, 1.6, 1.8, 2.0, 2.2$. }
\label{fig:t_ee_t}
\end{figure}

\begin{figure}
\centering
\includegraphics[width=85mm]{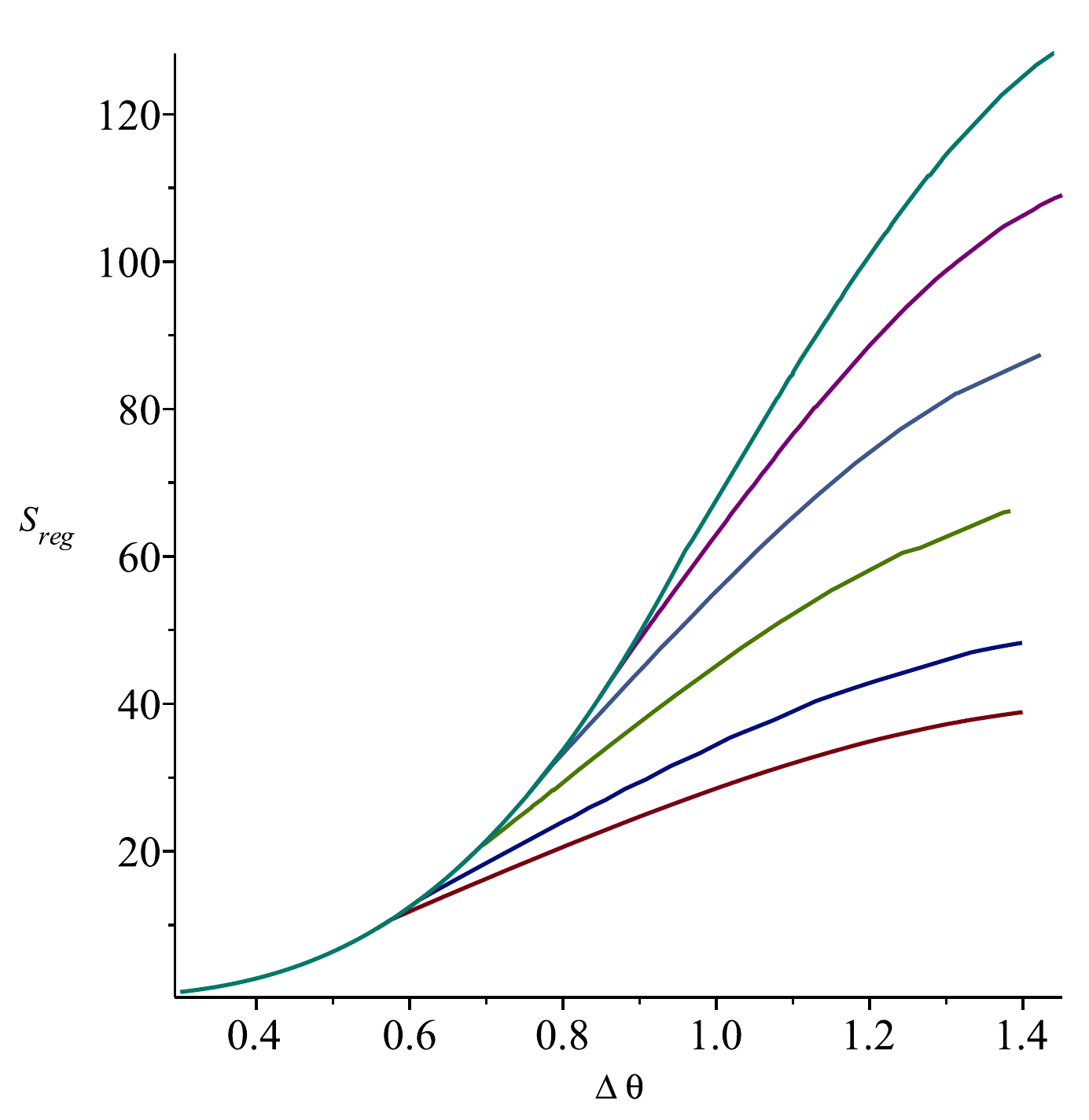}
\caption{Regulated entanglement entropy as a function of subregion size for $\tilde{T}=0.5, r_{\textnormal{H}} = 3 L_{AdS}, r_{\textnormal{max}} = 100$. Plots from bottom to top show $t/L_{AdS} = 0, 0.1, 0.2, 0.3, 0.4, 0.5$. }
\label{fig:t_ee_angle}
\end{figure}

The results of this computation are found in Figures \ref{fig:t_ee_t} and \ref{fig:t_ee_angle}. The results are qualitatively similar to the case of $d=2$ dimensions; in particular, for a boundary subregion of sufficiently large size, the entanglement entropy has a period of time dependence, during which the extremal surface probes the brane geometry.
However, whereas in $d=2$ the entanglement entropy was independent of the size of the boundary subregion whilst the minimal area surface was probing the brane, this is visibly no longer the case in $d=4$. This property was unique to $d=2$, where the area of the interior extremal surface was independent of the size of the subtended boundary region.

\section{Entanglement entropy: SYK model calculation}

Here we study a coupled-cluster generalization~\cite{Gu:2016sykchain} of the single SYK cluster consider in \cite{Kourkoulou:2017zaj}. The first step is to define the analog of boundary states for this model, which now include both spatial and internal degrees of freedom, and generalize the analysis of \cite{Kourkoulou:2017zaj}. We also present entanglement data obtained from exact diagonalization of a single cluster and two coupled clusters which corroborate the holographic entanglement calculations above.

Consider $LN$ Majorana fermions $\chi_{r,a}$ with $r=1,\cdots,L$ and $a=1,\cdots, N$ with $N$ even. The basic anticommutator is
\begin{equation}
    \{\chi_{r,a},\chi_{r',a'}\} = \delta_{r,r'} \delta_{a,a'}.
\end{equation}
The Majorana fermions are arranged in the Hamiltonian into $L$ clusters of $N$ Majoranas each with the clusters having only nearest neighbor interactions. The Hamiltonian is
\begin{equation}
    H = \sum_{r=1}^L \sum_{a<b<c<d} J_{abcd} \chi_{r,a} \chi_{r,b} \chi_{r,c} \chi_{r,d} + \sum_{r=1}^L \sum_{a<b, c<d} \tilde{J}_{abcd} \chi_{r,a}\chi_{r,b} \chi_{r+1,c} \chi_{r+1,d},
\end{equation}
assuming periodic boundary conditions. The couplings are Gaussian random variables with zero mean and variance
\begin{equation}
    \overline{J_{abcd}^2} = \frac{6 J_0^2}{N^3}
\end{equation}
and
\begin{equation}
    \overline{\tilde{J}_{abcd}^2} = \frac{J_1^2}{N^3}.
\end{equation}

The bare Euclidean 2-point function is
\begin{equation}
    \langle \chi_{r,a}(\tau) \chi_{r',a'}\rangle = \frac{1}{2} \sgn (\tau) \delta_{r,r'} \delta_{a,a'}.
\end{equation}
The dressing is the usual melonic large $N$ analysis, but here extended to the coupled chain~\cite{Gu:2016sykchain}. For our present purpose, the key point of this analysis is that the system possesses an emergent $O(N)^L$ symmetry at large $N$. Essentially, one can apply an independent $O(N)$ transformation acting on the $a$ index of $\chi_{r,a}$ at every site of the chain. This occurs because, ignoring a possible spin glass or localized phase, the $J$ and $\tilde{J}$ couplings can be treated as dynamical fields with a particular 2-point function, at large $N$.

A complete basis for the Hilbert space can be obtained as follows. For each pair of Majorana operators in a cluster, $\chi_{r,2k-1}$ and $\chi_{r,2k}$, define the complex fermion
\begin{equation}
    c_{r,k}= \frac{\chi_{r,2k-1}+i \chi_{r,2k}}{\sqrt{2}}.
\end{equation}
These fermions obey the usual algebra, $\{ c_{r,k},c^\dagger_{r',k'}\} = \delta_{r,r'} \delta_{k,k'}$. It is convenient to label the Hilbert space using the spin-like operator $\hat{s}_{r,k} = 1- 2 c_{r,k}^\dagger c_{r,k} = \pm 1$. In terms of the Majoranas, it is
\begin{equation}
   \hat{s}_{r,k} = 1-2c_{r,k}^\dagger c_{r,k} = -2 i \chi_{r,2k-1} \chi_{r,2k}.
\end{equation}
The mutual eigenbasis of all the $\hat{s}_{r,k}$ operators forms a complete basis denoted $|s \rangle$ and obeying
\begin{equation}
    \hat{s}_{r,k} | s\rangle = s_{r,k} |s \rangle.
\end{equation}
Note that the transformations which flip a particular even numbered $\chi$, such as taking $\chi_{r,2k}$, to $-\chi_{r,2k}$, also flips the eigenvalue of $\hat{s}_{s,k}$.

Now consider the imaginary time evolved $|s\rangle$ basis,
\begin{equation}
    |s,\beta \rangle = e^{-\beta H/2} | s\rangle.
\end{equation}
Let $Q_{r,k}$ denote the unitary which sends $\chi_{r,2k}$ to $-\chi_{r,2k}$. The idea of the analysis in \cite{Kourkoulou:2017zaj} is, roughly speaking, that the Hamiltonian is invariant under $Q_{r,k}$ at large $N$, so that when computing correlation functions one can use the relation
\begin{equation} \label{eq:syk_QH}
    Q_{r,k} e^{-\beta H/2} |s \rangle \sim e^{-\beta H/2} Q_{r,k} |s \rangle,
\end{equation}
though it is not literally true for fixed $J$ and $\tilde{J}$.

The goal is to analyze various physical properties in the states $|s,\beta \rangle$. The most basic object is the 2-point function,
\begin{equation}
    G_{r,a}(\tau;s,\beta) = \frac{\langle s,\beta |\chi_{r,a}(\tau) \chi_{r,a} |s,\beta \rangle}{\langle s,\beta | s,\beta\rangle }.
\end{equation}
Since each $\chi_{r,a}$ is mapped to $\pm \chi_{r,a}$ by $Q_{r,k}$, it follows from Eq.~\eqref{eq:syk_QH} that $G_{r,a}(\tau;s,\beta)$ is actually independent of $s$, at least to leading order at large $N$. Hence, even though the states $|s,\beta\rangle$ are not translation invariant in general, the 2-point function in state $|s,\beta\rangle$ is approximately translation invariant.

To determine the value of $G_{r,a}(\tau;s,\beta)$, first observe that the leading large $N$ part of $\langle s ,\beta| s,\beta\rangle $ is also independent of $s$ by virtue of Eq.~\eqref{eq:syk_QH}. Summing over $s$ gives
\begin{equation}
    \sum_{s} \langle s ,\beta | s,\beta\rangle = \text{Tr}(e^{-\beta H}) = Z(\beta),
\end{equation}
so since each term is approximately equal, it must be that
\begin{equation}
    \langle s,\beta | s,\beta \rangle \approx \frac{Z(\beta)}{\mathcal{D}}.
\end{equation}
This in turn implies that $G_{r,a}(\tau; s,\beta)$ must be given by the thermal answer at inverse temperature $\beta$ independent of $s$.

One property of particular interest is the entanglement entropy of subregions in the state $|s,\beta\rangle$. The $n$-th Renyi entropy of a subset $A$ of Majorana fermions in the normalized state
\begin{equation}
    \sigma(s,\beta) = \frac{|s,\beta\rangle \langle s,\beta|}{\langle s,\beta | s,\beta \rangle}
\end{equation}
is
\begin{equation}
    e^{-(n-1)S_n(A)} = \text{Tr}\left( \Pi^A_n \sigma(s,\beta)^{\otimes n} \right).
\end{equation}
Here $\Pi_n^A$ is a shift operator acting on the $n$ copies which swaps fermions from the set $A$ between the copies. It is defined for a single pair of Majoranas below. Crucially, it is invariant under the $Q_{r,k}$ transformation provided it is enacted in every copy (replica) simultaneously. Hence at the level of rigor we have been observing, it follows that the large $N$ part of the Renyi entropy of a collection $A$ in state $|s,\beta\rangle$ is independent of $s$.

The value of $S_n(A)$ is less clear. The same trick, summing over $s$, which showed that $G_{r,a}(\tau;s,\beta)$ was thermal does not work here because there are two copies of the state appearing. While the thermal Renyi entropy is one natural candidate, this cannot be true for all collections since the state is pure. At a minimum, non-thermality must occur when $A$ exceeds half the total system. However, it is certainly consistent to lose thermality for smaller sets, as this occurs in holographic calculations. To say more requires a detailed calculation of the Renyi entropy using the replicated path integral, which we defer to future work.

Note that, in the numerical data reported below, the entanglement entropy of subsystems is computed by first grouping fermions into pairs and performing a Jordan-Wigner transformation to a spin basis. The definition of entanglement in the spin basis is trivial, and moreover, one can show that the precise location of the Jordan-Wigner string does not effect the entropy calculation. This is because given two different strings, meaning two different mappings of fermion states to spin states, the two final sets of spin states are related by a local unitary. Hence as long a fixed fermion pairing is chosen to define the spins, the choice of string is actually irrelevant since entanglement entropy is invariant under local unitary transformations.

\subsection{Data for a single SYK cluster}

\begin{figure}
    \centering
    \includegraphics[width=.9\textwidth]{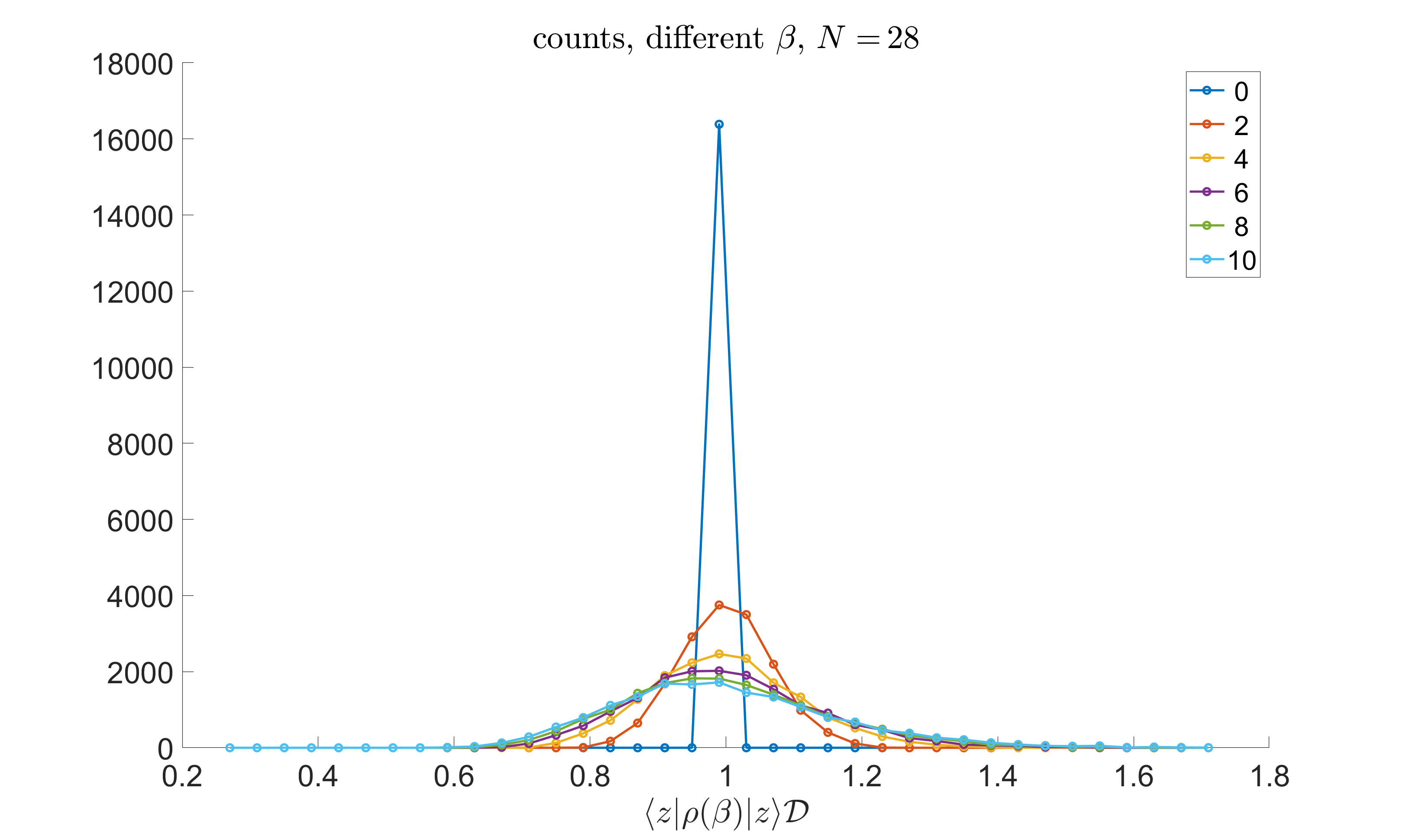}
    \caption{Histogram of $\langle s,\beta | s,\beta\rangle$ for $N=28$ Majorana fermions in a single SYK cluster ($L=1$). The different curves correspond to $\beta=0,\cdots,10$ in units with $J_0=1$. There is a strong concentration around the value predicted by the random model studied above.}
    \label{fig:syk_rhodiag_counts}
\end{figure}

Here data is presented for a single SYK cluster, $L=1$, for a variety of $N$ and $\beta$. Turning first to the diagonal matrix elements of the thermal state, Figure~\ref{fig:syk_rhodiag_counts} shows a histogram of $\langle s,\beta | s,\beta \rangle$ for all $s$ for an $N=28$ cluster. There is a clear concentration around the central value of $Z(\beta)/\mathcal{D}$ and some evidence of an emerging universal distribution at large $\beta$, although the data are also consistent with the distribution merely varying slowly with $\beta$.

\begin{figure}
    \centering
    \includegraphics[width=.9\textwidth]{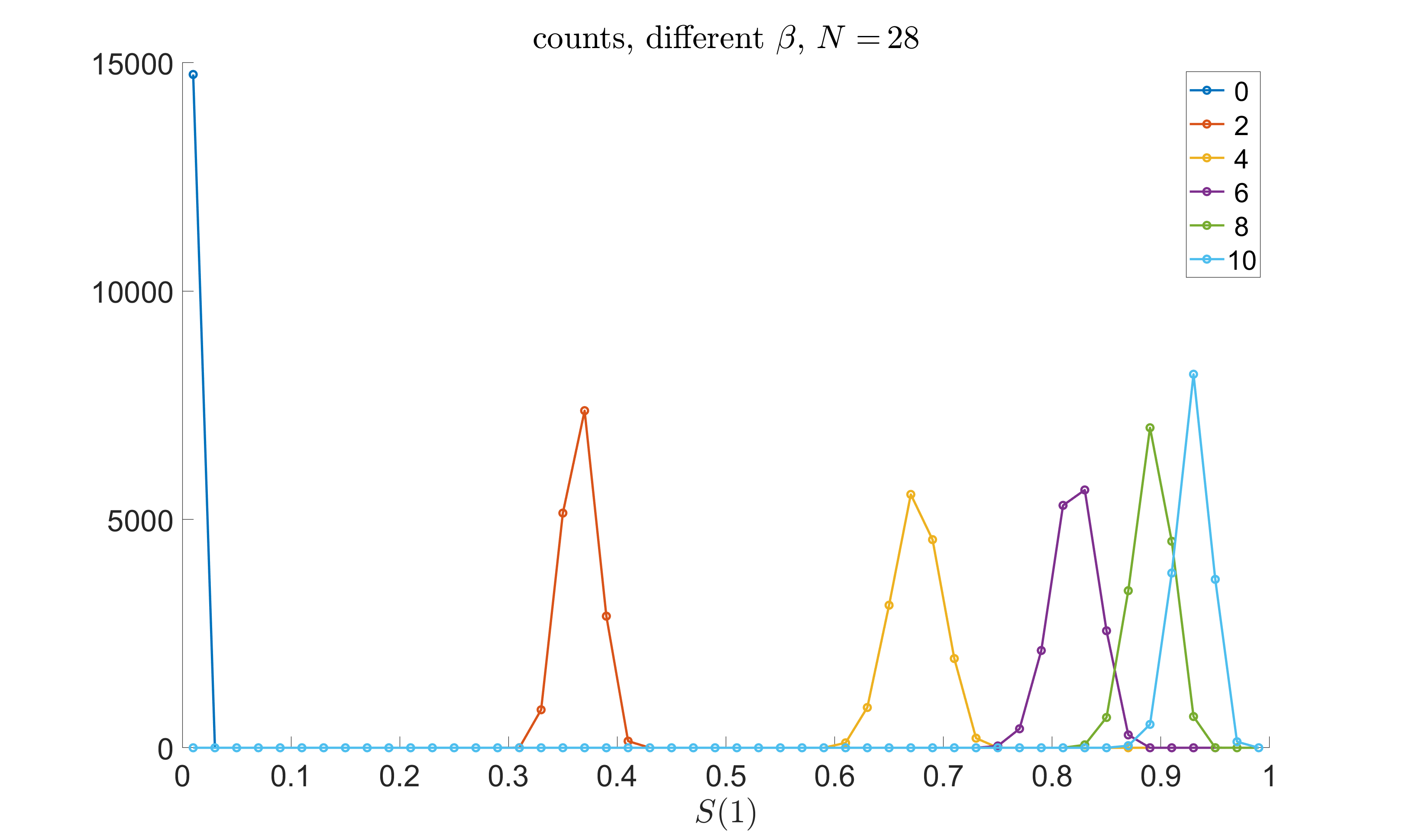}
    \caption{Histogram of the entropy of one pair of Majoranas for $N=28$ Majorana fermions in a single SYK cluster ($L=1$). The different curves correspond to $\beta=0,\cdots,10$ in units with $J_0=1$. As $\beta$ varies, the entropy increases from zero and remains reasonably peaked. As the average approaches one, the distribution appears to become more peaked, possibly indicating convergence to a value independent of $s$ at large $\beta$ and large $N$.}
    \label{fig:syk_S_counts}
\end{figure}

Turning to the entanglement of subsets of the Majoranas, Figure~\ref{fig:syk_S_counts} shows a histogram of the entanglement of the first site for various $\beta$s and $N=28$. As $\beta$ increases, the distribution appears to peak near one, although the width does not dramatically decrease with increasing $\beta$. An analysis of the data for smaller values of $N$ suggests that the distribution is also becoming sharper as $N$ increases.

\begin{figure}
    \centering
    \includegraphics[width=.9\textwidth]{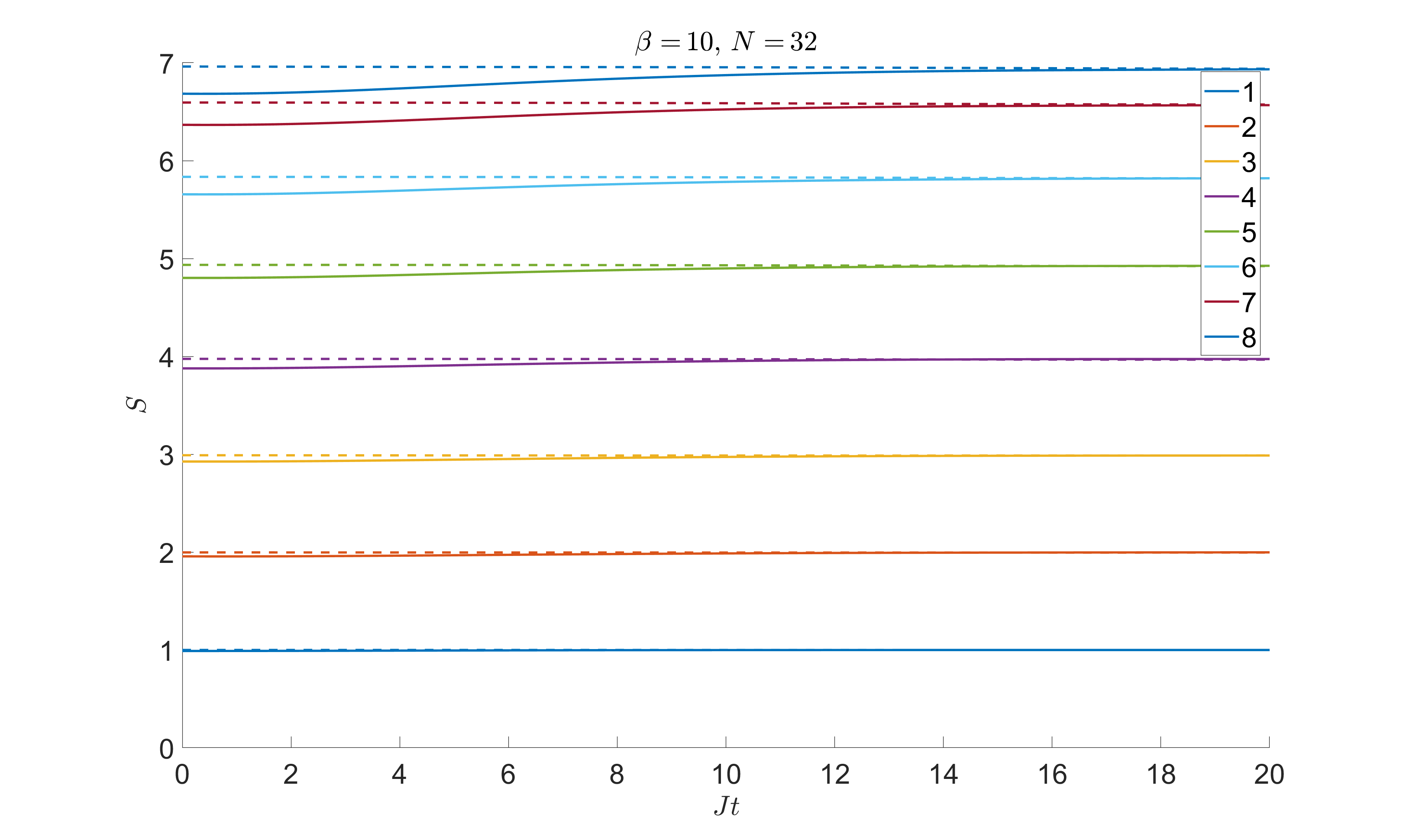}
    \caption{The solid lines are the entropies of different sized subsystems as a function of time for $N=32$ Majoranas in a single SYK cluster ($L=1$) with $\beta=10$. The dashed lines show the same subsystem entropies in a random state which has been evolved in imaginary time as a proxy for the thermal entropy. After a short time of order $\beta$, all subsystem entropies have reached their late time thermal values. }
    \label{fig:syk_S_time}
\end{figure}

Next we consider the time evolution of entanglement, with Figure~\ref{fig:syk_S_time} showing the time evolution of entanglement for a single state $s$ and $N=32$ fermions. For small subsystems, the entanglement entropy is close to the thermal value (obtained by imaginary time evolution acting on a random Hilbert space state) even at zero time. The result is similar to the holographic results, where it was found that small subsystems look exactly thermal to leading order in large $N$. By contrast, larger systems deviate from thermality at early time but quickly thermalize. Unlike the holographic calculations, there is no sharp transition as subsystem size is increased, but such a transition is not expected at finite $N$.

To show that such imaginary time evolved boundary have a thermal character for systems beyond SYK at large-$N$, Appendix~\ref{app:entgrowth} and Appendix~\ref{app:sqrt} contain simple spin systems where very rapid entanglement growth and other thermal properties of boundary states can be shown exactly.

\subsection{Data for two coupled clusters}

\begin{figure}
    \centering
    \includegraphics[width=.9\textwidth]{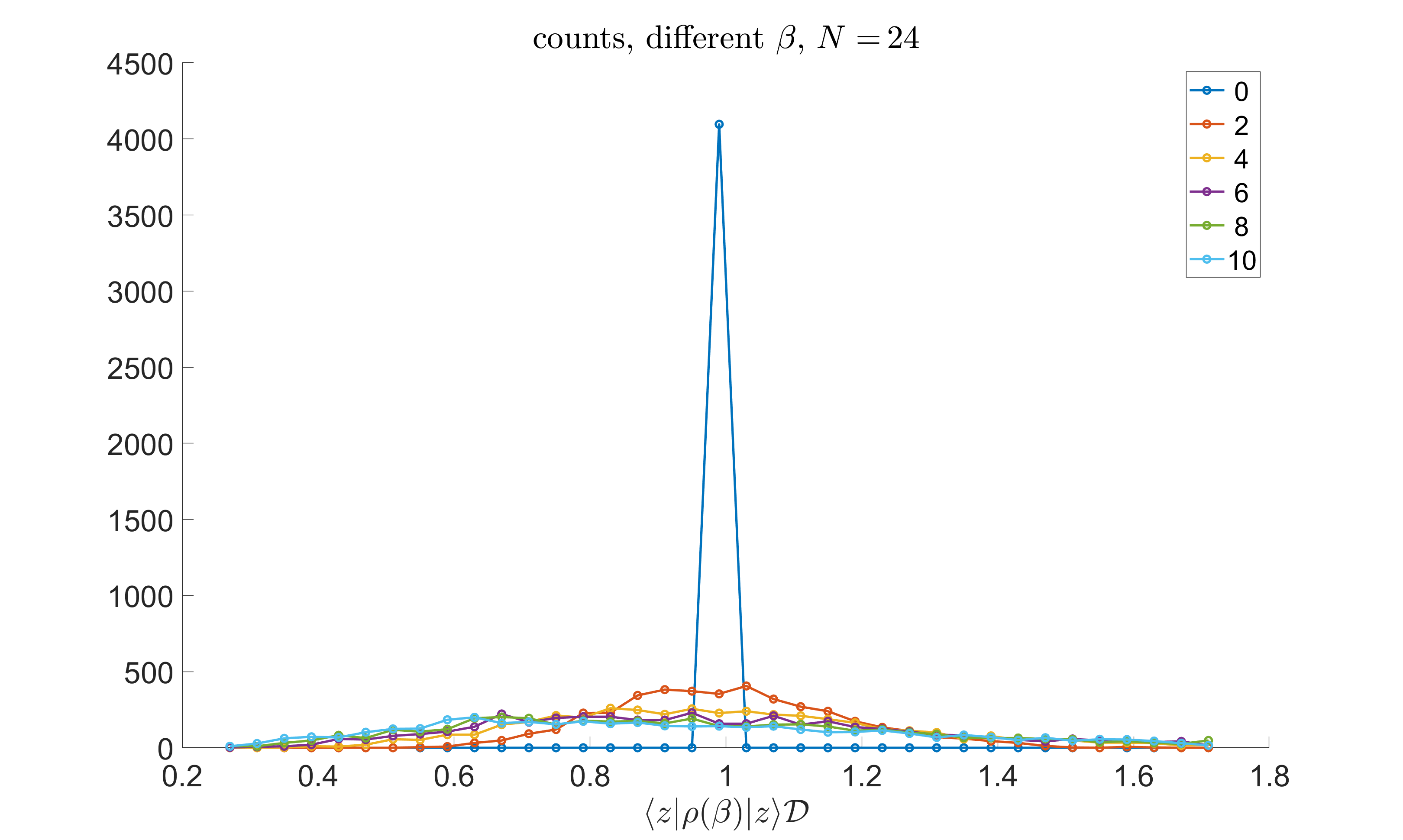}
    \caption{Histogram of $\langle s | \rho(\beta) | s \rangle \mathcal{D}$ for two coupled SYK clusters corresponding to $L=2$ and $N=12$. The different curves correspond to $\beta=0,\cdots,10$ in units with $J_0=1$.}
    \label{fig:2syk_rhodiag}
\end{figure}

\begin{figure}
    \centering
    \includegraphics[width=.9\textwidth]{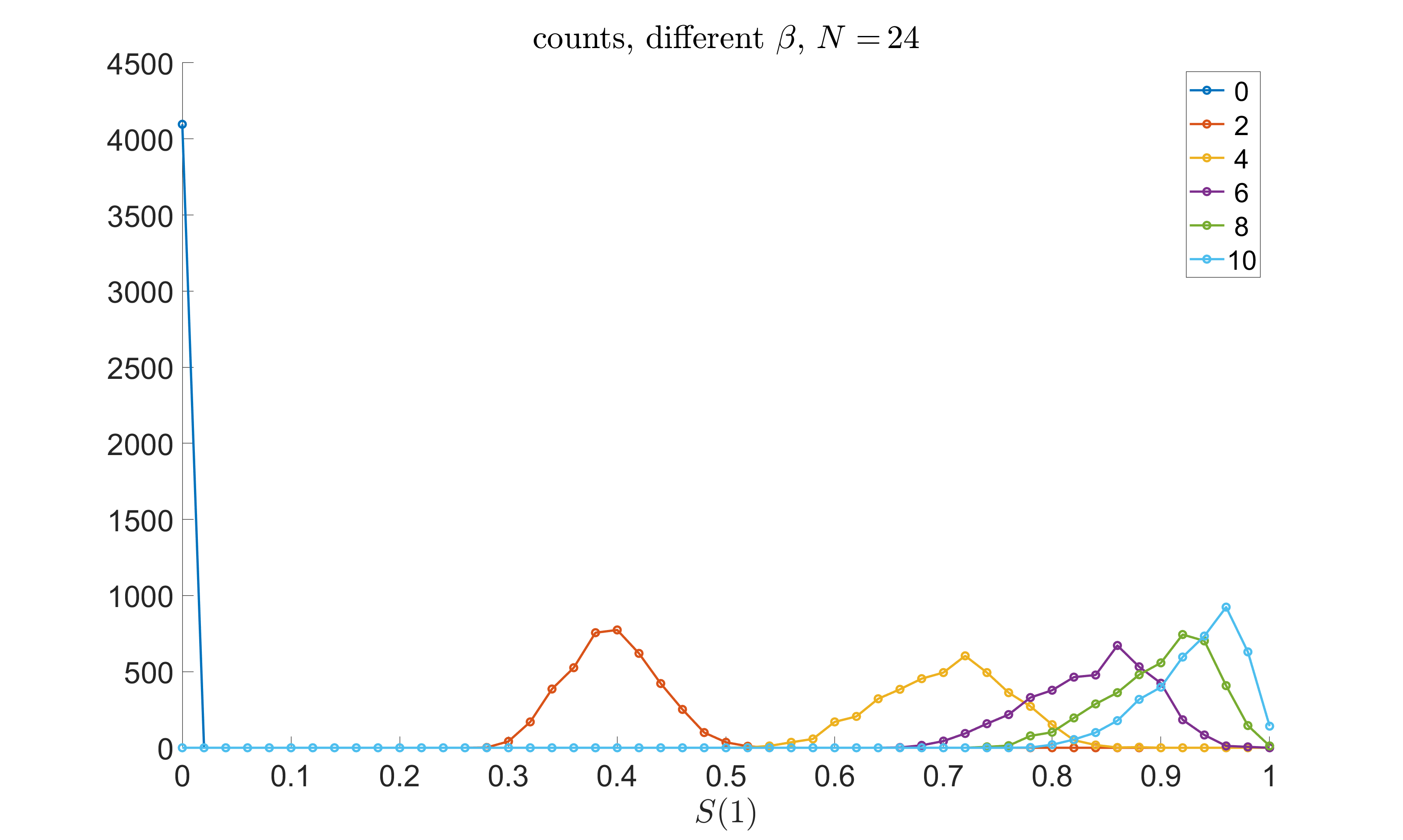}
    \caption{Histogram of the entropy of one cluster relative to thermal value for two coupled SYK clusters corresponding to $L=2$ and $N=12$. The different curves correspond to $\beta=0,\cdots,10$ in units with $J_0=1$.}
    \label{fig:2syk_S}
\end{figure}

\begin{figure}
    \centering
    \includegraphics[width=.9\textwidth]{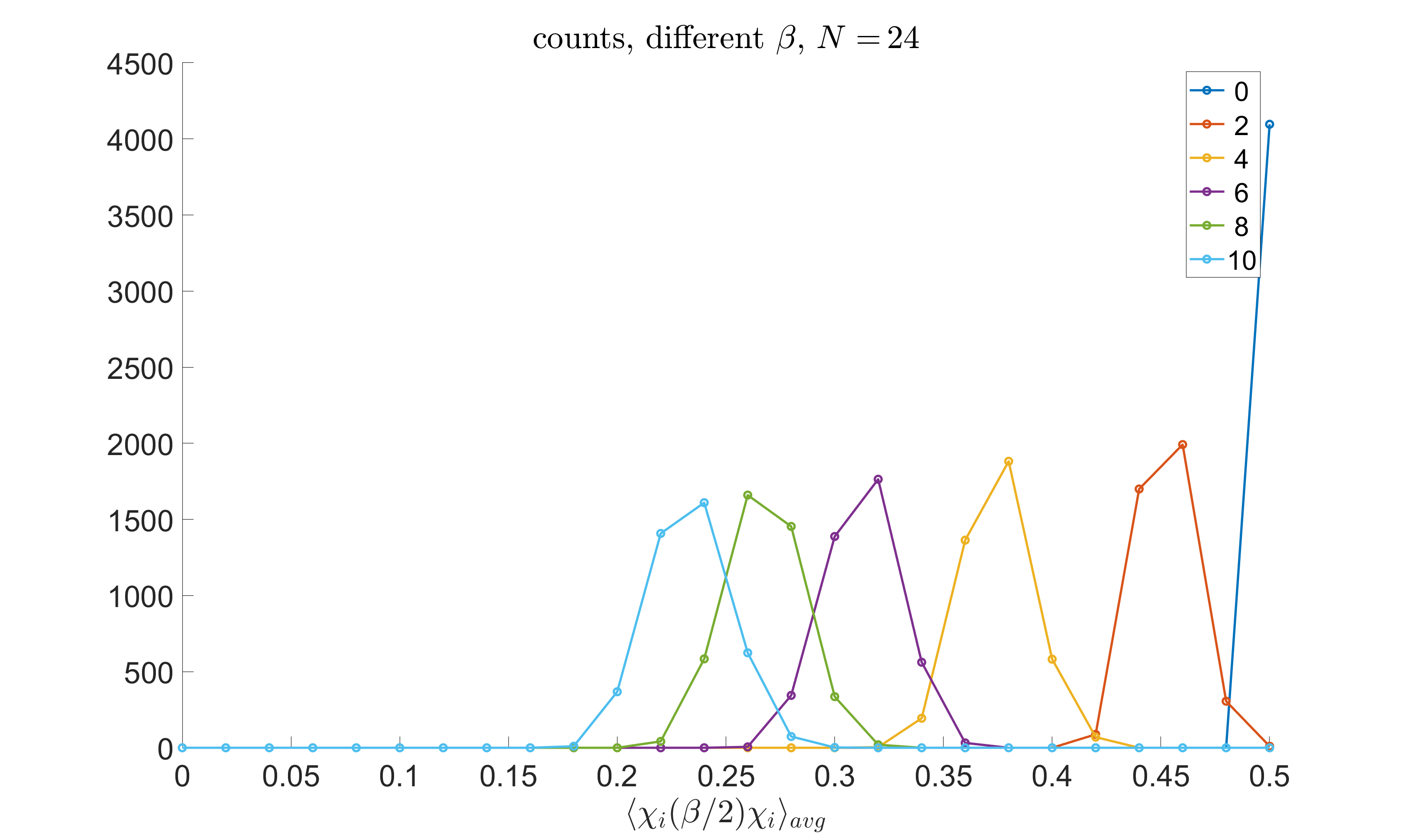}
    \caption{Histogram of TFD-like correlation averaged over fermions for two coupled SYK clusters corresponding to $L=2$ and $N=12$. The different curves correspond to $\beta=0,\cdots,10$ in units with $J_0=1$.}
    \label{fig:2syk_chi2_avg}
\end{figure}

The single cluster analysis can be repeated for two coupled clusters, with the caveat that adding a second cluster reduces the number of fermions that can be studied in each cluster. Figures~\ref{fig:2syk_rhodiag}, \ref{fig:2syk_S}, and \ref{fig:2syk_chi2_avg} show data for two coupled SYK clusters, $L=2$, with $N=12$ Majoranas in each cluster. Some similar features to the single cluster case are visible, although the necessarily smaller sizes induce larger finite size effects.

In Figure~\ref{fig:2syk_rhodiag} we see evidence that the diagonal matrix elements of the thermal density are beginning to concentrate near the value $Z(\beta)/\mathcal{D}$ predicted by the large-$N$ analysis. However, the distribution is considerably wider. One possible explanation is that the much smaller value of $N$ has led to much larger finite size effects. Figure~\ref{fig:2syk_S} shows a histogram of the entanglement of one cluster normalized to its thermal value. A similar kind of concentration effect near the thermal value is seen as $\beta$ is increased.

Finally, Figure~\ref{fig:2syk_chi2_avg} shows a thermofield double-like correlation averaged over all the fermions. Those data also show signs of concentrating near the thermal value, albeit with significant width to the distribution. It is plausible that this broadening is a finite size effect coming from the rather small value of $N$ on each cluster in the two cluster system.

We did not study time-evolution of entanglement for the two cluster system because the single cluster data is already a reasonable caricature of the holographic results and the numerics do not have enough spatial resolution to study in detail the dependence on spatially non-uniform boundary states. The above data for $L=2$ indicate that the thermal behavior of boundary states expected at large-$N$ is beginning to emerge for two coupled SYK clusters at quite modest $N$, but a definite conclusion is hard to make from the finite size numerical data.

In Appendix~\ref{app:sqrt} we exhibit a simple model with spatial locality where the thermality of simple correlators can be shown rigourously. Hence, evidence is accumulating that imaginary time evolved states across a broad class of models, including those with spatial locality, have a thermal character.

%\begin{figure}
%    \centering
%    \includegraphics[width=.9\textwidth]{n24_2cluster_chi2_counts.png}
%    \caption{Histogram of TFD-like correlation.}
%    \label{fig:2syk_chi2}
%\end{figure}

\subsection{Swap operator for fermions}

Given $n$ fermion modes, the shift operator, $\Pi_n$, is defined by $\Pi_n a_i \Pi_n^{-1} = a_{i+1}$ for $i<n$ and $\Pi_n a_n \Pi_n^{-1} = (-1)^{n-1} a_1$. Its meaning is obtained from its relation to Renyi entropies. Given a fermion density matrix $\rho = (1-p) + (2p-1) a^\dagger a$, the $n$-th Renyi entropy of $\rho$ is
\begin{equation}
    e^{-(n-1)S_n} = (1-p)^n + p^n.
\end{equation}
From the definition of $\Pi_n$ it follows that the empty state and the full state are mapped to themselves with no phase factor by $\Pi_n$. The factor of $(-1)^{n-1}$ is needed to ensure that the full state does not acquire a phase, since
\begin{equation}
    \Pi_n a_1^\dagger \cdots a_{n}^\dagger \Pi_n^{-1} = (-1)^{n-1} a_2^\dagger \cdots a_n^\dagger a_1^\dagger = a_1^\dagger \cdots a_n^\dagger.
\end{equation}
Every other state in the $a_i$ basis is mapped to an orthogonal state (obtained, up to a phase, by rearranging the occupation numbers). Hence the expectation value of $\Pi_n$ in the $n$-copy state is
\begin{equation}
     \text{Tr}\left( \Pi_n   \prod_{i=1}^n \rho(a_i) \right) = (1-p)^n + p^n,
\end{equation}
the desired Renyi entropy.

Now suppose each $a_i$ is written in terms of Majorana operators,
\begin{equation}
    a_i = \frac{\chi_i + i\tilde{\chi}_i}{\sqrt{2}},
\end{equation}
and consider the transformation $\tilde{\chi}_i \rightarrow - \tilde{\chi}_i$. This transformation maps $a_i$ to $a_i^\dagger$ and hence exchanges the empty and filled states. Moreover, it commutes with the transformation induced by $\Pi_f$, hence if the unitary $Q$ implements the sign inversion, then $Q \Pi_n Q^{-1} = \Pi_n$. For example, with two copies, $n=2$, the shift is
\begin{equation}
    \Pi_2 = e^{-\frac{\pi}{2} (a_1^\dagger a_2 - a_2^\dagger a_1)},
\end{equation}
which enacts $\Pi_2 a_1 \Pi_2^{-1} = a_2$ and $\Pi_2 a_2 \Pi_2^{-1} = -a_1$. Its Majorana representation is
\begin{equation}
    \Pi_2 = e^{-\frac{\pi}{2} (\chi_1 \chi_2 + \tilde{\chi}_1 \tilde{\chi}_2)},
\end{equation}
which is manifestly invariant under a sign flip of all $\tilde{\chi}_i$.

The generalization to many modes in a single copy is straightforward. The conclusion remains the same: the swap operator is invariant under the transformation $\chi_{i,\alpha}\rightarrow -\chi_{i,\alpha}$ provided it acts on all copies simultaneously.

\section{Holographic Complexity}

We have seen that the entanglement entropy for sufficiently large CFT subsystems can provide a probe of behind-the-horizon physics for our black hole microstates. In \cite{Susskind:2014rva} and \cite{Brown:2015bva}, a pair of additional probes capable of providing information behind the horizon were defined holographically and conjectured to provide a measure of the complexity of the CFT state.\footnote{For a more detailed exposition of definition and calculation of holographic complexity, see \cite{Carmi:2016wjl}.} The first, which we denote by $\mathcal{C}_V$, is proportional to the volume of the maximal-volume spacelike hypersurface ending on the boundary time slice at which the state is defined \cite{Susskind:2014rva}. The second, which we denote by $\mathcal{C}_A$, is proportional to the gravitational action evaluated on the spacetime region formed by the union of all spacelike hypersurfaces ending on this boundary time slice (called the Wheeler-deWitt patch for this time slice) \cite{Brown:2015bva}.

In this section, we explore the behaviour of both of these quantities as a function of time and the parameter $T$ for our microstates in the case $d=2$. We will see that while the late-time growth of both quantities is the same and matches the expectations for complexity, the time-dependence at early times is significantly different. This may provide some insight into the CFT interpretations for these two quantities.

\subsection{Calculation of $\mathcal{C}_V$ for $d=2$}

The volume-complexity for a CFT state defined on some boundary time slice is defined holographically as
\be
\mathcal{C}_V = \frac{V}{G l} \; ,
\label{eq:CVdef}
\ee
where $V$ is the volume of the maximal-volume co-dimension one bulk hypersurface anchored at the asymptotic CFT boundary on the time slice in question. Here, $l$ is a length scale associated to the geometry in question, taken here to be $L_\AdS$.  We will generally set $L_{AdS} = 1$ and make use of the $s, y$ coordinates defined in appendix \ref{app:sycoords}.

Consider the boundary time-slice corresponding to a particular time $s_0$ at the boundary. The maximal volume bulk hypersurface anchored here will wrap the circle direction and have some profile $s(y)$ in the other two directions. For a surface described by such a parametrization, the volume is
\be
V =2 \pi r_H \int dy {\cos(s) \over \cos^2(y)} \sqrt{ 1- \left({ds \over dy}\right)^2} \; .
\ee
Extremizing this gives
\be
{d^2 s \over dy^2} = \left(1 - \left({ds \over dy} \right)^2 \right)\left(\tan(s) - 2 \tan(y) {ds \over dy} \right) \; .
\ee
Maximizing volume also requires that the slice intersects the ETW brane normally,
\be
{ds \over dy} = 0 \qquad y = y_0 \; .
\ee
We regulate the volume by integrating up to $r_{\text{max}}=L/\epsilon$ in the Schwarzschild coordinates. We can subtract the regulated volume for pure AdS to obtain a result that is finite for $\epsilon \to 0$. This regulated volume for pure AdS (working in Schwarzschild coordinates with $f(r) = r^2 + 1$)
\bea
V_{AdS} &=& \int_0^{1 \over \epsilon} dr 2 \pi r \sqrt{{1 \over f(r)} - f(r)\left({dt \over dr} \right)^2} \cr
&=& 2 \pi \left[{1 \over \epsilon} - 1 + {\cal O}(\epsilon)\right]
\eea
In the $s-y$ coordinates, this maximum value corresponds to
\bea
y_{\text{max}} &=&  \arctan \left( e^{- r_H t} \sqrt{r_{\text{max}} - r_H \over r_{\text{max}} + r_H} \right) + \arctan \left( e^{r_H t} \sqrt{r_{\text{max}} - r_H \over r_{\text{max}} + r_H} \right) \cr
&=& \pi/2 - \epsilon {r_H \over \cosh(t r_H)} + {\cal O}(\epsilon^2)
\eea
The values of $s$ at the boundary are related to the original Schwarzschild time by
\be
t = {1 \over r_H} \ln(\tan(\pi/4 + s/2)) \; .
\ee
We find that there is a monotonic relationship between the intersection time $s_0$ of the maximal volume slice with the ETW brane and the Schwarzschild time of the maximal volume slice at the AdS boundary. A finite range $s_0 \in [-s_*,s_*]$ with $s_* < \pi/2$ maps to the full range $t \in [-\infty,\infty]$ of Schwarzschild time. We have that $s_* \to 0$ as $T \to 1$ or equivalently as $y_0$ (the brane location) approaches $-\pi/2$.

For $t=0$, the maximal volume slice is just the $s=0$ slice of the spacetime, and the subtracted volume is
\bea
V_{t=0} &=& 2 \pi r_H \int_{y_0}^{y_{\text{max}}} {dy \over \cos^2 y} - V_{AdS} \cr
&=& \lim_{\epsilon \to 0}\left[2 \pi r_H (\tan(y_{\text{max}}) - \tan(y_0)) - V_{AdS}(r_{\text{max}})\right] \cr
&=& 2 \pi ( 1 + r_H \tan |y_0|) \cr
&=& 2 \pi (1 + {r_H T \over \sqrt{1 - T^2}} )
\eea
It is actually convenient to subtract off the $2 \pi$ here and below, since the remaining volumes are all proportional to $r_H$. We will refer to this subtracted volume as $\Delta V$.

\begin{figure}
\centering
\includegraphics[width=120mm]{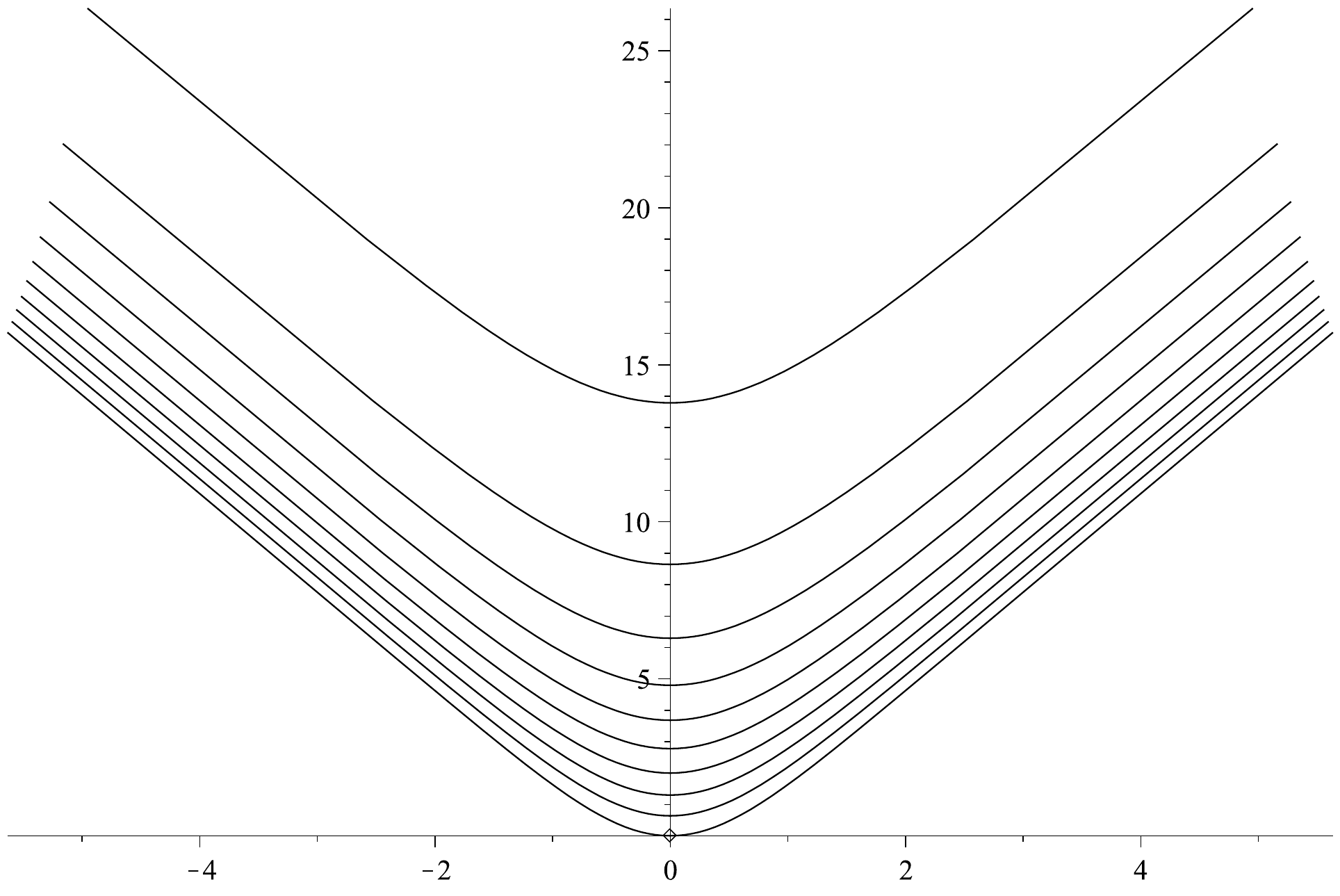}
\caption{Volume $\Delta V/r_H$ of maximal slice vs Schwarzschild time $r_H t_\infty$ for $T = k/10$, $k=0..9$ from bottom to top.}
\label{fig:volumes}
\end{figure}

We can numerically find the maximal volume slices and evaluate $\Delta V$ for different values of $s_0$ to understand how the volume depends on time. For each $s_0$ we calculate $t_\infty$, the Schwarzschild time where the slice intersecting the ETW brane at $s_0$ intersects the AdS boundary. The results for $\Delta V/r_H$ vs $t_\infty r_H$ are independent of $r_H$; these are plotted in figure \ref{fig:volumes}.

As a function of Schwarzschild time, the regulated volume increases smoothly to infinity as $t \to \infty$, with a linear increase in volume as a function of Schwarzschild time for late times. The slope is the same in all cases,
\be
{d V \over d t} \sim \pi r_H^2\;.
\ee
Using this result to compute the late time rate of change of volume-complexity, one finds:
\bea
\lim_{t\to\infty} \frac{d C_V}{d t} &=& \frac{\pi r_H^2}{G} \cr
						&=& 8 \pi M
\label{eq:CVlaterate}
\eea
where we have used the relation
\be
r_H^2 = 8 G M
\ee
between the horizon radius $r_H$ and the black hole mass $M$ for a non-rotating BTZ black hole.

The same slope can be obtained analytically as a lower bound by noting that in the future interior region, which can be described by Schwarzschild coordinates with\footnote{These are related to the $u,v$ coordinates by $u = e^{r r_H} \sqrt{r_H + t \over r_H - t}$, $v = e^{-r r_H} \sqrt{r_H + t \over r_H - t}$.}
\be
ds^2 = - {dt^2 \over r_H^2 - t^2} + (r_H^2 - t^2) dr^2 + t^2 d \theta^2 \; ,
\ee
with $t \in [-r_H,0]$,  there is an extremal volume surface $\Sigma$ described by
\be
t = -{\sqrt{2} \over 2} r_H \; .
\ee
This is a tube with constant radius $r_H/\sqrt{2}$. In the $u,v$ coordinates, this is $uv = (2- \sqrt{2})/(2 + \sqrt{2})$. From a time $t_\infty$ at the AdS boundary, we can consider a surface which lies along a future-directed lightlike surface $u = e^{r_H t_\infty}$ until the intersection with $\Sigma$ and then along $\Sigma$ until the intersection with the ETW brane. The part of this surface with $y > 0$ has volume
\be
V = \pi r_H^2 t_\infty + \pi r_H \ln\left({1 \over \sqrt{2} - 1} \right) \; .
\ee
This gives a lower bound for the maximal volume, and has the same time derivative as our result above.

The late time growth of $C_V$ is in line with earlier studies (e.g. \cite{Susskind:2014rva,Carmi:2017jqz}) of holographic complexity for black hole states (e.g. evolution of the two-sided black hole with forward time-evolution on both sides,) and has the same qualitative bulk explanation. We also see a monotonic increase for all $t>0$, as would generically be expected for the evolution of complexity in a generic state with less-than-maximal complexity.

\subsection{Calculation of $\mathcal{C}_A$ for $d=2$}

The action-complexity for a CFT state defined on some boundary time slice is defined holographically as
\be
\mathcal{C}_A = \frac{I_\mathcal{W}}{\pi \hbar};.
\label{eq:CAdef}
\ee
Here, $I_\mathcal{W}$ is the value of the gravitational action of the bulk theory when evaluated on some region $\mathcal{W}$. In particular, this region is the Wheeler-DeWitt patch anchored at the asymptotic boundary at the time slice in question. That is, $\mathcal{W}$ is the union of all the spatial slices anchored at this time slice. Again, in these calculations we will take $L_\AdS=1$.

\begin{figure}[h]
\centering
  \includegraphics[width=16cm]{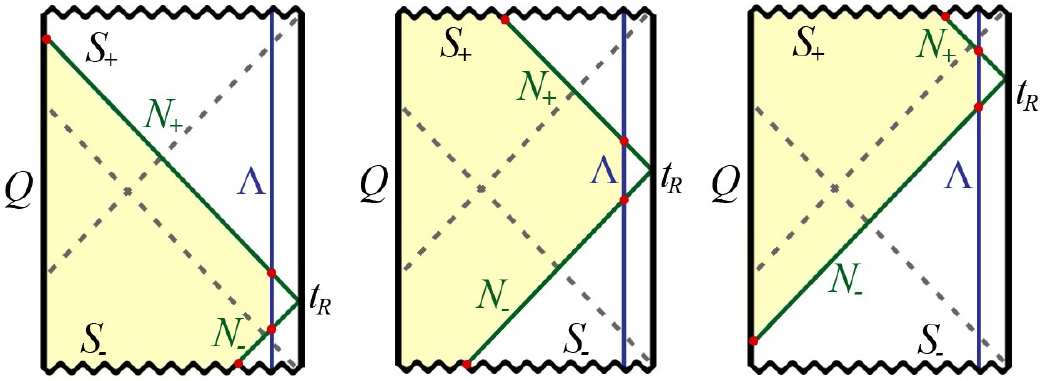}
  \caption{Penrose diagrams showing the Wheeler-DeWitt patch (shaded yellow) during each phase. Left-to-right: Phase i, Phase ii, Phase iii. The surface $\Lambda$ is used in calculations of the regulated action.}
  \label{fig:CAphases}
\end{figure}

As shown in figure \ref{fig:CAphases}, the boundary of the region $\mathcal{W}$ is comprised of different surfaces depending upon which asymptotic time slice we choose. To avoid conflating this boundary time with the bulk Schwarzschild time coordinate, let us refer to the time on the asymptotic CFT boundary as $t_R$ (and $s_R$ for the boundary time in $s,y$ coordinates). We find that there are three distinct phases depending on the time slice in question:
\bea
\textrm{Phase i: }&~~ s_R< -\arcsin(T) \cr
\textrm{Phase ii: }&~~ -\arcsin(T) < s_R < \arcsin(T) \cr
\textrm{Phase iii: }&~~ s_R > \arcsin(T)
\eea
This $s_R$ is related to the Schwarzschild boundary time, $t_R$, by:
\be
t_R = \frac{1}{r_H} \ln\left[\tan\left(\frac{\pi}{4} + \frac{s_R}{2}\right)\right]\;.
\ee
The Wheeler-DeWitt patches for each of these phases are depicted in the Penrose diagrams shown in figure \ref{fig:CAphases}. One should note that, due to the symmetry of our system, the results for the negative boundary times are related to those for the positive times by $t_R\to-t_R$. Hence, we only explicitly list here the results for the distinctly different phases: ii and iii.

The details of our calculations in this section may be found in appendix \ref{app:CAcalc}; here, we describe the results.
The action diverges as we integrate up to the asymptotic boundary, but we can define a finite quantity by subtracting off half of the action for the two-sided black hole at time $\tau = t_L + t_R = 0$ where $t_L$ and $t_R$ are the TFD's left and right boundary times respectively.\footnote{The asymptotic geometries are the same here, so the subtraction is unambiguous.} We will refer to this subtracted complexity as $\Delta \mathcal{C}_{A}$; results for the bare complexity with an explicit UV regulator may be found in the appendix.

\begin{figure}[h]
\centering
  \includegraphics[width=10cm]{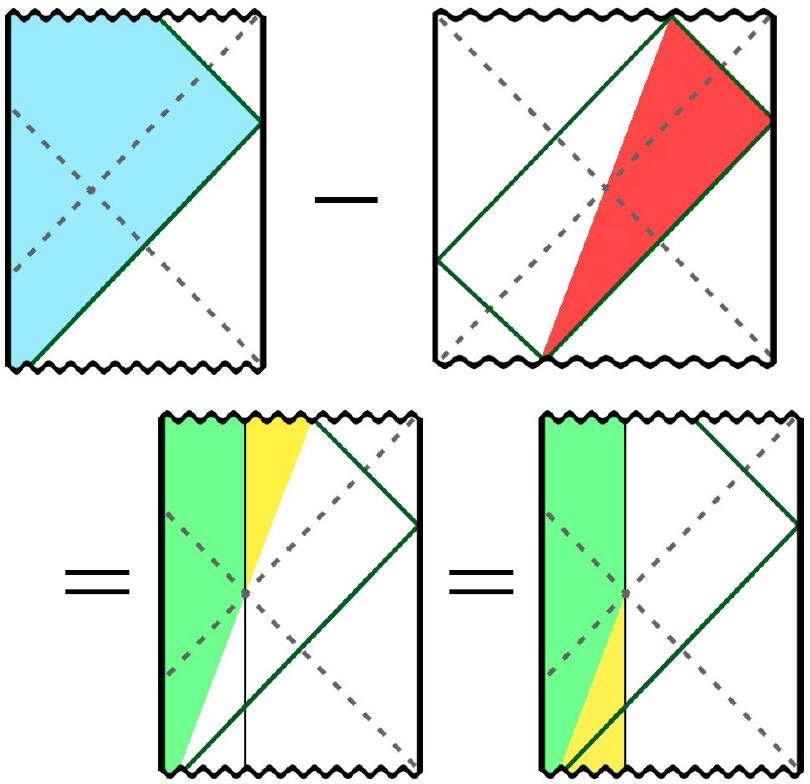}
  \caption{The geometric argument for why the complexity is constant during phase ii. The half TFD Wheeler-DeWitt patch (red) is subtracted from the phase ii patch (blue). The remaining region is broken into two pieces (green and yellow) that are rearranged to become the entire region behind the horizon. This ``proof'' is independent of boundary time.}
  \label{fig:CAproof}
\end{figure}

In phase ii, for times $-\arcsin(T) < s_R < \arcsin(T)$, we find the very simple result that
\bea
 \Delta \mathcal{C}_{A} &=& \mathcal{C}_{A}(t_R) - \frac{1}{2}\mathcal{C}_{TFD}(\tau=0) \cr
								&=& 0 \; .
\label{eq:CAphaseii}
\eea
We can understand this directly from the geometric argument shown in figure \ref{fig:CAproof}.

The complexity during phase iii, with the divergence subtracted in the same way as above, is found to simply be\footnote{We don't know if there is any reason for the ``entropic'' form of this result.}
\be
\Delta\mathcal{C}_{A}(s_R) = {r_H  \over 4 \pi G \hbar} {\sin(s_R - s_*) \over \cos s_*} \ln \left({\sin(s_R - s_*) \over \cos s_R}\right)
\ee
where $s_* = \arcsin(T)$ or equivalently\footnote{The results here include the null boundary counterterms first proposed in \cite{Lehner:2016vdi}.}
\be
 \Delta\mathcal{C}_{A}(t_R) =  \frac{r_H}{4 \pi G \hbar}
 \ln \bigg| \sqrt{1-T^2}\sinh(r_H t_R) - T \bigg| \left( \tanh(r_H t_R) - \frac{T \sech(r_H t_R)}{\sqrt{1-T^2}} \right)
 \label{eq:CAphaseiii} \; .
\ee

\begin{figure}[h]
\centering
  \includegraphics[width=13cm]{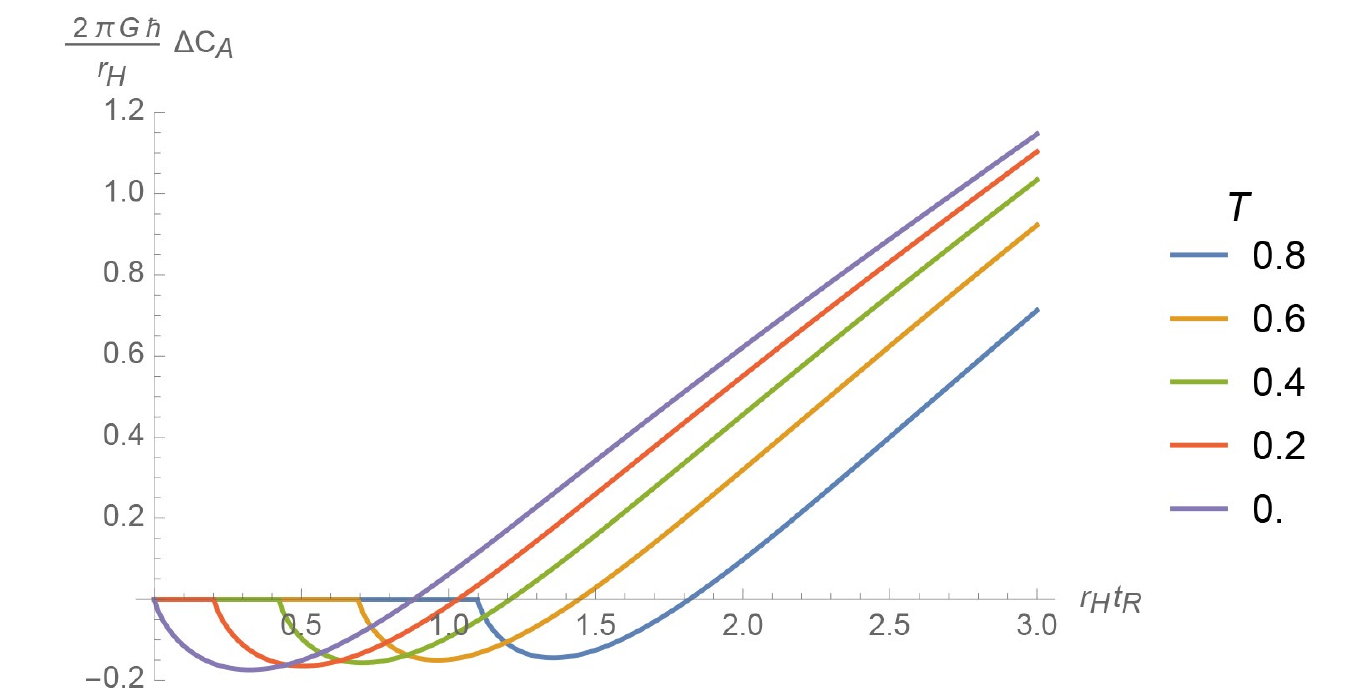}
  \caption{The regularised complexity during phases ii and iii, as a function of boundary time, for a selection of different brane tensions, $T$.}
  \label{fig:CAplot}
\end{figure}

In the $T\to0$ limit this result is simply the complexity for the BTZ geometry without any additional spacetime behind the horizon. Figure \ref{fig:CAplot} shows the regularized complexity for a range of ETW brane tensions. We see again the linear growth of complexity at late times, which takes the form
\be
\lim_{t_R\to\infty}  \frac{d \mathcal{C}_{A}}{d t_R} = \frac{2 M}{\pi \hbar} \; .
\ee
% Thus, the late time behavior is typical for a complexity, but the unchanging behavior for early times and the decrease after the transition are difficult to understand from the point of view of the evolution of complexity. In particular, it is somewhat surprising that the entanglement entropy is increasing for these times, indicating thermalization, while the action-complexity is constant or decreasing.
We see that both $\mathcal{C}_{V}$ and $\mathcal{C}_{A}$ grow linearly at late times, but exhibit different behaviour at early times. The volume-complexity increases smoothly from the time-symmetric surface $t = 0$, but the action-complexity is constant until one of the null boundaries defining the Wheeler-DeWitt patch intersects the ETW brane.
During the period that the action-complexity is constant, the entanglement entropy is increasing, indicating thermalization without complexity increase.
This is puzzling, but not impossible.
Alternatively, it may be that the action tracks the complexity well over large time scales but not during this early-time regime.

\section{Pure AdS analogue}

There is a close analogy between the maximally extended AdS-Schwarzschild black hole spacetime and pure AdS space divided into complementary Rindler wedges \cite{Czech:2012be}, where the two exterior regions correspond to the interiors of the two Rindler wedges, as shown in Figure \ref{ads-rindler}. In this section, we extend this analogy to describe states of a CFT on a half-sphere that are analogous to the black hole microstates considered in the main part of the paper. We specialize to 2+1 dimensions for simplicity.

\begin{figure}
\centering
\includegraphics[width=70mm]{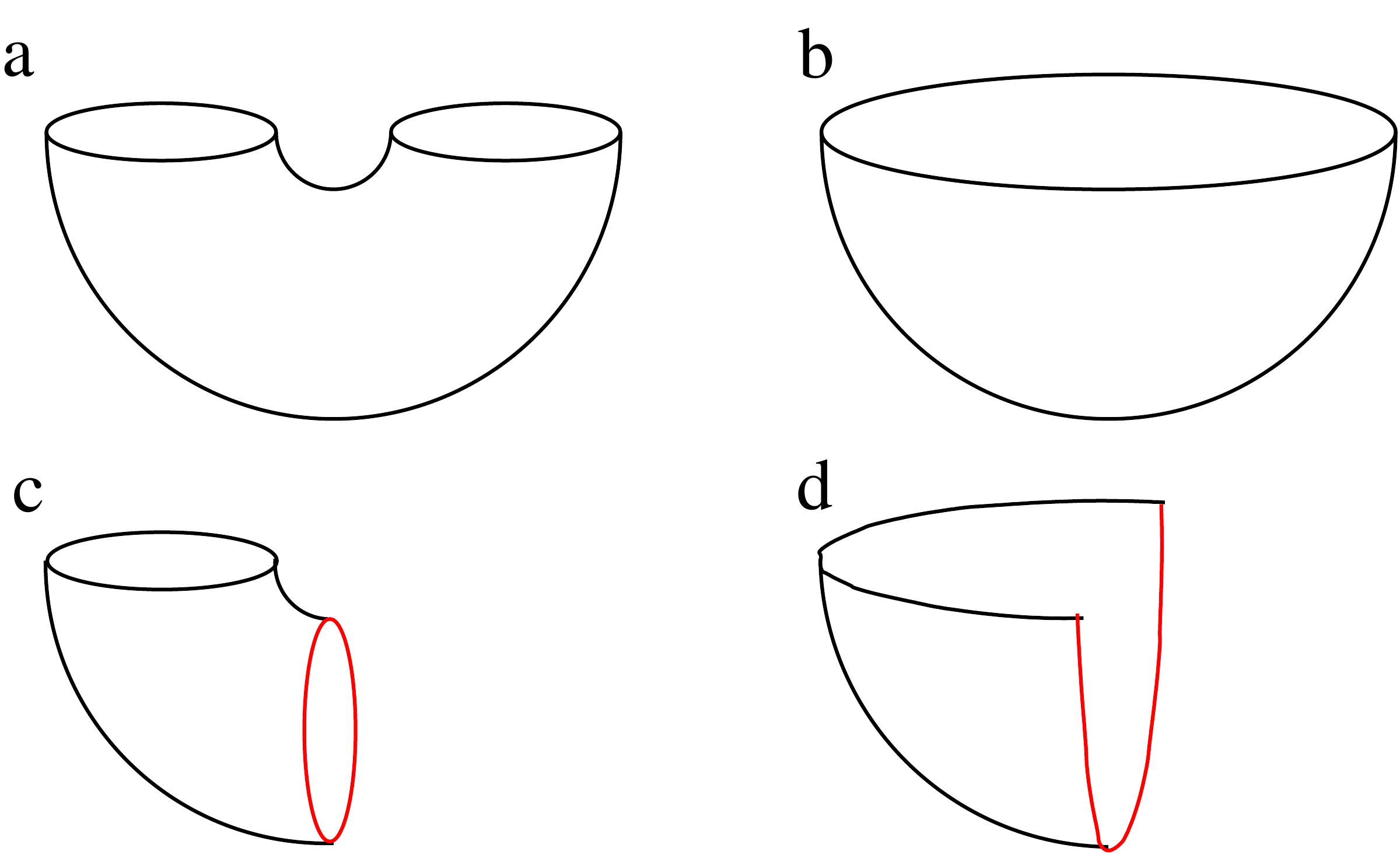}
\caption{Euclidean path integral geometries defining (a) thermofield double state of two CFTs (b) the vacuum state of a single CFT (c) a black hole microstate (d) a microstate for a half space. The red curves indicate BCFT boundary conditions.}
\label{fig:BCFTmicro}
\end{figure}

In the black hole story, the full geometry is described by two entangled CFTs, each in a thermal state. Our microstates are pure states of just one of these CFTs. For pure AdS, the geometry is described by a state in which the CFT degrees of freedom on two halves of a circle are entangled. The analog of a black hole microstate is a pure state of the CFT on a half circle (i.e. an interval). To make this fully well defined, we can place boundary conditions on the two ends of the interval, so that our CFT on a circle is replaced by a pair of BCFTs each on an interval. As discussed in \cite{VanRaamsdonk:2018zws}, we can define an entangled state of this pair of BCFTs whose dual geometry is a good approximation to the geometry of the original CFT state (inside a Wheeler-deWitt patch). Now, the analog of one of our black hole microstates is a pure state of one of these BCFTs that we can define using a path integral, as shown in figure \ref{fig:BCFTmicro}.

The path integral in Figure \ref{fig:BCFTmicro}d is equivalent via a conformal transformation to the path integral that defines the vacuum state of the BCFT on an interval. For this state, the corresponding geometry was described in \cite{Takayanagi:2011zk} and can be represented as a portion of the global AdS geometry ending on a static ETW brane, as shown in figure \ref{ads-rindler}. That figure also shows the Rindler wedges that are analogous to the two exterior regions in the maximally extended black hole geometry. We can see that (in the $T>0$ case) the ETW brane emerges from the past Rindler horizon in the second asymptotic region, reaches some maximum distance from the horizon, and then falls back in.

\subsection*{Explicit geometry}

\begin{figure}[ht]
\centering
\includegraphics[width=50mm]{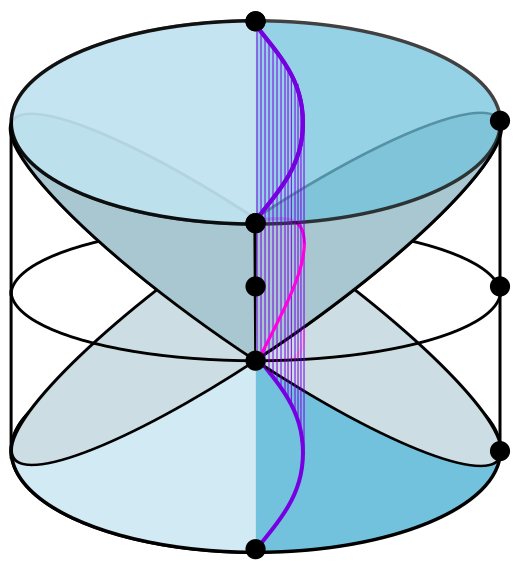}
\hspace{10pt}
\includegraphics[width=50mm]{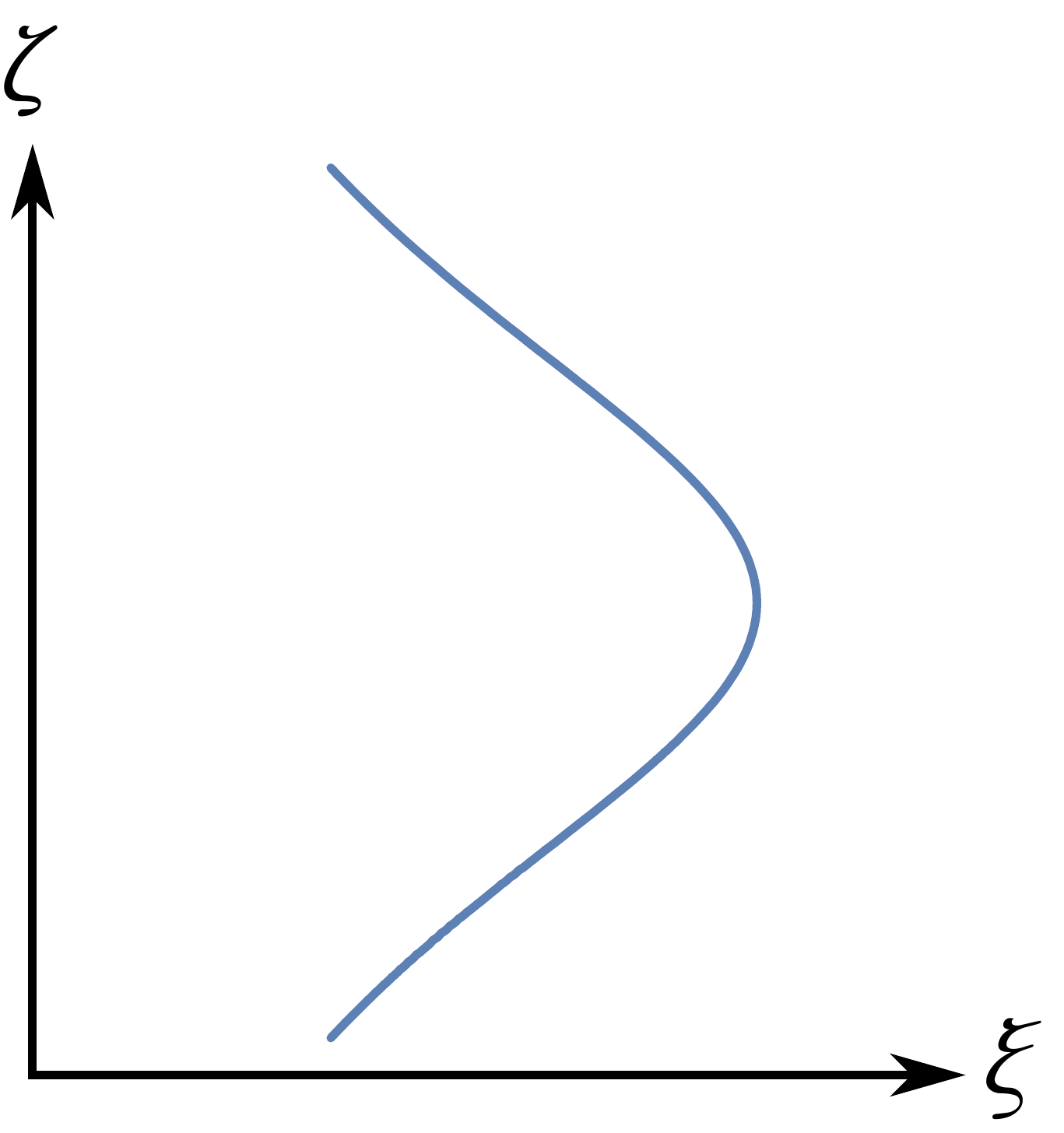}
\caption{Left: The ETW brane in global AdS. For $T > 0$ we have the geometry on the left of the brane. For $T>0$, we have the geometry on the right of the brane. Diagonal planar surfaces are Rindler horizons dividing the spacetime into complementary Rindler wedges plus past and future regions.  Right: dependence of the radial position parameter $\xi = \sqrt{r^2 - 1}$ on Schwarzschild time $\zeta$.}
\label{ads-rindler}
\end{figure}

To find the geometry associated with the BCFT vacuum state, it is simplest to consider a conformal frame where the interval on which the BCFT lives is $(-\infty,0]$. In this case, we recall from section 2 that in Poincar\'e coordinates
\be
ds^2 = {L^2 \over z^2} (-dt^2 + dz^2 + dx^2) \; ,
\ee
the vacuum geometry corresponds to the region $x/z < T/\sqrt{1-T^2}$ terminating with an ETW brane, as shown in figure \ref{fig:BoundEnt}. Passing to global coordinates via the transformations
\be
{L \over z} = \cosh(\rho) \cos(\tau) - \sinh(\rho) \sin(\theta) \qquad {x \over z} = \sinh(\rho)\cos(\theta) \qquad {t \over z} = \cosh(\rho) \sin(\tau) \; ,
\ee
the ETW brane locus becomes
\be
\sinh(\rho) \cos(\theta) = {T \over \sqrt{1 - T^2}} \;
\ee
in coordinates where the metric is
\be
ds^2 = L^2 (- \cosh^2\rho d \tau^2 + d \rho^2 + \sinh^2 \rho d \theta^2) \; .
\ee
Here, the brane is static in the global coordinates, extending to antipodal points at the boundary of AdS, as shown in figure \ref{ads-rindler}. In that figure, we see that from the point of view of one of the Rindler wedges, the brane

To make the analogy with the black hole more clear, we can now describe the ETW brane trajectory for $T > 0$ in a Rindler wedge, the analog of the second asymptotic region in the black hole case. Defining coordinates $(\chi,\zeta,r)$ from the Poincar\'e coordinates by
\be
{t \over L} = e^\chi \sinh(\zeta)  \sqrt{1-{1 \over r^2}} \qquad {x \over L} = e^\chi \sinh(\zeta) \sqrt{1-{1 \over r^2}} \qquad {z \over L} = e^\chi {1 \over r} \; ,
\ee
the Rindler wedge corresponding to the second asymptotic region takes the form of a Schwarzschild metric with non-compact horizon \cite{Emparan:1999gf},
\be
ds^2 = L^2(-(r^2 - 1)d \zeta^2 + {dr^2 \over r^2 - 1} + r^2 d \chi^2)\; ,
\ee
and the brane locus is simply
\be
\sqrt{r^2 - 1} \cosh(t) = {T \over \sqrt{1-T^2}} \; .
\ee
Note that this is precisely the same as the result (\ref{brane2D}) (setting $r_H = 1$). The reason is that the black hole geometry we considered previously is simply obtained from the present case by periodically identifying the $\chi$ direction. Thus, as in that case, for each time $t$, the ETW brane sits at a constant $r$ in the Schwarzschild picture, with $r(t)$ reaching a maximum at $t=0$.

%\begin{figure}[ht]
%\centering
%\includegraphics[width=80mm]{rindler.png}
%\caption{The Rindler coodinates in $\text{AdS}_3$: $\xi$ (acceleration parameter), $\chi$ (hyperbolic angle), $\zeta$ %(Rindler time). These only cover $\Delta t = \pi$ of global time and one Rindler wedge.}
%\label{rindler}
%\end{figure}

\subsubsection*{Entanglement calculations}

In analogy to the earlier result for BTZ black holes, the entanglement entropy of sufficiently large intervals in the BCFT can provide information about the geometry behind the Rindler horizon.

\begin{figure}
\centering
\includegraphics[width=30mm]{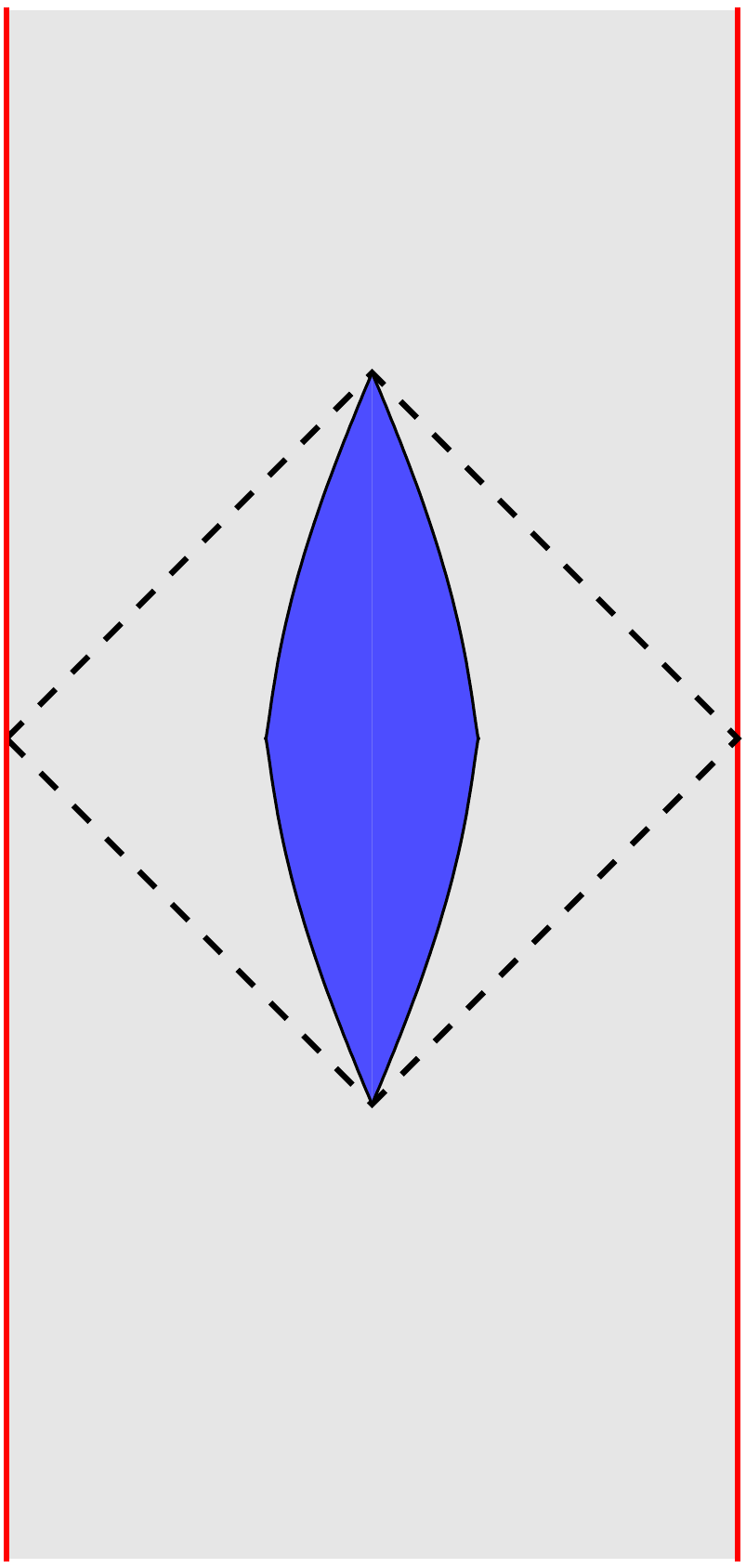}
\caption{Interval of fixed width in Schwarzschild time (blue shaded region) in the BCFT world-volume geometry.}
\label{fig:SchInterval}
\end{figure}

Using the standard CFT time in a conformal frame where we have a fixed distance between the two boundaries, the entanglement entropy for a connected boundary region is time-independent. However, to provide the closest analogy with our earlier calculations, we can instead consider the entanglement entropy of an interval of fixed width in the Schwarzschild spatial coordinate $\chi$, as shown in figure \ref{fig:SchInterval}.

We have seen that the geometry and the brane trajectory in the present case is mathematically identical to the black hole case for $r_H=1$ except that the $\chi$ coordinate is now non-compact. The compactness of $\theta$ did not enter into the  previous calculations of entanglement entropy, so all the calculations in section 3 apply here as well, and we can immediately jump to the result, that the entangling surface will probe behind the horizon when
\be
\sinh\left({\Delta \chi \over 2} \right) \ge \cosh(\zeta_0) \sqrt{1+T \over 1-T} \; .
\ee
Since $\chi$ is noncompact now, we have that for any time $\zeta_0$ and any $T$, we can always choose a large enough interval $\Delta \chi$ so that the entangling surface probes behind the horizon. The explicit expressions for entanglement entropy in the two phases are the same as those in section 3.1 (with $r_H = 1$).

Thus, if we unwrap the compact direction of the BTZ black hole, the ETW branes will be dual to boundary states on a spatial interval of pure $\text{AdS}_3$.
Our BTZ entanglement calculations carry over, implying that control of a suitably large boundary subregion should allow an observer to probe behind the Rindler horizon.

\section{Effective cosmological description?} \label{sec:cosmo}

We have seen in section 2 that the world-volume geometry of our ETW brane takes the form of a $d$-dimensional FRW spacetime. For the simple model with a constant tension ETW brane, the explicit metric was given in (\ref{FRW4}) for the case of a 3+1 dimensional ETW brane. Generally speaking, the physics on this brane does not provide a model of $d$-dimensional cosmology, since the gravitational physics is higher-dimensional. However, there is a vast literature on brane-world cosmology (see \cite{Maartens:2010ar} for a review) exploring scenarios where the physics of a $d$-dimensional brane embedded in a higher-dimensional spacetime does have an effective description as $d$-dimensional gravity coupled to matter. This requires gravity to ``localize'' to the brane, such that over a large range of distance scales gravitational interactions between matter on the brane are well-described by $d$-dimensional rather than higher dimensional gravity. In \cite{Randall:1999vf}, such localization was shown to occur for a brane which cuts off the UV region of an Anti-de-Sitter spacetime; this is known as the Randall-Sundrum II (RSII) brane-world scenario. In our geometries, the brane cuts off the UV in an asymptotically AdS spacetime (the AdS/Schwarzschild black hole). While this is globally different from pure AdS, it is expected that for appropriate values of $L_{AdS}$, $r_H$, and the ETW brane trajectory, the physics should be sufficiently similar to the pure-AdS case that gravity localization still occurs and we still have an effective $d$-dimensional description. Since our brane world-volume is that of a cosmological FRW spacetime, our model would then provide a microscopic description of $d$-dimensional brane-world cosmology.

\subsubsection*{Gravity localization in the Randall-Sundrum II model}

In this section, we will review the basic mechanism of gravity localization (see \cite{MANNHEIM} for a pedagogical introduction) and try to understand the requirements on the parameters in our model in order that an effective lower-dimensional description exists. In the Randall-Sundrum II model \cite{Randall:1999vf} we have an infinite extra dimension, but the bulk metric (for $d=4$) is now a warped product of the form
\begin{equation}
    ds_{5}^{2} = dz^{2} + e^{- 2A(z)} g_{\mu \nu}(x) dx^{\mu} dx^{\nu} \: , \qquad - \infty < z < \infty \: ,
\end{equation}
with a 3-brane placed at $z = 0$ and $\mathbb{Z}_{2}$-symmetry imposed in this coordinate.
In the original RSII model, one has warp factor $A(z) \sim | z| / \ell$; the bulk spacetime is then simply a slice of AdS$_{5}$ which is cut off in the UV by a 3-brane (referred to as a UV or Planck brane), with $\mathbb{Z}_{2}$-symmetry imposed about the brane. Tuning the brane tension against the bulk cosmological constant allows for a Poincar{\'e}-invariant brane metric $g_{\mu \nu}(x) = \eta_{\mu \nu}$. Randall and Sundrum found that, within this setup, one reproduces 4-dimensional Einstein gravity on the brane for distances much larger than the AdS radius $\ell$; for example, the gravitational potential on the brane is \cite{Garriga:1999yh}
\begin{equation}
    V(r) \approx \frac{G M}{r} \Big( 1 + \frac{2 \ell^{2}}{3 r^{2}}  \Big) \: .
\end{equation}
The reason for the localization is that the warp factor suppresses metric perturbations far from the brane, with $\ell$ the length scale on which this suppression occurs. Formally, one considers separable metric perturbations of the form $h_{\mu \nu} = \epsilon_{\mu \nu} \psi(z) \phi(x^{\mu})$, with $\phi(x^{\mu})$ an eigenstate of the 4-dimensional wave operator $\Box_{4} \phi = m^{2} \phi$; the linearized Einstein equations then reduce to an analogue Schr\"{o}dinger problem for $\psi(z)$, where the Schr\"{o}dinger ``energy'' determines the particle mass in the 4-dimensional description. The analysis reveals a massless `zero mode' wavefunction which localizes at the brane and exactly reproduces the 4-dimensional Newtonian potential; the continuum of massive ``KK modes'' provide corrections, but they are suppressed at the position of the brane due to a peak in the potential.

The localization phenomenon has been interpreted in the context of AdS/CFT \cite{Verlinde:1999fy, Gubser:1999vj, Giddings:2000mu, Hawking:2000kj, Giddings:2000ay, Deger:2000ir, deHaro:2000wj}, by the observation that the RSII model in a $d+1$-dimensional AdS bulk (and its curved-brane descendents in $d+1$-dimensional AAdS spacetimes) should be equivalent to a $d$-dimensional CFT with some UV cutoff coupled to dynamical gravity on the brane.\footnote{This doesn't provide a full microscopic description of the theory since the dynamical gravity is added in ``by hand'' to the cutoff CFT. In contrast, the CFT in our discussion corresponds to the asymptotic region on the far side of the black hole; this is an ordinary CFT with no dynamical gravity and thus can provide a microscopic description.}

\subsubsection*{Locally localized gravity}

Based on these results, it is natural to ask whether gravity localization extends to cases where we have an approximately AdS bulk cut off by a UV brane which is approximately Minkowski. In fact, there are some complications; for example, as noted by Karch and Randall in \cite{Karch:2000ct}, in the case of a brane with AdS$_{4}$ world-volume in global AdS$_{5}$, one no longer has a normalizable zero mode. This is because only part of the UV region of global AdS is excised by the introduction of an AdS$_{4}$ UV brane; a graviton at the brane can still tunnel toward the true boundary of AdS, where the warp factor blows up, so this geometry does not trap gravity at the brane. However, Karch and Randall showed that if we are close enough to the Minkowski situation, the time scale for this tunnelling is long, so that 4-dimensional Einstein gravity still provides a good approximation over sufficiently short time scales. This supports the more general idea that  localization of gravity should be a `local' phenomenon, which should not depend upon the behaviour of the warp factor far from some region of interest.

\subsubsection*{Branes in AdS/Schwarzschild}

The question relevant for us is whether one retains gravity localization when the bulk is modified through the introduction of a black hole, and the brane world-volume is allowed to be dynamical. The first question has been previously investigated \cite{Singh:2002ru, Singh:2002xa, Csaki:2000fc, Kiritsis:2006ua, Seahra:2005us, Clarkson:2005mg, Hebecker:2001nv}. Based on the work of Karch and Randall, one expects that if the brane is taken far enough from the black hole horizon, so that the nearby spacetime is approximately AdS, then the local character of gravity localization should allow for effective Einstein gravity on the brane, up to $\mathcal{O}(r_{\textnormal{H}} / r_{\textnormal{b}})$ corrections (where $r_{\textnormal{b}}$ is the position of the brane). The detailed analysis performed in \cite{Seahra:2005us, Clarkson:2005mg} for the case of an Einstein static (ES) brane-world (with $r_{\textnormal{b}} = \textnormal{const}$) in Schwarzschild AdS supports this conclusion. Our FRW branes are not static, but we expect similar qualitative behaviour during the period when the effective Hubble parameter is small compared with the AdS scale $H \equiv \dot{r}/r \ll \approx 1 / \ell$.

\subsubsection*{Implications for the constant-tension brane scenario}

Let is now apply these constraints to the geometries arising in the simple model with a constant tension ETW brane. We have seen that obtaining an effective four-dimensional description requires $r_{\textnormal{b}} \gg r_{\textnormal{H}}$ and $H \ll \ell$. In our setup, the maximum proper radial size of the brane, in the case of critical tension $T=1$, is given for $d=4$ by $r_{\textnormal{max}} = r_{\textnormal{H}} \sqrt{1 + \frac{r_{\textnormal{H}}^{2}}{\ell^{2}}}$;
thus, in order to have some regime for which $r_{\textnormal{b}} \gg r_{\textnormal{H}}$, we must consider a large black hole $r_{\textnormal{H}} \gg \ell$, and almost-critical tension $T \approx 1$. The requirement that $H \ll \ell$ will be satisfied for most of the evolution as long as the total proper time (\ref{Life4}) is large in AdS units. Again, this requires that $T$ is very close to 1.

Unfortunately, we recall that while the Lorentzian solutions for any value $T < 1$ (and even larger values for $d>2$) look physically reasonable, the corresponding Euclidean solutions for $d>2$ appear to make sense only for $T < T_* < 1$ since otherwise the ETW brane overlaps itself in the Euclidean picture (see Figure \ref{fig:EuclideanOverlaps}). The requirement $T < T_*$ would rule out a viable model with an effective four dimensional description since this required $r <  1.2876 r_H$. On the other hand, we had reason to question the validity of the simple holographic treatment in these cases.

To summarize, in the simplest toy model for how to treat the BCFT boundary conditions holographically, it does not seem possible to realize microstates for which the effective description of the ETW brane physics corresponds to a four-dimensional cosmology. However, it remains very interesting to understand whether this scenario for cosmology can be realized with more general effective actions that would correspond to a more complete treatment of the holographic BCFT physics.

\section{Discussion}

In this final section, we discuss a few possible generalizations and future directions.

For the specific examples in this paper, we have mainly considered geometries obtained by assuming the very simple holographic ansatz for how to model CFT boundary conditions holographically. In that model, the ETW brane is filling in for some more detailed microscopic physics. This could involve branes or orientifold planes of string/M theory, or geometrical features such as the degeneration of an internal manifold. Depending on the particular situation, a more realistic model might include additional terms in the brane action or couplings to additional bulk fields. As a particular example, scalar operators in a BCFT can have one-point functions growing as $~1/|x|^{2 \Delta}$ as the distance $x$ to the boundary decreases. This would correspond to having some extra scalar fields in the bulk, sourced by the ETW brane.\footnote{Some particular top-down examples of complete geometries dual to supersymmetric BCFT states have already been understood: see \cite{Chiodaroli:2011fn, Chiodaroli:2012vc, Gutperle:2012hy}.} In our context, this would lead to matter outside the black hole that falls into the horizon. Thus, the explicit geometries we have utilized should be viewed as simple examples that may elucidate the basic physics of more precise holographic duals for Euclidean-time-evolved boundary states. It will be interesting to flesh out the AdS/CFT correspondence for BCFTs more fully and explore the microstate geometries emerging from more general bulk effective actions. It will also be interesting to understand better the constraints on boundary conditions / boundary states for a given holographic CFT that lead to a fully geometrical bulk description.

Within the context of any particular choice of bulk effective action (e.g. the constant tension ETW brane model we used here), it is also interesting to understand which parameter values can be realized in some microscopic theory. For example, if there are microscopic models that realize (at least approximately) the simple ansatz, which values of the parameter $T$ arise from legitimate boundary conditions for a holographic CFT. For 1+1 dimensional CFTs, this is related to the question of which boundary entropies are possible. Some constraints have been discussed previously \cite{Friedan:2012jk}, but these do not apply for holographic models. An interesting result is that for the monster CFT, only positive values (or perhaps extremely small negative values) of $\log(g)$ (proportional to $\arctanh(T)$ in the holographic case) are allowed \cite{Friedan:2013bha}. If this extended to holographic theories, it would imply that only the case with an ETW brane behind the horizon is physical.

Another interesting generalization would be to consider states constructed in a similar way, but with boundary conditions that do not preserve conformal invariance. For example, we can have boundary conditions that correspond to boundary RG flows from one conformally invariant boundary condition to another. These may be represented by a more general class of ETW brane actions, and give rise to a wider variety of geometries. Finally, we can consider similar constructions in holographic theories which are not conformal, for example in holographic RG flow theories or in holographic theories derived from low-energy Dp-brane actions. For all these cases, we expect that the basic idea of probing behind-the-horizon physics via time-dependence of subsystem entanglement remains valid.

It would be very interesting to perform direct entanglement entropy calculations for Euclidean-time-evolved boundary states in specific CFTs, to see whether the results are qualitatively similar to those in our model calculation, and to generate microscopic examples of black hole microstates for which we can learn about the behind-the-horizon physics directly. Naively, this will be challenging in strongly coupled holographic CFTs, but perhaps even calculations for tractable non-holographic theories (such as large $c$ symmetric orbifold CFTs\footnote{We thank Volker Schomerus for this suggestion.}) will be enlightening. In unpublished work, we have already developed formulae for entanglement in imaginary time-evolved product states of non-interacting particles, so the orbifold calculation appears within reach. It may also be possible to perform direct calculations in holographic CFTs by assuming something about the structure of holographic BCFT correlators, similar to the calculations in \cite{Hartman:2014oaa, Anous:2016kss}. Another possibility we are pursuing is to develop a general calculational tools based on a randomness assumption, similar to the eigenstate thermalization hypothesis.

Finally, with a larger toolbox for studying holographic duals of Euclidean-time-evolved boundary states, it will be interesting to see if it is possible to realize any examples where gravity is localized on the ETW brane, or more generally, that the physics of the spacetime causally disconnected from the asymptotic boundary is effectively described by four dimensional cosmology. This would be very interesting whether or not such a cosmology can be made realistic, since there currently aren't any known complete, non-perturbative quantum descriptions of four-dimensional big bang cosmology, as far as we are aware. In our case, the CFT and the specific microstate would provide the complete description and allow (in principle) a calculation of the initial conditions for cosmology that should be used as inputs for the effective field theory description (also to be determined from the CFT/state) that would be valid at intermediate times.\footnote{If our approach can be realized, it would be similar in some ways to the Hartle and Hawking's 'no boundary' approach to cosmology,\cite{Hartle:1983ai} except that our Euclidean path integral is for a non-gravitational boundary theory, and the path integral itself is defined using a boundary. So one might call it the ``boundary-boundary-no-boundary'' approach.}
 Of course, these calculations would require a much better understanding of how black hole behind-the-horizon physics is encoded in a CFT.

One of the major challenges in coming up with candidates for quantum gravity theories capable of describing cosmology is that it is not even clear what the very basic mathematical framework could be. Usual examples of holography making use of conventional quantum systems describe spacetimes with some fixed asymptotic behavior. This is normally assumed to be incompatible with cosmological physics, so various qualitative ideas have been put forward for how to come up with something more general (see e.g. \cite{Alishahiha:2004md, Banks:2001px, VanRaamsdonk:2009ar, Nomura:2016ikr, Giddings:2018koz, Ito:2015mxa} for a variety of perspectives). However, to date, none of these has led to a complete model, or even a precise mathematical structure that could generalize the usual state-in-a-Hilbert-space of ordinary quantum mechanics. A likely possibility is that we have simply not yet stumbled across the right idea. But it is worth considering the alternative, that cosmology is somehow described by a conventional quantum system, just like the rest of physics. If this quantum system is related to gravity in the usual holographic way, we would need to understand how our cosmological observations could be compatible with fixed asymptotic behavior for the global spacetime. One of the most attractive features of our suggestion is that it gives a possible way to realize this, and thus, to describe cosmology with ordinary quantum mechanics.

\section*{Acknowledgments}

We would like to thank Tarek Anous, Stefano Antonini, Eliot Hijano, Andreas Karch, Alex May, Shiraz Minwalla, Volker Schomerus, and Tadashi Takayanagi for useful discussions. DW is supported by an International Doctoral Fellowship from the University of British Columbia. BGS and MVR are supported by the Simons Foundation via the It From Qubit Collaboration and a Simons Investigator Award (MVR). SC and MR  are supported by a Discovery grant from the Natural Sciences and Engineering Research Council of Canada.

\appendix

\section{Derivation of the microstate solutions}
\label{sec:etw-derivation}

In this appendix, we provide details of the calculations in section 2 for the geometries associated with Euclidean-time-evolved boundary states using the simple holographic prescription with a constant-tension ETW brane.

\subsubsection*{Action and equations of motion}

The physics of the bulk spacetime and ETW brane can be encoded in an action $I
= I_\text{bulk} + I_\text{ETW}$.
The first term $I_\text{bulk}$ is the usual Einstein-Hilbert term,
regularized by a Gibbons-Hawking term at the asymptotic boundary:
% , needed to regularise
% the boundary term which arises from variation:
\begin{equation}
  \label{I_M}
  I_\text{bulk} = \frac{1}{16\pi G}\int_{N_{AdS}} \D^{d+1}x\, \sqrt{-g}(R-2\Lambda) +
  I^\text{matter}_\text{bulk} + I_\text{GHY}.
\end{equation}
% and $I_N$ is the usual Gibbons-Hawking action needed to cancel the
% boundary term which arises from varying $I_M$.
% The cosmological constant is related to the AdS radius $L_\text{AdS}$
% by $\Lambda = -d(d-1)/2L_\text{AdS}^2$.
The action on the ETW brane $Q$ is a Gibbons-Hawking term, but for a
dynamical boundary metric,
\begin{equation}
  \label{I_Q}
  I_\text{ETW} = \frac{1}{8\pi G}\int_{Q_\text{ETW}} \D^{d-1}y\, \sqrt{-h}K +
  I^{\text{matter}}_{\text{ETW}},
\end{equation}
where $y^a$ are intrinsic coordinates on the brane, $h_{ab}$ is the
intrinsic brane metric, and $K_{ab}$ is the extrinsic curvature.
The extrinsic curvature is roughly the derivative the intrinsic metric
in the normal direction $n_\mu$.

More precisely,
\begin{equation}
  K_{ab} = n_{\mu;\nu}e^\mu_a e^\nu_b, \quad K = K_{ab}h^{ab}\quad e^\mu_a
  = \frac{\partial x^\mu}{\partial y^a}.\label{extrinsic}
\end{equation}
Stress-energy on the brane is defined as the variational derivative of
the brane matter action with respect to the intrinsic metric:
\begin{equation}
  T^\text{ETW}_{ab} = \frac{2}{\sqrt{-h}}\frac{\delta
    I^\text{matter}_\text{ETW}}{\delta h^{ab}}.\label{T^Q}
\end{equation}
Varying with respect to $g^{\mu\nu}$ and $h^{ab}$ \cite{Fujita:2011fp},
we obtain Einstein's equation in the bulk and the Neumann condition on the brane:
\begin{align}
R_{\mu\nu} - \frac{1}{2}Rg_{\mu\nu} & = 8\pi G
  T_{\mu\nu}^\text{bulk} - \Lambda g_{\mu\nu}\label{EOM-g} \\ K_{ab} - K h_{ab}
                                                        & = 8\pi G
  T_{ab}^\text{ETW}.\label{EOM-h}
\end{align}
% We will typically examine brane quantities $h_{ab}, K_{ab}$ in
% intrinsic coordinates, but it is easy to translate them into the
% extrinsic quantities \cite{Poisson2004}.
We will focus on \emph{constant tension} branes, with
\be
\label{const_tension}
8\pi G T^\text{ETW}_{ab} = (1-d) T h_{ab} \; ,
\ee
where the prefactor on the right hand side is chosen for convenience.

\subsubsection*{Comparison of the gravitational actions: details}

To establish the critical value $\tau_*(T)$ for $\tau_0$ below which the black hole phase dominates the path-integral, we need to compare the gravitational action for solutions from the two phases. For $d=2$, this calculation was carried out in \cite{Fujita:2011fp} (section 4) while studying the Hawking-Page type transition for BCFT on an interval. We now generalize this to arbitrary dimensions.

The Euclidean gravitational action is the sum of bulk and boundary contributions,
\be
I_E = -{1 \over 16 \pi G} \int d^{d+1} x \sqrt{g} (R - 2 \Lambda) - {1 \over 8 \pi G} \int d^d x \sqrt{h} (K - (d-1)T) \; .
\ee
For the solutions we consider, the bulk and boundary equations of motion (\ref{EOM-g}), (\ref{const_tension}) imply that
\be
R - 2 \Lambda = -2 d
\ee
and
\be
(K - (d-1)T) = T \; .
\ee
For geometries of Schwarzschild form, we have
\be
\sqrt{g} = r^{d-1}
\ee
and with the ETW brane parameterized by $\tau(r)$ given by (\ref{taur2}) or (\ref{taur3}) we get
\beas
\sqrt{h} &=& r^{d-1} \sqrt{f(r) \left({dr \over d \tau} \right)^2 + {1 \over f(r)}} \cr
&=& {r^{d-1} \over  \sqrt{f(r) -T^2 r^2}} \cr
&=& \pm {1 \over T} r^{d-2} f(r) {d \tau \over dr}
\eeas
where we have the $+$ or $-$ depending on whether $\tau$ is an increasing or decreasing function of $r$.

To regulate the actions, we integrate in each case up to $r_{max}$ corresponding to $z=\epsilon$ in Fefferman-Graham coordinates.

{\bf Pure AdS phase:} For the pure AdS phase (where $f(r) = r^2 + 1$), the bulk action gives
\be
{\omega_{d-1} \over 8 \pi G} \int_0^{r_{max}} dr d \cdot r^{d-1} (2 \tau (r))
\ee
where $\omega_{d-1}$ is the volume of a unit $d-1$ sphere and
\be
\tau(r) = \tau_0 + {\rm arcsinh}\left({T \over (r^2 + 1) \sqrt{1-T^2}}\right) \; .
\ee
Each component of the boundary action gives
\be
{\omega_{d-1} \over 8 \pi G} \int_0^{r_{max}} dr r^{d-2} f(r) {d \tau \over dr} \; .
\ee
Combining these, we have
\beas
I_E^{AdS} &=& {\omega_{d-1} \over 4 \pi G} \int_0^{r_{max}} dr \left[ d r^{d-1} \tau(r)  + r^{d-2} f(r) {d \tau \over dr} \right] \cr
&=& {\omega_{d-1} \over 4 \pi G} \left\{ r_{max}^d \tau(r_{max}) + \int_0^{r_{max}} dr r^{d-2} {d \tau \over dr} \right\}
\eeas
where $d \tau / dr $ can be read off from (\ref{taur3}).

{\bf Black hole phase:} For the black hole phase, we can write the bulk action as the full action for the Euclidean black hole up to $r = \hat{r}_{max}$ (generally not the same as $r_{max}$ -- see below) minus the action for the excised part. This gives
\be
{ \omega_{d-1} \over 8 \pi G} \int_{r_H}^{r_M} dr d \cdot r^{d-1} \beta - \int_{r_0}^{\hat{r}_{max}} dr d \cdot r^{d-1} 2 \tau(r)
\ee
where $\tau(r)$ is given in (\ref{taur2}). The brane action gives
\be
-{\omega_{d-1} \over 4 \pi G} \int_{r_0}^{\hat{r}_{max}} r^{d-2} f(r) {d \tau \over dr} \; ,
\ee
where in this case,
\be
f(r) = r^2 + 1 -{r_H^{d-2} \over r^{d-2}} (1 + r_H^2)
\ee
Combining everything, we get
\beas
I_E^{BH} &=& {\omega_{d-1} \over 4 \pi G} \int_{r_H}^{\hat{r}_{max}} dr d \cdot r^{d-1} {\beta \over 2} - \int_{r_0}^{\hat{r}_{max}} dr ( d \cdot r^{d-1} \tau(r) + r^{d-2} f(r) {d \tau \over dr}) \cr
&=& {\omega_{d-1} \over 4 \pi G} \left\{ \left. {\beta \over 2} r^d \right|_{r_H}^{\hat{r}_{max}} -  r^d \tau(r) \bigg\rvert_{r_0}^{\hat{r}_{max}} - \int_{r_0}^{\hat{r}_{max}} dr (r^{d-2} f(r) - r^d) {d \tau \over dr} \right\}
\eeas
where $\tau$ and $d \tau / dr$ can be read off from (\ref{taur2}).

{\bf Cutoff surface:} In order to compare the actions, we choose both $r_{max}$ and $\hat{r}_{max}$ to each correspond to the surface $z = \epsilon$ in Fefferman-Graham coordinates. In each case, the $z$ coordinate is related to the $r$ coordinate by
\be
{dz \over z} = {dr \over \sqrt{f(r)}}
\ee
with the integration constant fixed by demanding that $r \sim 1/z$ at leading order for small $z$. For the pure AdS case, this gives in any dimension
\be
r_{max} = {1 \over \epsilon} - {\epsilon \over 4}
\ee
while for the Euclidean black hole case, we get for example
\be
\hat{r}_{max}^{d=2} = {1 \over \epsilon} + {\pi^2 \epsilon \over 16 \tau_0^2} + {\cal O}(\epsilon^3)
\ee
for $d=2$ and
\be
\hat{r}_{max}^{d=4} = {1 \over \epsilon} - {\epsilon \over 4} + {1 \over 8} r_H^2(1 + r_H^2) \epsilon^3 + {\cal O}(\epsilon^5)
\ee
for $d=4$.

{\bf Action difference:}
We can now evaluate the difference
\be
I_E^{AdS}(T,\tau_0,\epsilon) - I_E^{BH}(T,\tau_0,\epsilon)
\ee
and take the limit $\epsilon \to 0$ in order to determine which solution has smaller action and gives rise to the classical geometry associated with the state.

As examples, we find that for $d=2$, we have
\be
\lim_{\epsilon \to 0} (I_E^{AdS}(T,\tau_0,\epsilon) - I_E^{BH}(T,\tau_0,\epsilon))  = {1 \over 2G} \left[ - {\rm arctanh}(T) - {\tau_0 \over 2} + {\pi^2 \over 8 \tau_0} \right] \; .
\ee
Thus, our states correspond to bulk black holes when
\be
\tau_0 < - {\rm arctanh}(T) + \sqrt{{\pi^2 \over 4} + {\rm arctanh}^2(T)} \; .
\ee
Here, we assume that the CFT is defined on a circle of length $2 \pi$. This critical value of $\tau_0$ decreases monotonically from $\tau_*(-1) = \infty$ to $\tau_*(0) = \pi/2$ to $\tau_*(1) = 0$, as shown in figure \ref{fig:Ttau}. This result agrees with the calculation of \cite{Fujita:2011fp} (reinterpreted for our context).

For $d=4$, it is most convenient to parameterize the action difference in terms of $r_H$ and $T$ since there can be more than one solution in the black hole phase with the same $T$ and $\tau_0$. We find that
\beas
&&\Delta I(r_H,T) \equiv {4 \pi G  \over \omega_3 } \lim_{\epsilon \to 0} (I_E^{AdS}(T,r_H,r_{max}(\epsilon)) - I_E^{BH}(T,r_H,\hat{r}_{max}(\epsilon))) \cr
&=& \left[{T \over 1-T^2} + {\rm arctanh}(T)\right] - \left[{1 \over 2} r_H^2( 1 + r_H^2) \tau_0(r_H) - {\pi r_H^5 \over 1 + 2 r_H^2} + {T r_0(r_H,T) \over \sqrt{1-T^2}} - I_4(rH,T)\right] \; ,
\label{DeltaI4}
\eeas
where (taking $f(r) = r^2 + 1 - r_H^2/r^2(1 + r_H^2)$ in the formulae below), $r_0(r_H,T)$ is defined as above by
\be
f(r_0) = T^2 r_0^2 \; ,
\ee
and $\tau_0(r_H,T)$ is defined as
\be
\tau_0(rH,T) = \int_{r_0}^\infty dr {Tr \over f(r) \sqrt{f(r) - T^2 r^2}}
\ee
and
\be
I_4(rH,T) = \int_{r0(r_H,T)}^\infty dr \left\{{Tr (r^2 - r_H^2 (1 + r_H^2)) \over f(r) \sqrt{f(r) - T^2 r^2}} - {T \over \sqrt{1-T^2}} \right\} \; .
\ee

Evaluating $\Delta I(r_H,T)$ for $T \ge 0$, we find that for $T < T_c \approx 0.37505$, the difference $\Delta I$ is positive for $r_H > r_H^*(T)$ where $r_H^*(T)$ increases monotonically from $r_H^* = 1$ at $T=0$ to $r_H^* = \infty$ at $T=T_c$. The corresponding value of $\tau_0$ decreases from $\pi/6$ at $T=0$ to 0 at $T = T_c$, as shown in figure \ref{fig:Ttau}. We note that in cases where there are two solutions in the black hole phase with the same $\tau_0$, the lowest action solution is always either the one with larger $r_H$ or the corresponding pure AdS phase solution.

\subsubsection{Lorentzian geometries: general $T$}

In this subsection, we discuss the Lorentzian solutions corresponding to general values of the parameter $T$. We recall that in terms of the proper time and the variable $L = \log(r)$ (where $r$ is the proper radius of the brane), the equation for the brane trajectory is
\be
\dot{L}^2 + V(L) = T^2
\ee
where
\be
V(L) = {f(r) \over r^2} = 1 + e^{- 2 L} - e^{-d (L - L_H)}(1 + e^{- 2 L_H}) \; .
\ee
So the trajectory $L(\lambda)$ is that of a particle in a one-dimensional potential $V(L)$ with energy $T^2$. These potentials were displayed in Figure \ref{fig:Veff}.

For $d = 2$, the potential is monotonically increasing and asymptotes to 1. The Lorentzian trajectories for $|T|<1$ all correspond to time-symmetric configurations where the brane emerges from the past singularity at $r=0$, reaches a maximum size $r_0 = r_H/\sqrt{1-T^2}$, and shrinks again to $r=0$ at the future singularity. These all have analytic continuations to Euclidean solutions as discussed above. For $T > 1$, there are no time-symmetric trajectories; the ETW brane size either increases from $r=0$ to $r=\infty$ or shrinks from $r=\infty$ to $r=0$. These do not come from analytically continued time-symmetric geometries, and we expect that they do not correspond to the types of states we have been discussing.

\begin{figure}[h]
  \vspace{10pt}
  \centering
  \includegraphics[scale=0.60]{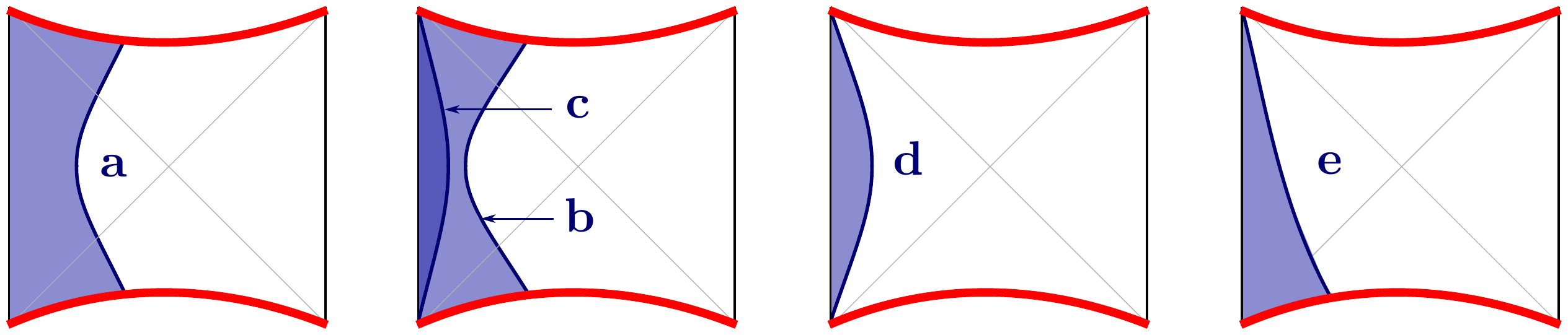}
  \caption{Lorentzian ETW branes for various values of $T$. }
  \label{fig:PenroseCases}
\end{figure}

For $d > 2$, the potential is monotonically increasing to some value $T_{\mathrm{crit}}^2 > 1$, where
\be
T_{\mathrm{crit}} = 1 + \left({2 \over d} \right)^{2 \over d-2} \left(1 - {2 \over d}\right){1 \over r_H^2 (1 + r_H^2)^{2 \over d-2}}
\ee
We have five classes of trajectories, as shown on the right in Figure \ref{fig:Veff}. The corresponding spacetimes are shown in Figure \ref{fig:PenroseCases}.

{\bf Case a: $0 < T < T_*$}

For this case, we have time-symmetric solutions which have analytic continuations to good Euclidean solutions corresponding to some finite positive Euclidean preparation time. These are the geometries that are most plausibly providing a holographic picture of the microstate geometries for some legitimate CFT states. The Lorentzian geometry takes the form in Figure \ref{fig:ETW}. The brane emerges from the past singularity, expands and enters the second asymptotic region and then shrinks, eventually falling into the future horizon. The maximum radius of the ETW brane is $r_0$ (the minimal radius in the Euclidean solution), realized at the time-symmetric point $t=0$. The entire trajectory covers some finite amount of proper time given by
\be
\lambda_{\mathrm{tot}} = 2 \int_0^{r_0} {dr \over \sqrt{T^2 r^2 - f(r)}} \; .
\ee
For $d=2$, this gives
\be
\lambda_{\mathrm{tot}}^{d=2} = {\pi L_{AdS} \over \sqrt{1-T^2}}
\ee
while for $d=4$, we get
\be
\lambda_{\mathrm{tot}}^{d=4}  = { L_{AdS} \over \sqrt{1-T^2}} \arccos \left({1 \over \sqrt{(1-T^2)(2 r_H^2 +1)^2 + T^2}}\right) \; .
\ee
The $d=3$ result is given in terms of elliptic integrals.

{\bf Case b: $1 < T < T_{\mathrm{crit}}$, small $r$ branch}

For this case, we have Lorentzian trajectories that are qualitatively similar to the previous case, but we recall that here the corresponding Euclidean solutions are not sensible (at least without some improvement of the model). It is possible that these Lorentzian solutions still correspond to some CFT states, but we do not have a clear argument for this.

{\bf Case c: $1 < T < T_{\mathrm{crit}}$, large $r$ branch}

For these solutions the ETW brane starts and ends at infinite size, shrinking to a minimum size at the time-symmetric point. We have an infinitely large portion of the second asymptotic region both in the past and the future, so it is unlikely these geometries correspond to pure states of a single CFT.

{\bf Case d: $T = T_{\mathrm{crit}}$}

In this case, we have Lorentzian brane trajectories at a constant radius, and the ETW brane geometry is the Einstein static universe. Here, the solutions retain the isometry present in the maximally extended black hole geometry and the physics of the CFT is time-independent. The Euclidean solutions in this case also have the brane at a constant radius, so the trajectory does not intersect the Euclidean boundary and does not seem likely to correspond to the class of states we have been discussing. However, it is interesting that the spacetime picture we have been discussing is similar to the proposal of \cite{deBoer:2018ibj} for the geometries dual to typical states, so perhaps the Lorentzian geometries in this case can serve as a model of the typical states. It is interesting that we are constrained to have the brane at one specific radius,
\be
{r \over r_H} = \left({d \over 2} \right)^{1 \over d-2}(1 + r_H^2)^{1 \over d-2} \; .
\ee

{\bf Case e: $T > T_{\mathrm{crit}}$}

For these case, there are no time-symmetric ETW brane trajectories, and we have an infinitely large portion of the second asymptotic region either in the past or the future, so it seems unlikely that these geometries correspond to pure states of a single CFT.

\section{Coordinate systems for $d=2$}
\label{app:sycoords}
In this appendix, we give the coordinate transformations relating $s-y$ coordinates in (\ref{SYmetric}) which cover the full maximally extended black hole geometry to the Schwarzschild coordinates.

We first go to Kruskal-type coordinates by defining
\be
\label{StoK}
r = r_H {1 - uv \over 1 + uv} \qquad t = {1 \over 2 r_H} \ln \left(-{u \over v} \right) \; .
\ee
In these coordinates, the metric becomes (here, we have set $L_\mathrm{AdS}=1$)
\be
ds^2 = -{4 du dv \over (1 + uv)^2} + r_H^2 {(1 - uv)^2 \over (1 + uv)^2} d \phi^2 \; .
\ee
These coordinates cover the whole extended spacetime. The two boundaries are at $uv = -1$, the singularities are at $uv = 1$, and the horizons are at $uv=0$. The relation to Schwarzschild coordinates in the second asymptotic region is given by (\ref{StoK}) with the replacement $u \leftrightarrow v$. To obtain the metric  (\ref{SYmetric}), we further define
\be
u = \tan(\alpha) \qquad v = \tan(\beta) \qquad s = \beta + \alpha  \qquad y = \alpha - \beta \; .
\ee

From (\ref{LorTraj}), the Lorentzian ETW brane trajectory in Schwarzschild coordinates for the second asymptotic region is given (in the case for $0 < T <1$) by
\be
t  = {1 \over r_H} {\rm arctanh} \left( \sqrt{r_H^2 - r^2(1-T^2)} \over T r_H \right) \; .
\label{etw-btz}
\ee

In the $u,v$ coordinates, we find that this becomes (setting $L=1$),
\be
T = {v - u \over \sqrt{1 + u^2} \sqrt{1 + v^2}} \; .
\ee
In the $s,y$ coordinates we get simply
\be
y = -\arcsin(T) \; .
\ee

\section{Imaginary time entanglement growth}
\label{app:entgrowth}

Imaginary time evolution can generate extremely rapid entanglement growth even if the Hamiltonian doesn't couple different degrees of freedom. This fact severely restricts any conceivable bound on entanglement growth under imaginary time dynamics.

Consider a decoupled Hamiltonian on $N$ spins of the form
\begin{equation}
    H= \sum_{r=1}^N \Delta \frac{1-\sigma^z_r}{2}.
\end{equation}
Spin up is identified with $0$ and spin down with $1$. The system is divided into two pieces, left $L$ and right $R$, with $N/2$ spins each.

Now define two states as follows. State one is the all down state, the highest energy state of $H$,
\begin{equation}
    |\psi_1\rangle = |1 \cdots 1\rangle.
\end{equation}
State two is an entangled Bell-type state obtained as an equal superposition of all states $|\psi_i \rangle_L \otimes |\psi_i \rangle_R$ where $|\psi_i \rangle$ is a product state with $S^z = 0$ (we assume $N/2$ is even). There are approximately $2^{N/2}$ such states (a significant fraction of the full left or right Hilbert space). Note that energy of state one is $N\Delta$ and the energy of state two is $N\Delta/2$.

The example is based on the superposition
\begin{equation}
    |\psi\rangle = \sqrt{1-\epsilon} |\psi_1 \rangle + \sqrt{\epsilon} |\psi_2\rangle,
\end{equation}
which can be prepared using a low depth quantum circuit. The entropy of $L$ or $R$ in this pure state is $N \epsilon/2$, so if $\epsilon$ is very small, then the entropy is very small. Now consider the imaginary time evolved state
\begin{equation}
    e^{-\beta H/2}|\psi\rangle.
\end{equation}
Up to an overall normalization, the effect is to exponentially re-weight states one and two in the superposition,
\begin{equation}
    e^{-\beta H/2} |\psi \rangle \propto \sqrt{1-\epsilon} |\psi_1 \rangle + \sqrt{\epsilon} e^{\beta N \Delta/2} |\psi_2 \rangle.
\end{equation}
The normalized state is
\begin{equation}
\frac{e^{-\beta H/2} |\psi \rangle}{\| e^{-\beta H/2} |\psi \rangle\|} = \sqrt{\frac{1-\epsilon}{1+ (e^{N \beta \Delta/2}-1)\epsilon}}|\psi_1\rangle +  \sqrt{\frac{\epsilon e^{N \beta \Delta/2}}{1+ (e^{N \beta \Delta/2}-1)\epsilon}} |\psi_2 \rangle.
\end{equation}
Hence the entropy as a function of $\beta$ is
\begin{equation}
    S = \frac{N}{2} \frac{\epsilon e^{N \beta \Delta/2}}{1+ (e^{N \beta \Delta/2}-1)\epsilon}.
\end{equation}
This formula yields extremely rapid entanglement growth; for example, if $\epsilon \sim 1/N$ so that the initial entanglement is of order a single bit, then the imaginary time evolution can generate $N$ bits of entanglement in an imaginary time of order $\frac{\ln N}{N \Delta}$.

If the ground state is also added to the superposition, then the entanglement depends on the relative size of the coefficients in the superposition. If the coefficients are roughly the same size, then the ground state will grow large much more rapidly than the middle energy states. In this case the entanglement may not ever become very large.

\section{Boundary states in a solvable model}
\label{app:sqrt}

By considering a simple model with a completely classical Hamiltonian, it is possible to rigorously establish some claims analogous to those made at large $N$ for the coupled SYK clusters.

Consider a classical Hamiltonian on $N$ qubits,
\begin{equation}
    H_c = \sum_{r,r'} J_{r,r'} \sigma^z_r \sigma^z_{r'},
\end{equation}
where classical means that the Hamiltonian is diagonal in a local product basis. One could add additional terms which are diagonal in the $\sigma^z_r$ basis without changing the subsequent story.

Now consider a generic product state $|x\rangle$ in the $\sigma^x_r$ basis. It obeys $\sigma^x_r | x\rangle = x_r | x\rangle$ with $x_r = \pm 1$. When expanding in the $z$ basis, these states are
\begin{equation}
    |x\rangle = \frac{1}{\sqrt{2^n}} \sum_z \left( \prod_r z_r^{\frac{1-x_r}{2}} \right) | z\rangle.
\end{equation}
Define the imaginary time-evolved states
\begin{equation}
    |x,\beta \rangle = e^{-\beta H_c /2} | x\rangle.
\end{equation}
The norm of these states is independent of $x$:
\begin{equation}
    \langle x,\beta | x,\beta\rangle = \sum_{z} \left( \prod_r z_r^{\frac{1-x_r}{2}} \right)^2 \langle z | e^{-\beta H_c } | z\rangle = Z_c(\beta),
\end{equation}
where $Z_c$ is the partition function associated with $H_c$. Similarly, one can show that any moment of $H_c$ in the state $|x,\beta \rangle$ is independent of $x$. More generally, any observable that is diagonal in the $\sigma^z_r$ basis has an expectation value in the state $|x,\beta\rangle$ that is independent of $x$ and given by the corresponding value in the classical statistical problem with weight $e^{-\beta \langle z |H_c | z\rangle}$.

Moreover, every state $|x,\beta \rangle $ is related to every other state $|x',\beta \rangle$ by a local unitary transformation. More precisely, we have
\begin{equation}
|x' ,\beta \rangle = \prod_{r=1}^N \left(\sigma^z_r \right)^{\frac{1-{x_r x'_r}}{2}} |x,\beta\rangle.
\end{equation}
This shows that every state $|x,\beta\rangle$ has the same entanglement for every spatial subregion independent of $x$. In particular, even though the states $|x,\beta\rangle$ need not be translation invariant, all the entanglement entropies are if the Hamiltonian $H_c$ is.

Finally, by tuning $\beta H_c$ to a classical statistical critical point or into an ordered phases, it follows that imaginary time evolution can generate long-range correlations after only a ``finite depth'' imaginary time evolution. This is in stark contrast to the situation with real time dynamics, in which long-range correlations must be established slowly starting from a short-range correlated state due to causality restrictions. In fact, in one dimension Araki has established an imaginary time analog of the Lieb-Robinson bound in which operators are allowed to expand exponentially fast~\cite{araki1969}.

\section{Details of the Action-Complexity Calculation}
\label{app:CAcalc}
As can be seen in figure \ref{fig:CAphases}, the Wheeler-DeWitt patch during each phase is defined by two null hypersurfaces, $N_{+}$ and $N_{-}$, anchored at the asymptotic boundary. Whether these null surfaces intersect the future/past singularity ($S_{+}$/$S_{-}$), or the ETW brane ($Q$), determines which phase is being considered. The problem of calculating the gravitational action on a region with boundaries is a well studied one (see \cite{Lehner:2016vdi} for a comprehensive review), and generically we will have terms corresponding to: the enclosed region, the region's boundaries, and the joints where boundaries meet non-smoothly. Here we breakdown each of these terms and state the results before and after the null boundary counter-term is included.

The first term that one must consider is the Einstein-Hilbert action evaluated on the Wheeler-DeWitt patch. In the $s,y$ coordinates this amounts to computing:
\bea
I_{EH} &=& \frac{1}{16 \pi G} \int_\mathcal{W} d^{d+1} x \sqrt{-g} (R - 2 \Lambda)  \cr
	&=& -\frac{r_H}{4 \pi G} \int_\mathcal{W} d s ~d y~ d \theta~ \sec^3(y) \cos(s)
\eea
This term diverges during all phases, since we are integrating all the way out to the asymptotic boundary. As such, a regulator surface $\Lambda$ is introduced to classify the divergence. In the the $s,y$ coordinates $\Lambda$ is the hypersurface defined by:
\be
\Lambda:~~y=\pi/2 - \delta
\ee
In the limit $\delta\to0$ we simply recover our asymptotic boundary. Another common cutoff method is to set the Schwarzschild radius to some maximum value, i.e.:
\be
\Lambda:~~r=\frac{l_\AdS}{\delta'}
\ee
Working with $l_\AdS = 1$, one can convert back and forth between the two cutoff schemes via the relation:
\be
\delta' = \frac{\sin(\delta)}{r_h \sech(r_H t_R)}
\ee

One may then ask what the contribution to the action is from this boundary $\Lambda$ itself. In general, a non-null boundary, $\mathcal{B}$, contributes a Gibbons-Hawking-York (GHY) term to the action:
\be
I_{GHY} = \frac{1}{8 \pi G} \int_\mathcal{B} d^{d} x \sqrt{|h|} K
\label{eq:appGHY}
\ee
Here, we must be careful to choose the orientation of each hypersurface consistently so that the relative sign of each action contribution is correct. For the hypersurface $\Lambda$, a unit one-form normal is chosen to be:
\be
\mathbf{n}_\Lambda = \frac{1}{\sin(\delta)} dy
\ee
Using this the extrinsic curvature is then calculated to be:
\be
K_\Lambda =2 \cos(\delta)
\ee
Solving for the induce metric on $\Lambda$ then putting this all into (\ref{eq:appGHY}) gives the action contribution:
\bea
I_\Lambda &=& \frac{r_H}{4 \pi G} \frac{\cot(\delta)}{\sin(\delta)} \int_\Lambda ds~ d\theta~ \cos(s) \cr
		&=& \frac{r_H}{\delta G} \sech(r_H t_R) + O(\delta)
\eea
This term is present during all three phases.

Next we will consider the contribution due to the ETW brane. The integration limits will be different depending on the phase, however the form of the action is always the same:
\be
I_Q = \frac{1}{8 \pi G} \int_Q d^{d} x \sqrt{|h|} ( K_Q - T )
\ee
This corresponds to the GHY term for the hypersurface plus a matter action. Here, a simplistic matter action for the brane is considered, with the matter Lagrangian being assumed to be a constant parametrized by the brane tension $T$ (this follows the approach outlined in \cite{Fujita:2011fp}) The unit normal one-form for $Q$ is chosen to be:
\be
\mathbf{n}_Q = - \frac{1}{\sqrt{1-T^2}} d y
\ee
Solving for the extrinsic curvature and induced metric we find:
\be
I_Q = \frac{r_H}{8 \pi G} \frac{T}{1-T^2} \int_Q ds ~ d\theta~\cos(s)
\ee

The only remaining non-null hypersurfaces to consider are the past and future singularities at $s=\pm \frac{\pi}{2}$. Calculating the contribution here slightly tricky: the induced metric on $S_\pm$ vanishes and the extrinsic curvature $K_\pm$ diverges. However, if we instead considers a hypersurfaces at $s=\text{constant}$ then we can compute the integrate explicitly. When doing this, one finds that in the limit $s \to \pm \frac{\pi}{2}$ the measure and extrinsic curvature actually combine to give a finite, regulator independent, integrand. The unit normal one-forms to the singularities are chosen to be:
\be
\mathbf{t}_\pm = \sec(y) ds
\ee
The measure for a constant $s$ surface is
\be
\sqrt{|h|} = r_H \cos(s)\sec^2(y)
\ee
and the extrinsic curvature is:
\be
K_\pm = \pm \tan(s) \cos(y)
\ee
Note that the sign difference here is due to the orientation of $S_\pm$. Combining this together, and taking the limit $s \to \pm \frac{\pi}{2}$, we write our GHY term for each singularity respectively as:
\be
I_{S_\pm} =  \frac{r_H}{8 \pi G}  \int_{S_\pm} d y~ d\theta~ \sec(y)
\ee
This corresponds to a total contribution during phase ii of:
\be
\text{Phase ii:} ~~I_{S_{+}} + I_{S_{-}} =  \frac{r_H}{2 G} \arctanh(T)
\ee
During phase iii the contribution is
\be
\text{Phase iii:} ~~I_{S_{+}}  =  \frac{r_H}{4 G} \left( r_H t_R + \arctanh(T) \right)
\ee
Notice that this calculation did not take into account any nonclassical effects. One might expect the divergences coming from the introduction of higher order curvature terms not to cancel away here. These stringy corrections have not been considered here, however in principle on could introduce a regulator surface in the same manner done for the asymptotic boundary in order to classify these divergences.\footnote{Some related calculations can be found in \cite{Cai:2016xho}, wherein the Gauss-Bonnet-AdS black hole is considered.}

We now move onto the discussion of the null hypersurfaces $N_+$ and $N_-$. These surfaces are defined by the equations:
\bea
N_+:~ s &=& -y + 2 \arctan\left(e^{r_H t_R}\right) \cr
N_-:~  s &=& +y - 2 \arccot\left(e^{r_H t_R}\right)
\eea
The null normal one-forms for these surfaces are chosen to be:\footnote{For brevity, we omit the derivations of these quantities. A thorough examination of null hypersurfaces can be found in \cite{poisson2004relativist}.}
\be
\mathbf{k}_\pm = \alpha_\pm ( \pm d s + d y)
\ee
Here, $\alpha_+$ and $\alpha_-$ are normalization constants. We also endow each null hypersurface with coordinates ($\lambda_\pm, \theta$), where $\theta$ is the angular BTZ coordinate and $\lambda_\pm$ is given by:
\be
\lambda_\pm = \frac{1}{\alpha_\pm} \tan(y)
\ee
Altogether, this constitutes an affine parametrization for the null hypersurfaces. I.e., they solve the affine geodesic equation:
\bea
k_{\alpha; \beta} k^\beta &=& \kappa k_\alpha \cr
					&=& 0
\eea
Thus, we see that for this parametrization the constant $\kappa=0$. The boundary term for a null hypersurface is typically given by:
\be
I_{N_\pm} = - \frac{1}{8 \pi G} \int_{N_\pm} d \lambda d^{d-1}\theta \sqrt{\gamma} \kappa
\ee
However, since we have chosen an affine parametrization this contribution vanishes.

Next we consider the joints between each of these boundary surfaces. In principle we have joints where $N_\pm$ intersect $S_\pm$, $Q$ and $\Lambda$, as well as non-null joints (of the type proposed in \cite{Hayward:1993my}) at $S_\pm \cap Q$. However, one finds that the joint terms at $S_\pm \cap Q$ and at $N_\pm \cap S_\pm$ all vanish. The only non-zero joint terms are from intersections of the null surfaces with the regulator surface and the ETW brane. These are joints between null and timelike hypersurfaces and so correspond to action contributions of the form:
\bea
I_{\text{joints}} &=& \frac{1}{8 \pi G} \int_\Sigma d^{d-1}x \sqrt{\sigma} a \cr
a &=& \epsilon \ln | \mathbf{k} \cdot \mathbf{n} | \cr
\epsilon &=& - \sign(\mathbf{k} \cdot \mathbf{n}) \sign( \mathbf{k} \cdot \hat{t})
\eea
Here $\mathbf{k}$ and $\mathbf{n}$ are the normal one-forms to the null and timelike surfaces respectively, and $\hat{t}$ is some auxiliary unit vector tangent to the timelike hypersurface. $\Sigma$ is the co-dimension two hypersurface that is the intersection between the two boundaries. Computing the contributions for $N_\pm \cap \Lambda$, one finds that in all phases we have:
\be
I_{N_+ \cap \Lambda} + I_{N_- \cap \Lambda} =\frac{r_H}{4 G} \bigg\{ \frac{2 \sech(r_H t_R)}{\delta} \ln\left(\frac{1}{\sqrt{\alpha_+ \alpha_-} \delta}\right)
					+  \tanh(r_h t_R) \ln\left(\frac{\alpha_-}{\alpha_+}\right)+ O(\delta) \bigg\}
\ee
Similarly, one can compute the action contribution for the intersections $N_\pm \cap Q$. These turn out to be:
\be
I_{N_\pm \cap Q} = - \frac{r_H}{4 G} \ln(\alpha_\pm \sqrt{1-T^2})  \left(\pm\tanh(r_H t_R) +\frac{T}{\sqrt{1-T^2}} \sech(r_h t_R)\right)
\ee
Where the term for $N_+ \cap Q$ is only present during phase i and the term for  $N_+ \cap Q$ only appears in phase iii.

Unfortunately, if we were to combine together all of the terms above we would find that the resulting action is dependent on $\alpha_+$ and $\alpha_-$. This isn't ideal as the quantity we find is not invariant under different choices of the parametrization of each null surface. Recently, it has been suggested that a counter-term be introduced to the gravitational action in order to cancel this dependence on $\alpha_+$ and $\alpha_-$:\footnote{This counter-term was first proposed in \cite{Lehner:2016vdi} and has since been discussed throughout the literature. Some more thorough exploration of this counter-term can be found in \cite{Reynolds:2016rvl} and \cite{Parattu:2015gga}.}
\bea
I_{counter} &=& - \frac{1}{8 \pi G} \int_\mathcal{B} d\lambda~ d^{d-1} \theta \sqrt{\gamma} \Theta \ln|L \Theta| \cr
\Theta &=&  \frac{1}{\sqrt{\gamma}} \frac{\partial \gamma}{\partial \lambda}
\eea
Where we introduce such a term for each null boundary $\mathcal{B}$. Here, $\gamma$ corresponds to the null hypersurface's metric. Just as the complexity$=$volume conjecture was only defined up to some relative length scale, this counter-term depends on an arbitrary length scale $L$. For the purposes of this analysis, we will simply choose to set $L=L_\AdS = 1$.\footnote{Some easy to interpret graphs are provided in appendices of \cite{Carmi:2017jqz} that show the effects of changing the value of this length scale.} For $N_+$ the counter-term takes the form:
\bea
I_{N_+} &=& - \frac{r_H}{4 G} \alpha_+ \sech(r_H t_R) \int_{N_+} d \lambda~ \ln \bigg| \frac{\alpha_+ \sech(r_H t_R)}{\alpha_+ \sech(r_H t_R) \lambda -\tanh(r_H t_R)}\bigg| \cr
	&=& \left. \frac{r_H}{4 G} \sech(r_H t_R) \left(\sinh(r_H t_R) - \alpha_+ \lambda \right) \left(1 + \ln\bigg|\frac{\alpha_+}{\sinh(r_H t_R) - \alpha_+ \lambda}\bigg| \right) \right\vert^{\lambda_f}_{\lambda_i}
\eea
Where, $\lambda_i = N_+ \cap \Lambda$ during every phase, and $\lambda_f = N_+ \cap Q$ during phase i or $N_+ \cap S_+$ otherwise. Similarly, the counterterm for $N_-$ can be calculated using:
\bea
I_{N_-} &=&  - \frac{r_H}{8 G} \alpha_- \sech(r_H t_R) \int_{N_-} d \lambda~ \ln \bigg|\frac{\alpha_- \sech(r_H t_R)}{\alpha_- \sech(r_H t_R)\lambda +\tanh(r_H t_R)}\bigg| \cr
&=&		- \left. \frac{r_H}{4 G} \sech(r_H t_R) \left(\sinh(r_H t_R) + \alpha_- \lambda \right) \left(1 + \ln\bigg|\frac{\alpha_-}{\sinh(r_H t_R) + \alpha_- \lambda}\bigg| \right) \right\vert^{\lambda_f}_{\lambda_i}
\eea
With  $\lambda_i = N_- \cap \Lambda$ during each phase, and $\lambda_f$ being $N_-\cap Q$ during phase iii or $N_-\cap S_-$ otherwise. Both of these integrals result in many terms, so we will refrain from including them here.

With all of the individual contributions to the action in place, all that remains is to combine them all together in accordance with the phases depicted in figure \ref{fig:CAphases} and use equation (\ref{eq:CAdef}) to calculate the complexity. In doing this, many, many terms cancel, resulting in the simple expressions stated in equations (\ref{eq:CAphaseii}) and (\ref{eq:CAphaseiii}) (the phase iii result is stated with the divergence already subtracted).

\bibliographystyle{JHEP}
\bibliography{BHcosmo}

\end{document}